\documentclass[a4paper,11pt]{article}

\usepackage{amssymb,amsmath,hyperref}
\hypersetup{
	colorlinks=true,
	linktoc=page,
	citecolor=blue,
	linkcolor=blue,
	urlcolor=blue} 
\usepackage[inner=2cm,outer=2cm,top=2cm,bottom=2cm]{geometry}
\usepackage{graphicx}

\usepackage{cite} 

\usepackage{tikz}
\usetikzlibrary{arrows.meta} 

\usepackage{enumitem}

\newcommand{\Deq}{\overset{\mathrm{D}}{\simeq}}

\newcommand{\agl}[2]{\langle#1 #2 \rangle}
\newcommand{\sqr}[2]{\lbrack #1 #2 \rbrack}
\newcommand{\tlambda}{\widetilde{\lambda}}

\begin{document}


\begin{flushright}
	QMUL-PH-22-05\\
	SAGEX-22-18-E\\
\end{flushright}

\vspace{20pt}

\begin{center}

	{\Large \bf Amplitude bases in generic EFTs}

	\vspace{25pt}

	{\mbox {\sf Stefano~De~Angelis}}
	\vspace{0.5cm}

	\begin{center}
		{\small \em
			Centre for Theoretical Physics\\
			Department of Physics and Astronomy\\
			Queen Mary University of London\\
			Mile End Road, London E1 4NS, United Kingdom
		}
	\end{center}

	\vspace{40pt}

	{\bf Abstract}

\end{center}

\vspace{0.3cm}

\noindent
We present for the first time an efficient algorithm to find a basis of kinematically independent structures built of (massless and massive) spinor helicity variables in four dimensions. This method provides a classification of independent contact terms for the scattering amplitudes with generic masses, spins, and multiplicity in any effective field theory (EFT). These contact terms are in one-to-one correspondence with a complete set of irrelevant operators in the EFT. As basic applications of our method, we classify the $D^{2n} F^4$ contact terms in SU$(N)$ Yang-Mills theory for $n\leq 8$, dimension-six operators involving five $W^\pm$, $Z$ and $\gamma$ vector bosons, and spin-tidal effective interactions for spin-1 massive particles in gravitational theories.

\vfill
\hrulefill
\newline
\vspace{-1cm}
{\tt\footnotesize s.deangelis@qmul.ac.uk}

\setcounter{page}{0}
\thispagestyle{empty}
\newpage

\tableofcontents


\section{Introduction}

On-shell methods \cite{Dixon:1996wi,Dixon:2013uaa,Elvang:2013cua,Henn:2014yza,Cheung:2017pzi} have proved to be extremely powerful in the study of effective field theories (EFTs). Among other applications, in recent years, they appeared as prominent tools in the study of Standard Model Effective Field Theories (SMEFT) and gravitational binary systems.

For instance, such methods have been applied to compute the anomalous dimension in generic EFTs \cite{Caron-Huot:2016cwu,EliasMiro:2020tdv,Baratella:2020lzz,Jiang:2020mhe,Bern:2020ikv,Baratella:2020dvw,AccettulliHuber:2021uoa,Baratella:2021guc,EliasMiro:2021jgu}, to understand the patterns of zeros in the one-loop anomalous dimension matrix for dimension-six operators in the SMEFT \cite{Cheung:2015aba,Bern:2019wie,Jiang:2020rwz} and Wilson coefficients when integrating out heavy modes \cite{Rose:2022njd}, and the (non-)interference of tree-level SMEFT and SM amplitudes \cite{Azatov:2016sqh}. In the study of gravitational binary systems, scattering amplitude techniques allowed to compute the conservative two-body scattering dynamics up to $\mathcal{O}(G^4)$ in all orders in velocity \cite{Neill:2013wsa,Bjerrum-Bohr:2013bxa,Cachazo:2017jef,Cheung:2018wkq,Kosower:2018adc,Bern:2019nnu,Bern:2019crd,Parra-Martinez:2020dzs,Bjerrum-Bohr:2021din,Brandhuber:2021eyq,Bern:2021dqo}, and to lower orders in $G$ including spin effects \cite{Guevara:2017csg,Cheung:2018wkq,Chung:2019duq,Maybee:2019jus,Bern:2020buy,Kosmopoulos:2021zoq,Chiodaroli:2021eug,Aoude:2021oqj,Haddad:2021znf}, tidal effects \cite{Cheung:2020sdj,Haddad:2020que,Bern:2020uwk,Kim:2020dif,Aoude:2020ygw} and higher-derivative interactions \cite{Brandhuber:2019qpg,AccettulliHuber:2020oou}.

In the context of EFTs at weak coupling, the on-shell programme provides us with an alternative to the standard Lagrangian EFTs, where the information of the theory beyond the IR physics is encoded by adding irrelevant operators. In particular, it exists a one-to-one correspondence between irrelevant independent operators and minimal amplitudes, \textit{i.e.} terms in the low-energy expansion of the S-matrix which are polynomial in the kinematic variables (and are linear in the Wilson coefficients). This means that from the S-matrix perspective the classification and enumeration of independent operators \cite{Grzadkowski:2010es,Lehman:2015via,Henning:2015alf,Henning:2015daa,Henning:2017fpj} are equivalent to finding a basis of kinematically independent polynomial structures in the amplitudes \cite{Cheung:2016drk,Shadmi:2018xan}. From these ``building blocks'', higher-point tree-level amplitudes can be constructed from knowledge of factorisation channels, using recursion relations \cite{Britto:2005fq,Risager:2005vk,Arkani-Hamed:2008bsc,Cohen:2010mi,Cheung:2015ota,Falkowski:2020aso,AccettulliHuber:2021uoa}, and loop-level amplitudes through generalised unitarity techniques \cite{Bern:1994cg,Bern:1994zx,Britto:2004nc,Bern:2004cz,Mastrolia:2008jb,Forde:2007mi,Badger:2008cm}.

The classification programme was carried out mainly in four dimensions where we take advantage of the simplicity of the massless \cite{DeCausmaecker:1981jtq,Berends:1981uq,Kleiss:1985yh,Xu:1986xb} and massive \cite{Arkani-Hamed:2017jhn} spinor helicity formalism. In particular, recent works on various aspects of this classification and applications to the SMEFT are \cite{Ma:2019gtx,Aoude:2019tzn,Durieux:2019eor,Chowdhury:2019kaq,Falkowski:2019zdo,Durieux:2019siw,Chakraborty:2020rxf,Li:2020gnx,Li:2020xlh,Durieux:2020gip,Falkowski:2020fsu,Li:2022tec,Balkin:2021dko}. Generic techniques to classify fully massless interactions have been worked out in these works, while a general method for massive particles with any mass, spin, and multiplicity is still lacking. The authors of \cite{Durieux:2019eor,Durieux:2020gip} introduced a \textit{unbolding/bolding} procedure to classify massive structures from their massless counterparts, and they presented a strategy to tackle this problem for four-point contact terms and particles with spin $S\leq 1$. An alternative strategy valid for any multiplicity and any spin has been presented in \cite{Dong:2021yak}, but this method does not allow for a clear one-to-one correspondence between independent structures and operators.

In this paper, we present an alternative method to classify generic contact terms in four-dimension, for any mass, spin and multiplicity $n$ (with $n\geq 4$, the classification for $n=3$ is known \cite{Benincasa:2007xk,Costa:2011mg,Arkani-Hamed:2017jhn,Durieux:2019eor}), starting from the graph method presented in \cite{AccettulliHuber:2021uoa}, strengthen by the \textit{unbolding/bolding} procedure of \cite{Durieux:2019eor,Durieux:2020gip} which we implement at the level of graphs. In this algorithm, structures involving powers of any mass are always regarded kinematically independent of structures with fewer powers of mass. This allows for a clear one-to-one correspondence between independent contact terms and operators in massive EFTs. In the following, we skip an introductory section on the correspondence between operators and contact terms, for which we refer to the papers mentioned above.

The paper is organised as follows. In Section~\ref{sec:massless} we review the graph method introduced in \cite{AccettulliHuber:2021uoa}, giving more details on the algorithm and its implementation. In Section~\ref{sec:massive} we will extend the method to structures involving massive spinor variables. In Section~\ref{sec:applications}, we work out explicitly three examples: the classification of $D^{2n} F^4$ operators in SU$(N)$ Yang-Mills theories, dimension-six interactions between (charged and uncharged) massive and massless vectors, and quadratic-in-spin (spin-)tidal interactions in gravity. Finally, in Appendix~\ref{sec:spinorhelicity} we review the spinor helicity formalism and fix our conventions, and in Appendix~\ref{sec:massivetwistors} we present a strategy to evaluate (massive) spinor structures on rational kinematics, based on momentum twistors. Appendix~\ref{sec:intoplanar} shows some technical details on the algorithms for planar graphs.

The algorithms presented in the article have been implemented in the \texttt{Mathematica} package \href{https://github.com/StefanoDeAngelis/MassiveEFT-Operators}{\texttt{MassiveGraphs}}, which makes use of the \href{https://github.com/StefanoDeAngelis/SpinorHelicity}{\texttt{SpinorHelicity}} package. Some of the functions are already publicly available and an \href{https://github.com/StefanoDeAngelis/MassiveEFT-Operators/blob/main/Example.nb}{\texttt{Example.nb}} notebook has been provided.

\section{The massless basis}
\label{sec:massless}

\subsection{Kinematic structures from graphs}
\label{sec:kinematics}

A simple way to find all possible Lorentz invariant structures in four dimensions is to identify them with an oriented graph with two types of edges, where each vertex is associated with a particle, and the edges correspond to angles (red) or squares (blue) ${\rm SL}(2,\mathbb{C})$ invariants, as shown in Figure~\ref{fig:2graphexample}\footnote{This figure have been taken from reference \cite{AccettulliHuber:2021uoa} and modified.}. The orientation of the edges then keeps track of the ordering in the spinor brackets and thus provides potential minus signs.

\begin{figure}[t]
	\begin{center}
		\begin{tikzpicture}[scale=2.5,>=Stealth]

			\node at (-0.05,-0.05) {$4$};
			\node at (-0.05,1.05) {$1$};
			\node at (1.05,1.05) {$2$};
			\node at (1.05,-0.05) {$3$};

			\draw [fill] (0,0) circle [radius = 0.02];
			\draw [fill] (1,0) circle [radius = 0.02];
			\draw [fill] (0,1) circle [radius = 0.02];
			\draw [fill] (1,1) circle [radius = 0.02];

			\draw [thick,<-,blue](0,0) -- (0,1);

			\draw [thick,->,blue](0,1) to [out=-10,in=-170] (1,1);
			\draw [thick,->,blue](0,1) to [out=10,in=-190] (1,1);

			\draw [thick,<-,blue](1,1) to [out=-100,in=100] (1,0);
			\draw [thick,->,red](1,1) to [out=-80,in=80] (1,0);

			\draw [thick,<-,red](0,1) -- (1,0);
			\draw [thick,<-,red](0,1) to [out=-55,in=145] (1,0);
			\draw [thick,<-,red](0,1) to [out=-35,in=125] (1,0);

		\end{tikzpicture}
		\hspace{1cm}
		\begin{tikzpicture}[scale=1.6,>=Stealth]

			\node at (-0.08,-0.08) {$5$};
			\node at (-0.08,1.08) {$1$};
			\node at (0.95,1.43) {$2$};
			\node at (1.58,0.5) {$3$};
			\node at (0.95,-0.43) {$4$};

			\draw [fill] (0,0) circle [radius = 0.02];
			\draw [fill] (0,1) circle [radius = 0.02];
			\draw [fill] (0.95,1.3) circle [radius = 0.02];
			\draw [fill] (1.45,0.5) circle [radius = 0.02];
			\draw [fill] (0.95,-0.3) circle [radius = 0.02];

			\draw [thick,->,blue](0.95,1.3) -- (0,0);

			\draw [thick,->,blue](0.41,0.45) -- (0.95,-0.3);
			\draw [thick,blue](0,1) -- (0.315,0.56);
			\draw [thick,blue] (0.41,0.45) arc (-60:150:0.08);
			\draw [thick,->,blue](0.66,0.77) -- (1.45,0.5);
			\draw [thick,blue](0,1) -- (0.51,0.82);
			\draw [thick,blue] (0.66,0.77) arc (-30:165:0.08);

			\draw [thick,->,red](0.95,1.3) -- (1.45,0.5);
			\draw [thick,->,red](1.45,0.5) -- (0.95,-0.3);

		\end{tikzpicture}
	\end{center}
	\caption{The graph associate to the kinematic structures $\textcolor{red}{\agl{2}{3}\agl{3}{1}^3} \textcolor{blue}{\sqr{1}{2}^2 \sqr{1}{4} \sqr{3}{2}}$ and $\textcolor{red}{\agl{2}{3}\agl{3}{4}} \textcolor{blue}{\sqr{1}{3} \sqr{1}{4} \sqr{2}{5}}$ respectively.}
	\label{fig:2graphexample}
\end{figure}

The valency of each vertex is given by two natural numbers $v^{i}=(v^i_a,v^{i}_s)$ such that $v^i_s-v^i_a =2 h_i$ is the helicity of the $i^{\rm th}$ particle. In general, we consider polynomial structures with an arbitrary number of momentum insertions $n_\partial \geq 0$. Each momentum in the structure can be assigned to any of the $n$ states, which increases the valency of the corresponding vertex by $(1,1)$. Then the number of momenta associated with each vertex is $\min\{v^i_a,v^i_s\}$. Moreover, for reasons which will become clear in the next section, it is crucial to consider a circular embedding for our graphs, \textit{i.e.} we take all the vertices to be ordered points on a circle.

We can associate to each graph a couple of $n\times n$ adjacency matrices $(\mathbf{A},\mathbf{S})$, whose elements are non-negative integer numbers: each element $A_{i j}\geq 0$ (or $S_{i j}\geq 0$) indicates the number of red (or blue) edges going from the $i^{\rm th}$-vertex to the $j^{\rm th}$-vertex. Finally, there is a trivial map $\mathbb{M}$ from the adjacency matrices to the corresponding polynomial spinor structure:
\begin{equation}
	\label{eq:graphTOspinors}
	\mathbb{M}(\mathbf{A},\mathbf{S}) = \prod_{i,j=1}^n\, \agl{i}{j}^{A_{i j}}\, \sqr{i}{j}^{S_{i j}}\ .
\end{equation}
This map is in general non-invertible, because the spinor brackets are antisymmetric ($\agl{i}{j}=-\agl{j}{i}$, $\sqr{i}{j} = - \sqr{j}{i}$). Then we restrict without lost of generality to upper-half triangular adjacency matrices ($A_{i j} = 0$ and $S_{i j}= 0$ if $i \geq j$), in order to make the correspondence between polynomial structures and graphs one-to-one.

At this point, we are interested in finding a basis of structures that are independent up to Schouten identity and momentum conservation. Notice that the former acts separately on the angle and square invariants, while the latter mixes the two structures. In the following sections, we will show how to deal with these identities in terms of the graphs mentioned above.

\subsection{Schouten identities}

Schouten identities for angle and square brackets read
\begin{equation}\label{eq:Schouten}
	\begin{aligned}
		\agl{i_1}{i_2}\agl{i_3}{i_4} + \agl{i_2}{i_3}\agl{i_1}{i_4} + \agl{i_3}{i_1}\agl{i_2}{i_4} & = 0\ , \\
		\sqr{i_1}{i_2}\sqr{i_3}{i_4} + \sqr{i_2}{i_3}\sqr{i_1}{i_4} + \sqr{i_3}{i_1}\sqr{i_2}{i_4} & = 0\ .
	\end{aligned}
\end{equation}
Thinking of the kinematic structures in terms of graphs, specifically using the circular embedding already mentioned, if $(i_1,i_2,i_3,i_4)$ are arranged in cyclic order, we notice that $\agl{i_1}{i_3}\agl{i_2}{i_4}$ correspond to intersecting edges in the graph, while $\agl{i_1}{i_2}\agl{i_3}{i_4}$ and $\agl{i_2}{i_3}\agl{i_1}{i_4}$ do not (the same is true for the square brackets). This is illustrated in Figure~\ref{fig:Schouten} \footnote{This figure has been taken from reference \cite{AccettulliHuber:2021uoa} and modified.}. In a generic graph, this relation can be applied recursively a finite number of times until we end up with a sum over graphs that do not have any crossings. It is then clear that a basis for kinematic structures that are independent under Schouten identity can be obtained by classifying all \textit{planar graphs} associated with a polynomial kinematic structure.

\begin{figure}[t]
	\begin{center}
		\begin{tikzpicture}[scale=1.8,>=Stealth]

			\node at (-0.08,-0.08) {$4$};
			\node at (-0.08,1.08) {$1$};
			\node at (1.08,1.08) {$2$};
			\node at (1.08,-0.08) {$3$};
			\node at (1.25,0.5) {$=$};

			\draw [fill] (0,0) circle [radius = 0.02];
			\draw [fill] (1,0) circle [radius = 0.02];
			\draw [fill] (0,1) circle [radius = 0.02];
			\draw [fill] (1,1) circle [radius = 0.02];

			\draw (1,0) [<-,thick,blue] -- (0.55,0.45);
			\draw [thick, blue] (0.45,0.55) arc (135:-45:0.07);
			\draw [thick, blue] (0.45,0.55) -- (0,1);
			\draw [->,thick,blue](1,1) -- (0,0);

		\end{tikzpicture}
		\begin{tikzpicture}[scale=1.8,>=Stealth]

			\node at (-0.08,-0.08) {$4$};
			\node at (-0.08,1.08) {$1$};
			\node at (1.08,1.08) {$2$};
			\node at (1.08,-0.08) {$3$};
			\node at (1.25,0.5) {$+$};

			\draw [fill] (0,0) circle [radius = 0.02];
			\draw [fill] (1,0) circle [radius = 0.02];
			\draw [fill] (0,1) circle [radius = 0.02];
			\draw [fill] (1,1) circle [radius = 0.02];

			\draw [<-,thick,blue](1,0) -- (1,1);
			\draw [->,thick,blue](0,1) -- (0,0);

		\end{tikzpicture}
		\begin{tikzpicture}[scale=1.8,>=Stealth]

			\node at (-0.08,-0.08) {$4$};
			\node at (-0.08,1.08) {$1$};
			\node at (1.08,1.08) {$2$};
			\node at (1.08,-0.08) {$3$};

			\draw [fill] (0,0) circle [radius = 0.02];
			\draw [fill] (1,0) circle [radius = 0.02];
			\draw [fill] (0,1) circle [radius = 0.02];
			\draw [fill] (1,1) circle [radius = 0.02];

			\draw [->,thick,blue](0,1) -- (1,1);
			\draw [->,thick,blue](1,0) -- (0,0);

		\end{tikzpicture}
	\end{center}
	\caption{Graphical representation of the relation $\textcolor{blue}{\sqr{1}{3}\sqr{2}{4}} = \textcolor{blue}{\sqr{1}{4}\sqr{2}{3}} + \textcolor{blue}{\sqr{1}{2}\sqr{3}{4}}$. Then Schouten identities are equivalent to untying crossings for both the two kinds of edges (red and blue) in the graph.}
	\label{fig:Schouten}
\end{figure}

The planarity of the graph translates into a sharp condition on the adjacency matrices associated to it:
\begin{equation}
	\label{eq:planarity}
	\mathrm{if} \ A_{i j} \neq 0\ , \qquad
	A_{k l} \overset{!}{=} 0 \ \ \mathrm{for} \ \ i+1\leq k \leq j-1 ,\ j+1\leq l \leq n\ ,
\end{equation}
for any $i=1,\dots , n$ and $j= i+2,\dots ,n-1$.
In particular, the number of crossings of a graph is given by the sum
\begin{equation}
	n_\times = \sum_{i=1}^{n-1} \sum_{j=i+1}^{n} A_{i j} \sum_{l=j+1}^n \sum_{k=i+1}^{j-1} A_{k l}\ ,
\end{equation}
where the extrema of the sums are consistent with the restriction to upped-half triangular adjacency matrices.

This also provides an algorithmic method to write any non-planar structure in terms of planar ones. Indeed, if we consider a graph $(\mathbf{A},\mathbf{S})$ for which, for example, the red edges are non-planar, \textit{i.e.} if in \eqref{eq:planarity} any of the $A_{k l} \neq 0$, then we can recursively untie the corresponding crossing(s) using
\begin{equation}
	\label{eq:untyingcrossing}
	\mathbb{M}(\mathbf{A},\mathbf{S}) = \mathbb{M}(\mathbf{A}+\mathbf{E}^{(i\,j)}_{(k\,l)},\mathbf{S}) + \mathbb{M}(\mathbf{A}+\mathbf{F}^{(i\,j)}_{(k\,l)},\mathbf{S})\ ,
\end{equation}
where
\begin{align}
	E^{(i\,j)}_{(k\,l),\, a b} & = - \delta_{a,i} \, \delta_{b,j} - \delta_{a,k} \, \delta_{b,l} + \delta_{a,i} \, \delta_{b,k} + \delta_{a,j} \, \delta_{b,l}
	\, ,\\
	F^{(i\,j)}_{(k\,l),\, a b} & = - \delta_{a,i} \, \delta_{b,j} - \delta_{a,k} \, \delta_{b,l} + \delta_{a,i} \, \delta_{b,l} + \delta_{a,j} \, \delta_{b,k}
	\, .
\end{align}
Applying such decomposition a finite number of times, every non-planar structure can be written as a linear combination of planar ones. A proof of this statement is given in Appendix~\ref{sec:intoplanar}.

This algorithm can be used to find a basis of $\mathrm{SU}(2)$ singlets in the tensor product of any finite-dimensional representation of the $\mathrm{SU}(2)$ group. In fact, any tensor that transforms in a representation $\mathbf{2\, q+1}$ can be written as a totally symmetric tensor $T^{a_1\, \cdots \, a_q} = T^{(a_1\, \cdots \, a_q)}$, where $a_i$ are indices in the fundamental of $\mathrm{SU}(2)$. The singlets are given by contractions of any product of tensors of this kind with $\epsilon_{a_1 a_2} = - \epsilon_{a_2 a_1}$ and the Schouten identities are equivalent to
\begin{equation}
	\epsilon_{[a_1 a_2} \epsilon_{a_3] a_4} = 0\ .
\end{equation}
Then to any tensor $T$ we can associate a vertex with valency $q$, and the edges correspond to contractions of two $\mathrm{SU}(2)$ indices through an $\epsilon_{a_i a_j}$. Any loop is then automatically zero because
\begin{equation}
	T^{a_1 \cdots a_i \cdots a_j \cdots a_n} \epsilon_{a_i a_j} = 0\ .
\end{equation}
This observation is useful if we want to apply our method to select a basis of independent $\mathrm{SU}(2)$ gauge structures (see for example \cite{AccettulliHuber:2021uoa}), or when we will consider polynomial structures with masses involved because the little group for massive particles in four dimensions is exactly $\mathrm{SU}(2)$. This could also be applied to the Lorentz group $\mathrm{SL}(2,\mathbb{C})$, because the finite-dimensional representations are in one-to-one correspondence with those of $\mathrm{SU}(2)\times \mathrm{SU}(2)$.

\subsection{Momentum conservation}

Momentum conservation is more subtle and does not have a clear graph-based interpretation. On the other hand, the classification above allows for a massive simplification and the conditions to find a basis of spinor structures independent, up to both Schouten identity and momentum conservation, are easy to implement.

We can take into account most of the relations coming from momentum conservation simply by excluding the momentum of the $n^{\rm th}$-particle from the previous assignment. Then the $n^{\rm th}$-vertex will have valency $(\frac{|h_n|+h_n}{2},\frac{|h_n|-h_n}{2})$\footnote{Fully eliminating the momentum of the $n^{\rm th}$-particle is a matter of choices and in principles we could choose any other particle.}. This is equivalent to imposing the constraint
\begin{equation}
	\label{eq:momentumN}
	p_n = - \sum_{i = 1}^{n-1} p_i\ ,
\end{equation}
and we can \textit{discard} from our basis any graph whose adjacency matrix \textit{does not} satisfy the conditions
\begin{equation}
	\label{eq:momcons1adjacency}
	A_{i n} = 0 \ \ \ \ \mathrm{or}\ \ \ \ S_{j n} = 0\ ,
\end{equation}
for any $i,j=1,\dots , n-1$.

However, this is not the end of the story, because there are in general $n$ additional momentum conservation conditions that do not involve any insertions of the momentum of the $n^{\rm th}$ particles:
\begin{align}
	\label{eq:momcons2}
	0 =
	\begin{cases}
		\sum\limits_{j=1}^{n-1} \agl{i}{j} \sqr{j}{n} & h_n>0 \\[.5em]
		\sum\limits_{j=1}^{n-1} \agl{n}{j} \sqr{j}{i} & h_n<0 \\
	\end{cases}
\end{align}
which are a consequence of the equation of motion for free particles $p_{n\, \alpha \dot{\alpha}}\, \tlambda_{n}^{\dot{\alpha}}=\, 0 \, = \lambda^{\alpha}_n\, p_{n\, \alpha \dot{\alpha}}$. If $h_n=0$ the valency of the $n^{\rm th}$ vertex is $(0,0)$ and there is only one additional condition to consider:
\begin{equation}
	\label{eq:momcons1}
	\left(\sum_{i = 1}^{n-1} p_{i \, \alpha \dot{\alpha}}\right)^2 =\sum_{i=1}^{n-2} \sum_{j=i+1}^{n-1} s_{i j} = p_n^2 = 0\ .\\
\end{equation}

As already noticed in the previous section, Schouten identities do not change the valency of vertices in the multigraph, so they do not change the number of momenta associated with each vertex. Then, we have to find a set of elements in our planar basis which can be written as a linear combination of the others via momentum conservation. Once we have discarded all the polynomial structures in which we find the momentum of the $n^{\rm th}$-particle, we need to carefully discard the structures that \textit{maximise} their appearance in conditions \eqref{eq:momcons2} and \eqref{eq:momcons1}. Since the edges $(1,n)$ and $(n-1,n)$ are always planar, the natural choice is to isolate terms where either $p_1$ or $p_{n-1}$ appears
and to write the additional momentum conservation conditions as
\footnote{
This is an actual choice between momenta of the $1^{\mathrm{st}}$ and the $n-1^{\mathrm{th}}$ momenta. We could choose an equivalent basis by writing \eqref{eq:momcons2} as
\begin{align*}
	\begin{array}{ll}
		\begin{cases}
			\agl{i}{1} \sqr{1}{n} = - \sum\limits_{j=2}^{n-1} \agl{i}{j} \sqr{j}{n} \\
			\agl{1}{(n-1)} \sqr{(n-1)}{n} = - \sum\limits_{j=2}^{n-2} \agl{1}{j} \sqr{j}{n}
		\end{cases} & h_n>0\ , \\[1.5em]
		\begin{cases}
			\agl{n}{1} \sqr{1}{i} = - \sum\limits_{j=2}^{n-1} \agl{n}{j} \sqr{j}{i} \\
			\agl{n}{(n-1)} \sqr{(n-1)}{1} = - \sum\limits_{j=2}^{n-2} \agl{n}{j} \sqr{j}{1}
		\end{cases} & h_n<0\ . \\
	\end{array}
\end{align*}
}
\begin{align}
	\label{eq:momconsCond}
	\begin{array}{ll}
		\begin{cases}
			\agl{i}{(n-1)} \sqr{(n-1)}{n} = - \sum\limits_{j=1}^{n-2} \agl{i}{j} \sqr{j}{n} \\
			\agl{(n-1)}{1} \sqr{1}{n} = - \sum\limits_{j=2}^{n-2} \agl{(n-1)}{j} \sqr{j}{n}
		\end{cases}                                         & h_n>0\ ,                                                       \\[1em]
		\hspace{0.35cm}s_{1\, n-1} = - \sum\limits_{j=2}^{n-2} s_{1 j} - \sum\limits_{i=2}^{n-2} \sum\limits_{j=i+1}^{n-1} s_{i j} & h_n=0\ , \\[1em]
		\begin{cases}
			\agl{n}{(n-1)} \sqr{(n-1)}{i} = - \sum\limits_{j=1}^{n-2} \agl{n}{j} \sqr{j}{i} \\
			\agl{n}{1} \sqr{1}{(n-1)} = - \sum\limits_{j=2}^{n-2} \agl{n}{j} \sqr{j}{(n-1)}
		\end{cases}                                         & h_n<0\ .                                                       \\
	\end{array}
\end{align}
Notice that also the structures $\agl{(n-1)}{1} \sqr{1}{n}$ (for $h_n>0$, or $s_{1\, n-1}$ and $\agl{n}{1} \sqr{1}{n-1}$ in the other cases) are \textit{always} planar, because any source of non-planarity would come from $\agl{i}{n}$ invariants, which are always excluded by our choice of eliminating any momentum insertions of the $n^{\rm th}$-particle. In other words, the valency of the $n^{\rm th}$-vertex is $(h_n,0)$ and $(0,h_n)$ for $h_n \leq 0$ and $h_n>0$, respectively.

The conditions on the adjacency matrices for the polynomial structures to be in our basis are trivial. We are going to write them in the case $h_n< 0$ for simplicity:
\begin{equation}
	\label{eq:momcons2adjacency}
	\begin{split}
		A_{n-1\, n} = 0 \ \ \ &\mathrm{or}\  \ \ S_{i\, n-1} = 0\ ,\\
		A_{1\, n} = 0 \ \ \ &\mathrm{or}\ \ \ S_{1\, n-1} = 0\ ,\\
	\end{split}
\end{equation}
Moreover, equations \eqref{eq:momconsCond} provide an algorithmic way of writing linear relations of the structures in terms of the elements of our basis.

\subsection{A summary of the algorithm}
\label{sec:masslessAlgorithm}

In this section, we elaborate on the algorithms that follow from the considerations discussed in the previous sections. In particular, we present, step-by-step, how to find a basis of kinematically independent minimal amplitudes associated with a given particle content (or field content of the associated marginal operators) and a specified mass dimension. In our discussion, we will avoid complications coming from colour structures, which have been treated carefully elsewhere in the literature, \textit{e.g.} \cite{Henning:2015alf,Li:2020gnx}.
\begin{enumerate}
	\item We start with an initial trivial assignment of nodes valencies determined by the field content of the operators we want to consider. In this step, we choose the arbitrary ordering of the particles in the circular embedding.
	\item Accordingly to their mass dimension, such operators can have a number $n_\partial$ of derivatives. These derivatives correspond to momenta insertions in the associated minimal amplitudes. Then, we must consider all the permutations of the partitions of $n_\partial$ momentum insertions into at most $n-1$ integers. By doing so, we have already taken into account the conditions \eqref{eq:momcons1adjacency} coming from momentum conservation, \textit{i.e.} we exclude any insertion of $n^{\rm th}$ momentum.
	\item Each momentum insertion modifies the valency of nodes, as explained in \ref{sec:kinematics}. Then, we have a set of possible valency assignations for the graphs and we need to generate the corresponding structures which are kinematically independent:
	      \begin{enumerate}
		      \item We classify all the planar graphs with the valency assignment just found.
		      \item From this set of graphs, we exclude those \textit{not} satisfying the conditions \eqref{eq:momcons2adjacency}.
		      \item Using the map $\mathbb{M}$, we write down our basis of kinematically independent spinor structures\footnote{Each of the previous steps can be visualised in the \texttt{Mathematica} package \texttt{MassiveGraphs}, using the function \texttt{UniformMassStructures} (which perform the classification) and setting the option \texttt{Echos} to \texttt{True}.}.
	      \end{enumerate}
	\item Operators may involve multiple insertions of the same field, \textit{i.e.} we have identical particles in the minimal amplitude. For a detailed discussion see \cite{Henning:2017fpj,Fonseca:2019yya}. In these cases, the set of kinematically independent structures does not correspond to an independent basis of EFT operators. In practice, we consider all the previously classified independent structures and we act on them with a proper Young projector over the labels of the identical states, as explained for example in Section 3.2.2 of reference \cite{AccettulliHuber:2021uoa}. When summing over permutations we introduce terms which are not elements of our planar basis and we need an algorithm to write them as a linear combination of such elements\footnote{This second part of the algorithm has not been made publicly available in the \texttt{MassiveGraphs} code yet. On the other hand, an older version (valid only for purely massless structures) can be found in the \texttt{Mathematica} package \href{https://github.com/StefanoDeAngelis/SMEFT-operators}{\texttt{HelicityStructures}} and the function is called \texttt{AllIdentities}.}.
	      \begin{enumerate}
		      \item The inverse map $\mathbb{M}^{-1}$ gives the graphs associated with such structures.
		      \item We apply recursively \eqref{eq:untyingcrossing} (both for the angles and squares invariants) a finite number of times to write such graph as a sum of planar terms.
		      \item We might end up with terms which do not satisfy \eqref{eq:momcons1adjacency} or \eqref{eq:momcons2adjacency}. Such terms must also be decomposed in our basis and the graph operations, equivalent to \eqref{eq:momentumN}, is
		            \begin{equation}
			            \mathbb{M}(\mathbf{A},\mathbf{S}) = - \sum_{k = 1}^{n-1} \mathbb{M}(\mathbf{A}+\mathbf{G}^{(i)}_{(k)},\mathbf{S} + \mathbf{G}^{(j)}_{(k)})\ ,
		            \end{equation}
		            where
		            \begin{equation}
			            G^{(i)}_{(j),\,a b} = - \delta_{a,i} \delta_{b,n} + \delta_{a,i} \delta_{b,j}\ .
		            \end{equation}
		            Similarly we take into account the relations \eqref{eq:momconsCond}. Obviously, such operations never introduce negative powers of the Lorentz invariant structures, by construction.
		      \item After applying momentum conservation identities, the terms in the sum might not be all planar and we have to recursively apply \eqref{eq:untyingcrossing} again.
	      \end{enumerate}
	\item After this decomposition we end up with linear combinations of terms in the chosen basis and transforming properly under permutations of the particles.
	\item Finally we check whether there is a linear relation between such terms.
\end{enumerate}

\subsection{Checking the algorithm}
\label{sec:checks}

We performed several non-trivial checks on the algorithm just presented.
\begin{enumerate}
	\item The procedure seems to rely a lot on the cyclic order chosen for the vertices of the graphs and the momenta which we want to eliminate (using momentum conservation and equation of motion). Different choices correspond simply to different but equivalent bases for the kinematic structures. We checked that the number of elements in the basis does not depend on these choices in many non-trivial examples, involving several particles, also with higher helicity, and a high number of momenta insertions.
	\item Generating all the graphs we classify all the corresponding structures. We verified numerically (on rational kinematic, as explained in Appendix~\ref{sec:massivetwistors}) that the relations we find through our algorithm are correct and that they are all.
	      \begin{enumerate}
		      \item In particular, once the basis $\{ \mathbf{b}_i\}_{i=1,\dots , N}$ is generate, we might ask whether additional identities exist, which have not been considered in our approach. If such relation exist, we can find rational non-zero coefficients $\{\alpha_i\}_{i = 1,\dots ,N}$ such that
			  \begin{equation}
			      \sum_{i = 1}^N \alpha_i \mathbf{b}_i = 0\ .
		      \end{equation}
		      We generate $N$ independent rational kinematics and evaluate the RHS of the previous equation. By doing so we obtain a homogeneous linear system of $N$ equations in $N$ variables and, if its solution is $\alpha_i = 0$ $\forall i$, we checked the full independence of the elements of the basis.
		      \item In a similar way we can check completeness. It is easy to generate all the graph corresponding to the helicity assignments and distribution of momentum insertions $\{ \mathbf{c}_i\}_{i=1,\dots , M}$ and we want to find numerically the rational coefficients $\{\beta_{i j}\}_{i = 1,\dots ,N,\, j = 1, \dots , M}$ such that
		      \begin{equation}
				  \mathbf{c}_{i} = \sum_{j = 1}^N \beta_{i j} \mathbf{b}_j\ .
			  \end{equation}
			  We evaluate $N+1$ times both LHS and RHS on random rational kinematics and we obtain an inhomogeneous linear system of $M\times (N+1)$ equations in $M\times N$ variables. If a solution exists, we have verified completeness. We can also check the solution against the linear relations found from the graphic decomposition described in detail in the previous section.
	      \end{enumerate}
	We always found that the bases were complete and their elements independent in several non-trivial cases. The linear relations match with the results of the graphic decomposition.
\end{enumerate}

\section{The massive basis}
\label{sec:massive}

The classification of independent structures in massive theories involves more technical considerations, but a generalisation of the method presented above for fully massless theories is possible. The sources of such additional complications are two:
\begin{enumerate}
	\item The little group structures\ ,
	\item The equations of motion involving mass terms\ .
\end{enumerate}
Indeed, when classifying irrelevant interactions for massive theories, we have to be careful about the mass ordering of the independent structures, \textit{i.e.} we should not consider the operators $\mathcal{O}_\Delta$ and $\mathcal{O}_{\Delta+2} = m_i^2 \mathcal{O}_\Delta$ as independent when listing operators of dimension $\Delta$ and $\Delta +2$, for example.

In Section~\ref{sec:massless}, we identified a basis of structures $\mathcal{B} = \{\mathbf{b}_i\}$ such that any other combination of spinors with the proper mass dimension and helicity configuration can be written as a linear combination of $\mathbf{b}_i$'s.

When dealing with massive particles, we fix the \textit{helicity category}\footnote{The helicity category \cite{Durieux:2020gip} of a minimal amplitude is the helicity configurations of the structures obtained \textit{unbolding} the massive spinors \cite{Durieux:2019eor,Arkani-Hamed:2017jhn}.} and the mass dimension of the structures. We can identify a set of structures such that any element within the above-mentioned helicity category can be written as a linear combination in this basis. On the other hand, the latter will contain terms proportional to $m_i$, $\widetilde{m}_i$ (through the equation of motion) and $p_i^2$, which are redundant when we exploit the correspondence between polynomial kinematic structures and irrelevant operators.

Then, for a specified helicity category $(S_1,\dots ,S_n)$ and mass dimension $\Delta$, we will identify different bases that are relevant for different purposes:
\begin{itemize}
	\item \textit{Kinematic basis}: any spinorial structures within the chosen helicity category and mass dimension can be written as a linear combination of the terms in the basis. This basis contains also structures in different helicity categories, multiplied by powers of the masses. Such basis is the relevant one when we test our method numerically on rational kinematics as explained in Section~\ref{sec:checks} or we are interested in building an ansatz for amplitudes (for example, see \cite{AccettulliHuber:2021uoa}).
	\item \textit{Helicity category basis}: any term proportional to $m_i$ or $\widetilde{m}_i$ is ruled out. This gives a basis of structures that are kinematically independent of each other modulo identities across different helicity categories. This basis is the relevant one when we classify independent minimal amplitudes in order of the mass dimension of the corresponding EFT operators. For example, when classifying terms of mass dimension $\Delta$, any terms proportional to $m_i$ or $\widetilde{m}_i$ have already been considered with arbitrary coefficients in the basis for terms with mass dimension $\Delta - 1$. This will allow us to work effectively up to terms proportional to any power of the masses.
\end{itemize}

\subsection{The Massive Little Group}
When dealing with combinations of massive spinors, we must distinguish between spinors whose little group indices are contracted and those with free indices (to which we will refer as \textit{free spinors}). Since we are interested in structures that transform under irreducible representations of the little group, the free indices associated with a particle will automatically be fully symmetric\footnote{For those spinors we will use the bold notation introduced in \cite{Arkani-Hamed:2017jhn}.}. This distinction suggests that we have to associate different vertices in the graph with each momentum (\textit{momentum vertices}) and free spinor (\textit{spin vertex}).

\begin{figure}[t]
	\begin{center}
		\begin{tikzpicture}[scale=1.8,>=Stealth]

			\node at (-0.08,-0.08) {$k$};
			\node at (-0.08,1.08) {$i$};
			\node at (1.1,1.1) {$p_i$};
			\node at (1.08,-0.08) {$j$};
			\node at (1.25,0.5) {$=$};

			\draw [fill] (0,0) circle [radius = 0.02];
			\draw [fill] (1,0) circle [radius = 0.02];
			\draw [fill] (0,1) circle [radius = 0.02];
			\draw [fill] (1,1) circle [radius = 0.02];

			\draw (1,0) [<-,thick,blue] -- (0.55,0.45);
			\draw [thick, blue] (0.45,0.55) arc (135:-45:0.07);
			\draw [thick, blue] (0.45,0.55) -- (0,1);
			\draw [->,thick,blue](1,1) -- (0,0);
			\draw [dashed,thick,red] (1,1) -- (.5,0.75);

		\end{tikzpicture}
		\begin{tikzpicture}[scale=1.8,>=Stealth]

			\node at (-0.08,-0.08) {$k$};
			\node at (-0.08,1.08) {$i$};
			\node at (1.1,1.1) {$p_i$};
			\node at (1.08,-0.08) {$j$};
			\node at (1.25,0.5) {$+$};

			\draw [fill] (0,0) circle [radius = 0.02];
			\draw [fill] (1,0) circle [radius = 0.02];
			\draw [fill] (0,1) circle [radius = 0.02];
			\draw [fill] (1,1) circle [radius = 0.02];

			\draw [<-,thick,blue](1,0) -- (1,1);
			\draw [->,thick,blue](0,1) -- (0,0);
			\draw [dashed,thick,red] (1,1) -- (.5,0.75);

		\end{tikzpicture}
		\begin{tikzpicture}[scale=1.8,>=Stealth]

			\node at (-0.08,-0.08) {$k$};
			\node at (-0.08,1.08) {$i$};
			\node at (1.1,1.1) {$p_i$};
			\node at (1.08,-0.08) {$j$};

			\draw [fill] (0,0) circle [radius = 0.02];
			\draw [fill] (1,0) circle [radius = 0.02];
			\draw [fill] (0,1) circle [radius = 0.02];
			\draw [fill] (1,1) circle [radius = 0.02];

			\draw [->,thick,blue](0,1) -- (1,1);
			\draw [->,thick,blue](1,0) -- (0,0);
			\draw [dashed,thick,red] (1,1) -- (.5,0.75);

		\end{tikzpicture}
	\end{center}
	\begin{center}
		\begin{tikzpicture}[scale=1.2,>=Stealth]

			\node at (-0.1,-0.1) {$k$};
			\node at (-0.12,1.1) {$i$};
			\node at (0.95,1.45) {$p_i$};
			\node at (1.64,0.5) {$p_i$};
			\node at (0.95,-0.47) {$j$};
			\node at (1.95,0.5) {$=$};

			\draw [fill] (0,0) circle [radius = 0.02];
			\draw [fill] (0,1) circle [radius = 0.02];
			\draw [fill] (0.95,1.3) circle [radius = 0.02];
			\draw [fill] (1.45,0.5) circle [radius = 0.02];
			\draw [fill] (0.95,-0.3) circle [radius = 0.02];

			\draw [->,thick,red] (0.95,1.3) -- (0.95,-0.3);
			\draw [->,thick,red] (0.87,0.3) -- (0,0);
			\draw [thick,red] (1.45,0.5) -- (1.015,0.35);
			\draw [thick,red] (0.8497,0.293) arc (180:30:0.11);

			\draw [dashed,thick,blue](0.95,1.3) -- (0.5,0.5);
			\draw [dashed,thick,blue](1.45,0.5) -- (0.8,0);

		\end{tikzpicture}
		\begin{tikzpicture}[scale=1.2,>=Stealth]

			\node at (-0.1,-0.1) {$k$};
			\node at (-0.12,1.1) {$i$};
			\node at (0.95,1.45) {$p_i$};
			\node at (1.64,0.5) {$p_i$};
			\node at (0.95,-0.47) {$j$};
			\node at (2,0.5) {$+$};

			\draw [fill] (0,0) circle [radius = 0.02];
			\draw [fill] (0,1) circle [radius = 0.02];
			\draw [fill] (0.95,1.3) circle [radius = 0.02];
			\draw [fill] (1.45,0.5) circle [radius = 0.02];
			\draw [fill] (0.95,-0.3) circle [radius = 0.02];

			\draw [->,thick,red] (0.95,1.3) -- (0,0);
			\draw [->,thick,red] (1.45,0.5) -- (0.95,-0.3);

			\draw [dashed,thick,blue](0.95,1.3) -- (0.5,0.5);
			\draw [dashed,thick,blue](1.45,0.5) -- (0.8,0);

		\end{tikzpicture}
		\begin{tikzpicture}[scale=1.2,>=Stealth]

			\node at (-0.1,-0.1) {$k$};
			\node at (-0.12,1.1) {$i$};
			\node at (0.95,1.45) {$p_i$};
			\node at (1.64,0.5) {$p_i$};
			\node at (0.95,-0.47) {$j$};

			\draw [fill] (0,0) circle [radius = 0.02];
			\draw [fill] (0,1) circle [radius = 0.02];
			\draw [fill] (0.95,1.3) circle [radius = 0.02];
			\draw [fill] (1.45,0.5) circle [radius = 0.02];
			\draw [fill] (0.95,-0.3) circle [radius = 0.02];

			\draw [->,thick,red] (0.95,1.3) -- (1.45,0.5);
			\draw [->,thick,red] (0.95,-0.3) -- (0,0);

			\draw [dashed,thick,blue](0.95,1.3) -- (0.5,0.5);
			\draw [dashed,thick,blue](1.45,0.5) -- (0.8,0);

		\end{tikzpicture}
	\end{center}
	\caption{Schouten identities w.r.t. LG indices \eqref{eq:SchoutenLG} are again equivalent to untying crossings of edges anchored to the spin and momentum vertices associated with the same particle.}
	\label{fig:LG}
\end{figure}
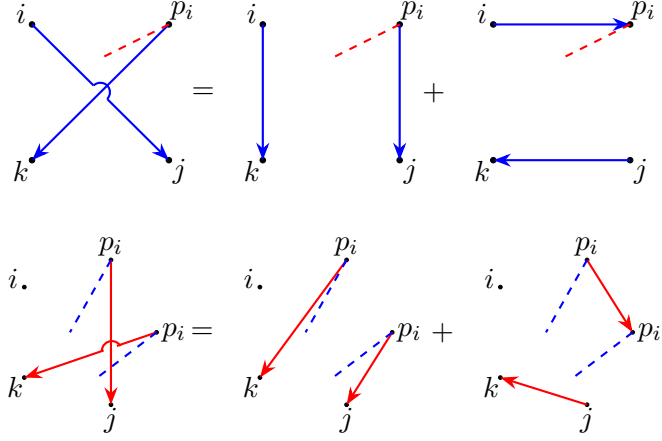

Schouten identities for the little group can involve either a momentum or a free spinor ($p_{i \alpha \dot{\alpha}}\, \widetilde{\lambda}^I_{i \dot{\beta}}$ or $p_{i \alpha \dot{\alpha}}\, \lambda^I_{i \beta}$) or two momenta $p_{i \alpha \dot{\alpha}}\, p_{i \beta \dot{\beta}}$. We will show in detail that Schouten identities are again equivalent to untying the crossing of two edges, both anchored to momentum vertices and/or to the spin vertices. Indeed, considering the combinations mentioned above, with $p_{i \alpha \dot{\alpha}} = \lambda_{i \alpha}^I \lambda_{i \dot{\alpha} I}$ and antisymmetrising over two little group group indices, we find
\begin{equation}
	\label{eq:SchoutenLG}
	\begin{split}
		p_{i \alpha \dot{\alpha}}\, \widetilde{\lambda}^I_{i \dot{\beta}} &= p_{i \alpha \dot{\beta}}\, \widetilde{\lambda}^I_{i \dot{\alpha}} - \epsilon_{\dot{\alpha} \dot{\beta}} p_{i \alpha \dot{\gamma}}\, \widetilde{\lambda}^{I \dot{\gamma}}_{i}\ ,\\
		p_{i \alpha \dot{\alpha}} p_{i \beta \dot{\beta}} &= p_{i \beta \dot{\alpha}} p_{i \alpha \dot{\beta}} - \epsilon_{\alpha \beta} p_{i \gamma \dot{\alpha}} p_{i \dot{\beta}}^{\ \gamma}\ ,
	\end{split}
\end{equation}
and their ``conjugates", which are identical to the relations one would find applying the antisymmetrisation directly over the ${\rm SL}(2,\mathbb{C})$ indices. We show their graph representation in Figure~\ref{fig:LG}.

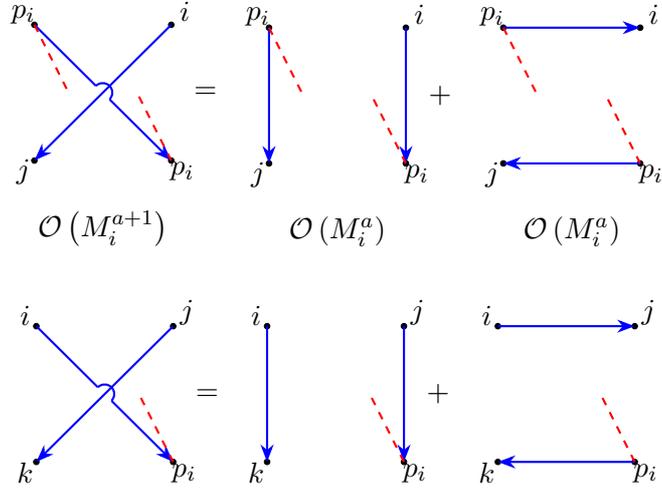
\begin{figure}[t]
	\begin{center}
		\begin{tikzpicture}[scale=1.8,>=Stealth]

			\node at (-0.08,-0.08) {$j$};
			\node at (-0.08,1.08) {$p_i$};
			\node at (1.1,1.1) {$i$};
			\node at (1.08,-0.08) {$p_i$};
			\node at (1.25,0.5) {$=$};
			\node at (0.5,-0.5) {$\mathcal{O}\left(M_i^{a+1}\right)$};

			\draw [fill] (0,0) circle [radius = 0.02];
			\draw [fill] (1,0) circle [radius = 0.02];
			\draw [fill] (0,1) circle [radius = 0.02];
			\draw [fill] (1,1) circle [radius = 0.02];

			\draw (1,0) [<-,thick,blue] -- (0.55,0.45);
			\draw [thick, blue] (0.45,0.55) arc (135:-45:0.07);
			\draw [thick, blue] (0.45,0.55) -- (0,1);
			\draw [->,thick,blue](1,1) -- (0,0);

			\draw [dashed,thick,red] (0,1) -- (0.25,0.5);
			\draw [dashed,thick,red] (1,0) -- (0.75,0.5);

		\end{tikzpicture}
		\begin{tikzpicture}[scale=1.8,>=Stealth]

			\node at (-0.08,-0.08) {$j$};
			\node at (-0.08,1.08) {$p_i$};
			\node at (1.1,1.1) {$i$};
			\node at (1.08,-0.08) {$p_i$};
			\node at (1.25,0.5) {$+$};
			\node at (0.5,-0.5) {$\mathcal{O}\left(M_i^{a}\right)$};

			\draw [fill] (0,0) circle [radius = 0.02];
			\draw [fill] (1,0) circle [radius = 0.02];
			\draw [fill] (0,1) circle [radius = 0.02];
			\draw [fill] (1,1) circle [radius = 0.02];

			\draw [<-,thick,blue](1,0) -- (1,1);
			\draw [->,thick,blue](0,1) -- (0,0);

			\draw [dashed,thick,red] (0,1) -- (0.25,0.5);
			\draw [dashed,thick,red] (1,0) -- (0.75,0.5);

		\end{tikzpicture}
		\begin{tikzpicture}[scale=1.8,>=Stealth]

			\node at (-0.08,-0.08) {$j$};
			\node at (-0.08,1.08) {$p_i$};
			\node at (1.1,1.1) {$i$};
			\node at (1.08,-0.08) {$p_i$};
			\node at (0.5,-0.5) {$\mathcal{O}\left(M_i^{a}\right)$};

			\draw [fill] (0,0) circle [radius = 0.02];
			\draw [fill] (1,0) circle [radius = 0.02];
			\draw [fill] (0,1) circle [radius = 0.02];
			\draw [fill] (1,1) circle [radius = 0.02];

			\draw [->,thick,blue](0,1) -- (1,1);
			\draw [->,thick,blue](1,0) -- (0,0);

			\draw [dashed,thick,red] (0,1) -- (0.25,0.5);
			\draw [dashed,thick,red] (1,0) -- (0.75,0.5);

		\end{tikzpicture}
	\end{center}
	\begin{center}
		\begin{tikzpicture}[scale=1.8,>=Stealth]

			\node at (-0.08,-0.08) {$k$};
			\node at (-0.08,1.08) {$i$};
			\node at (1.1,1.1) {$j$};
			\node at (1.08,-0.08) {$p_i$};
			\node at (1.25,0.5) {$=$};

			\draw [fill] (0,0) circle [radius = 0.02];
			\draw [fill] (1,0) circle [radius = 0.02];
			\draw [fill] (0,1) circle [radius = 0.02];
			\draw [fill] (1,1) circle [radius = 0.02];

			\draw (1,0) [<-,thick,blue] -- (0.55,0.45);
			\draw [thick, blue] (0.45,0.55) arc (135:-45:0.07);
			\draw [thick, blue] (0.45,0.55) -- (0,1);
			\draw [->,thick,blue](1,1) -- (0,0);

			\draw [dashed,thick,red] (1,0) -- (0.75,0.5);

		\end{tikzpicture}
		\begin{tikzpicture}[scale=1.8,>=Stealth]

			\node at (-0.08,-0.08) {$k$};
			\node at (-0.08,1.08) {$i$};
			\node at (1.1,1.1) {$j$};
			\node at (1.08,-0.08) {$p_i$};
			\node at (1.25,0.5) {$+$};

			\draw [fill] (0,0) circle [radius = 0.02];
			\draw [fill] (1,0) circle [radius = 0.02];
			\draw [fill] (0,1) circle [radius = 0.02];
			\draw [fill] (1,1) circle [radius = 0.02];

			\draw [<-,thick,blue](1,0) -- (1,1);
			\draw [->,thick,blue](0,1) -- (0,0);

			\draw [dashed,thick,red] (1,0) -- (0.75,0.5);

		\end{tikzpicture}
		\begin{tikzpicture}[scale=1.8,>=Stealth]

			\node at (-0.08,-0.08) {$k$};
			\node at (-0.08,1.08) {$i$};
			\node at (1.1,1.1) {$j$};
			\node at (1.08,-0.08) {$p_i$};

			\draw [fill] (0,0) circle [radius = 0.02];
			\draw [fill] (1,0) circle [radius = 0.02];
			\draw [fill] (0,1) circle [radius = 0.02];
			\draw [fill] (1,1) circle [radius = 0.02];

			\draw [->,thick,blue](0,1) -- (1,1);
			\draw [->,thick,blue](1,0) -- (0,0);

			\draw [dashed,thick,red] (1,0) -- (0.75,0.5);

		\end{tikzpicture}
	\end{center}
	\caption{The notion of mass ordering requires that momentum vertices succeed (or precede) the spin vertex, as dictated by the on-shell conditions $p_{i \alpha \dot{\alpha}} \widetilde{\lambda}_i^{\dot{\alpha} I} = \widetilde{m}_i \lambda_{i \alpha}^{I}$ and $p_{i \alpha \dot{\alpha}} p_{i}^{\dot{\alpha} \beta} = M_i^2 \delta_\alpha^\beta$.}
	\label{fig:massordering}
\end{figure}

We should emphasise a crucial point: since we are looking for a basis of polynomial structures with a \textit{well-defined notion of mass ordering}, momentum vertices must be all succeeding (or all preceding) the spin vertex. Indeed, we want the combinations which are proportional to higher powers of the mass to be always planar. This is not guaranteed if the condition mentioned above is lifted, as shown in Figure~\ref{fig:massordering}.

\subsection{Equations of Motion}
\label{sec:equationofmotion}

So far we ﬁnd a proliferation of vertices, each particle is associated with a vertex carrying both helicity weight and momentum insertions for massless particles and only spin weight for a massive one. Besides, in the latter case, we need to add a vertex next to the spin vertex of the corresponding particle for each insertion of massive momenta. The order of momentum vertices among themselves is irrelevant, even though it seems crucial when we are dealing with planar graphs only.

In this section, we will show how the Dirac equation for the spinors allows mapping our problem to a finite set of fully massless classifications. Then, before dealing with momentum conservation identities, we simplify our problem by using the graph equivalent of the \textit{unbolding-bolding} procedure presented in \cite{Durieux:2020gip}.

\begin{figure}[t]
	\centering
	\begin{tikzpicture}[scale=1.4,>=Stealth]

		\node at (-0.08,-0.08) {$4$};
		\node at (-0.08,1.08) {$1$};
		\node at (1.1,1.1) {$2$};
		\node at (1.08,-0.08) {$3$};

		\draw [fill] (0,0) circle [radius = 0.02];
		\draw [fill] (1,0) circle [radius = 0.02];
		\draw [fill] (0,1) circle [radius = 0.02];
		\draw [fill] (1,1) circle [radius = 0.02];

		\draw [->,red] (0,1) to [out=-100,in=100] (0,0);
		\draw [->,red] (0,1) to [out=-80,in=80] (0,0);
		\draw [->,red] (0,1) to [out=10,in=170] (1,1);
		\draw [->,blue] (0,1) to [out=-10,in=-170] (1,1);
		\draw [->,blue] (0,1) to [out=-35,in=125] (1,0);
		\draw [->,blue] (0,1) to [out=-50,in=145] (1,0);

	\end{tikzpicture}
	\begin{tikzpicture}[scale=1.4,>=Stealth]

		\node at (-0.08,-0.08) {$4$};
		\node at (-0.08,1.08) {$1$};
		\node at (1.1,1.1) {$2$};
		\node at (1.08,-0.08) {$3$};

		\draw [fill] (0,0) circle [radius = 0.02];
		\draw [fill] (1,0) circle [radius = 0.02];
		\draw [fill] (0,1) circle [radius = 0.02];
		\draw [fill] (1,1) circle [radius = 0.02];

		\draw [->,red] (0,1) to [out=10,in=170] (1,1);
		\draw [->,blue] (0,1) to [out=-10,in=-170] (1,1);
		\draw [->,blue] (1,1) to [out=-80,in=80] (1,0);
		\draw [->,blue] (1,1) to [out=-100,in=100] (1,0);
		\draw [->,red] (1,1) to [out=-125,in=35] (0,0);
		\draw [->,red] (1,1) to [out=-145,in=55] (0,0);

	\end{tikzpicture}
	\begin{tikzpicture}[scale=1.4,>=Stealth]

		\node at (-0.08,-0.08) {$4$};
		\node at (-0.08,1.08) {$1$};
		\node at (1.1,1.1) {$2$};
		\node at (1.08,-0.08) {$3$};

		\draw [fill] (0,0) circle [radius = 0.02];
		\draw [fill] (1,0) circle [radius = 0.02];
		\draw [fill] (0,1) circle [radius = 0.02];
		\draw [fill] (1,1) circle [radius = 0.02];

		\draw [->,red] (0,1) to [out=10,in=170] (1,1);
		\draw [->,blue] (0,1) to [out=-10,in=-170] (1,1);
		\draw [->,blue] (0,1) -- (1,0);
		\draw [->,blue] (1,1) -- (1,0);
		\draw [->,red] (0,1) -- (0,0);
		\draw [->,red] (1,1) -- (0,0);

	\end{tikzpicture}
	\begin{tikzpicture}[scale=1.4,>=Stealth]

		\node at (-0.08,-0.08) {$4$};
		\node at (-0.08,1.08) {$1$};
		\node at (1.1,1.1) {$2$};
		\node at (1.08,-0.08) {$3$};

		\draw [fill] (0,0) circle [radius = 0.02];
		\draw [fill] (1,0) circle [radius = 0.02];
		\draw [fill] (0,1) circle [radius = 0.02];
		\draw [fill] (1,1) circle [radius = 0.02];

		\draw [->,red] (0,1) to [out=-100,in=100] (0,0);
		\draw [->,red] (0,1) to [out=-80,in=80] (0,0);
		\draw [->,red] (1,1) to [out=-100,in=100] (1,0);
		\draw [->,blue] (0,1) to [out=-35,in=125] (1,0);
		\draw [->,blue] (0,1) to [out=-50,in=145] (1,0);
		\draw [->,blue] (1,1) to [out=-80,in=80] (1,0);

	\end{tikzpicture}\\
	\begin{tikzpicture}[scale=1.4,>=Stealth]

		\node at (-0.08,-0.08) {$4$};
		\node at (-0.08,1.08) {$1$};
		\node at (1.1,1.1) {$2$};
		\node at (1.08,-0.08) {$3$};

		\draw [fill] (0,0) circle [radius = 0.02];
		\draw [fill] (1,0) circle [radius = 0.02];
		\draw [fill] (0,1) circle [radius = 0.02];
		\draw [fill] (1,1) circle [radius = 0.02];

		\draw [->,red] (0,1) -- (0,0);
		\draw [->,red] (0,1) -- (1,1);
		\draw [->,red] (1,0) -- (0,0);
		\draw [->,blue] (0,1) to [out=-35,in=125] (1,0);
		\draw [->,blue] (0,1) to [out=-50,in=145] (1,0);
		\draw [->,blue] (1,1) -- (1,0);

	\end{tikzpicture}
	\begin{tikzpicture}[scale=1.4,>=Stealth]

		\node at (-0.08,-0.08) {$4$};
		\node at (-0.08,1.08) {$1$};
		\node at (1.1,1.1) {$2$};
		\node at (1.08,-0.08) {$3$};

		\draw [fill] (0,0) circle [radius = 0.02];
		\draw [fill] (1,0) circle [radius = 0.02];
		\draw [fill] (0,1) circle [radius = 0.02];
		\draw [fill] (1,1) circle [radius = 0.02];

		\draw [->,red] (0,1) -- (1,1);
		\draw [->,red] (1,1) -- (0,0);
		\draw [->,red] (1,0) -- (0,0);
		\draw [->,blue] (0,1) -- (1,0);
		\draw [->,blue] (1,1) to [out=-80,in=80] (1,0);
		\draw [->,blue] (1,1) to [out=-100,in=100] (1,0);

	\end{tikzpicture}
	\begin{tikzpicture}[scale=1.4,>=Stealth]

		\node at (-0.08,-0.08) {$4$};
		\node at (-0.08,1.08) {$1$};
		\node at (1.1,1.1) {$2$};
		\node at (1.08,-0.08) {$3$};

		\draw [fill] (0,0) circle [radius = 0.02];
		\draw [fill] (1,0) circle [radius = 0.02];
		\draw [fill] (0,1) circle [radius = 0.02];
		\draw [fill] (1,1) circle [radius = 0.02];

		\draw [->,red] (0,1) -- (0,0);
		\draw [->,red] (1,1) -- (0,0);
		\draw [->,red] (1,1) -- (1,0);
		\draw [->,blue] (0,1) -- (1,0);
		\draw [->,blue] (1,1) to [out=-70,in=70] (1,0);
		\draw [->,blue] (1,1) to [out=-110,in=110] (1,0);

	\end{tikzpicture}\\\rule{10cm}{.4pt}
	\vspace{.4cm}

	\begin{tikzpicture}[scale=0.9,>=Stealth]

		\node at (-0.15,-0.15) {$4$};
		\node at (-0.15,1.15) {$1$};
		\node at (1,1.55) {$p_1$};
		\node at (1.85,1.15) {$p_1$};
		\node at (1.85,0.05) {$2$};
		\node at (1,-0.55) {$3$};

		\draw [fill] (0,0) circle [radius = 0.02];
		\draw [fill] (0,1) circle [radius = 0.02];
		\draw [fill] (1.6,0) circle [radius = 0.02];
		\draw [fill] (1.6,1) circle [radius = 0.02];
		\draw [fill] (0.8,-0.35) circle [radius = 0.02];
		\draw [fill] (0.8,1.35) circle [radius = 0.02];

		\draw [->,blue] (0,1) -- (0.8,-0.35);
		\draw [->,blue] (0.8,1.35) -- (0.8,-0.35);
		\draw [->,red] (0,1) -- (0,0);
		\draw [->,red] (0.8,1.35) -- (0,0);
		\draw [->,blue] (1.6,1) to [out=-70,in=70] (1.6,0);
		\draw [->,red] (1.6,1) to [out=-110,in=110] (1.6,0);

	\end{tikzpicture}
	\begin{tikzpicture}[scale=0.9,>=Stealth]

		\draw [thick] (-0.4,-0.8) -- (-0.4,1.75) -- (2.15,1.75) -- (2.15,-0.8) -- (-0.4,-0.8);

		\node at (-0.15,-0.15) {$4$};
		\node at (-0.15,1.15) {$1$};
		\node at (1,1.55) {$2$};
		\node at (1.85,1.15) {$p_2$};
		\node at (1.85,0.05) {$p_2$};
		\node at (1,-0.55) {$3$};

		\draw [fill] (0,0) circle [radius = 0.02];
		\draw [fill] (0,1) circle [radius = 0.02];
		\draw [fill] (1.6,0) circle [radius = 0.02];
		\draw [fill] (1.6,1) circle [radius = 0.02];
		\draw [fill] (0.8,-0.35) circle [radius = 0.02];
		\draw [fill] (0.8,1.35) circle [radius = 0.02];

		\draw [->,blue] (1.6,1) -- (0.8,-0.35);
		\draw [->,blue] (1.6,0) -- (0.8,-0.35);
		\draw [->,red] (1.6,1) -- (0,0);
		\draw [->,red] (1.6,0) -- (0,0);
		\draw [->,blue] (0,1) to [out=45,in=-165] (0.8,1.35);
		\draw [->,red] (0,1) to [out=10,in=-145] (0.8,1.35);

	\end{tikzpicture}
	\begin{tikzpicture}[scale=0.9,>=Stealth]

		\node at (-0.15,-0.15) {$4$};
		\node at (-0.15,1.15) {$1$};
		\node at (1,1.55) {$p_1$};
		\node at (1.85,1.15) {$2$};
		\node at (1.85,0.05) {$p_2$};
		\node at (1,-0.55) {$3$};

		\draw [fill] (0,0) circle [radius = 0.02];
		\draw [fill] (0,1) circle [radius = 0.02];
		\draw [fill] (1.6,0) circle [radius = 0.02];
		\draw [fill] (1.6,1) circle [radius = 0.02];
		\draw [fill] (0.8,-0.35) circle [radius = 0.02];
		\draw [fill] (0.8,1.35) circle [radius = 0.02];

		\draw [->,blue] (0,1) -- (0.8,-0.35);
		\draw [->,blue] (1.6,0) -- (0.8,-0.35);
		\draw [->,red] (0,1) -- (0,0);
		\draw [->,red] (1.6,0) -- (0,0);
		\draw [->,blue] (0.8,1.35) to [out=-45,in=165] (1.6,1);
		\draw [->,red] (0.8,1.35) to [out=-10,in=145] (1.6,1);

	\end{tikzpicture}
	\begin{tikzpicture}[scale=0.95,>=Stealth]

		\node at (-0.08,-0.08) {$4$};
		\node at (-0.08,1.08) {$1$};
		\node at (0.95,1.43) {$p_1$};
		\node at (1.58,0.5) {$2$};
		\node at (0.95,-0.43) {$3$};

		\draw [fill] (0,0) circle [radius = 0.02];
		\draw [fill] (0,1) circle [radius = 0.02];
		\draw [fill] (0.95,1.3) circle [radius = 0.02];
		\draw [fill] (1.45,0.5) circle [radius = 0.02];
		\draw [fill] (0.95,-0.3) circle [radius = 0.02];

		\draw [->,red] (0,1) -- (0,0);
		\draw [->,red] (0.95,1.3) -- (0,0);
		\draw [->,blue] (0,1) -- (0.95,-0.3);
		\draw [->,blue] (0.95,1.3) -- (0.95,-0.3);
		\draw [->,blue] (1.45,0.5) to [out=-135,in=70] (0.95,-0.3);
		\draw [->,red] (1.45,0.5) to [out=-110,in=50] (0.95,-0.3);

	\end{tikzpicture}\\
	\begin{tikzpicture}[scale=0.95,>=Stealth]

		\node at (-0.08,-0.08) {$4$};
		\node at (-0.08,1.08) {$1$};
		\node at (0.95,1.43) {$p_1$};
		\node at (1.58,0.5) {$2$};
		\node at (0.95,-0.43) {$3$};

		\draw [fill] (0,0) circle [radius = 0.02];
		\draw [fill] (0,1) circle [radius = 0.02];
		\draw [fill] (0.95,1.3) circle [radius = 0.02];
		\draw [fill] (1.45,0.5) circle [radius = 0.02];
		\draw [fill] (0.95,-0.3) circle [radius = 0.02];

		\draw [->,red] (0,1) -- (0,0);
		\draw [->,red] (0.95,-0.3) -- (0,0);
		\draw [->,red] (0.95,1.3) -- (1.45,0.5);
		\draw [->,blue] (0,1) -- (0.95,-0.3);
		\draw [->,blue] (0.95,1.3) -- (0.95,-0.3);
		\draw [->,blue] (1.45,0.5) -- (0.95,-0.3);

	\end{tikzpicture}
	\begin{tikzpicture}[scale=0.95,>=Stealth]

		\node at (-0.08,-0.08) {$4$};
		\node at (-0.08,1.08) {$1$};
		\node at (0.95,1.43) {$2$};
		\node at (1.64,0.5) {$p_2$};
		\node at (0.95,-0.43) {$3$};

		\draw [fill] (0,0) circle [radius = 0.02];
		\draw [fill] (0,1) circle [radius = 0.02];
		\draw [fill] (0.95,1.3) circle [radius = 0.02];
		\draw [fill] (1.45,0.5) circle [radius = 0.02];
		\draw [fill] (0.95,-0.3) circle [radius = 0.02];

		\draw [->,red] (0,1) -- (0.95,1.3);
		\draw [->,red] (0.95,-0.3) -- (0,0);
		\draw [->,red] (1.45,0.5) -- (0,0);
		\draw [->,blue] (0,1) -- (0.95,-0.3);
		\draw [->,blue] (0.95,1.3) -- (0.95,-0.3);
		\draw [->,blue] (1.45,0.5) -- (0.95,-0.3);

	\end{tikzpicture}
	\begin{tikzpicture}[scale=0.95,>=Stealth]

		\draw [thick] (-0.4,-0.65) -- (-0.4,1.65) -- (1.9,1.65) -- (1.9,-0.65) -- (-0.4,-0.65);

		\node at (-0.08,-0.08) {$4$};
		\node at (-0.08,1.08) {$1$};
		\node at (0.95,1.43) {$2$};
		\node at (1.64,0.5) {$p_2$};
		\node at (0.95,-0.43) {$3$};

		\draw [fill] (0,0) circle [radius = 0.02];
		\draw [fill] (0,1) circle [radius = 0.02];
		\draw [fill] (0.95,1.3) circle [radius = 0.02];
		\draw [fill] (1.45,0.5) circle [radius = 0.02];
		\draw [fill] (0.95,-0.3) circle [radius = 0.02];

		\draw [->,red] (0,1) -- (0,0);
		\draw [->,red] (0.95,1.3) -- (0,0);
		\draw [->,blue] (0,1) -- (0.95,-0.3);
		\draw [->,blue] (0.95,1.3) -- (0.95,-0.3);
		\draw [->,blue] (1.45,0.5) to [out=-135,in=70] (0.95,-0.3);
		\draw [->,red] (1.45,0.5) to [out=-110,in=50] (0.95,-0.3);

	\end{tikzpicture}
	\caption{We considered the planar graphs associated terms in the helicity category $(1^{1_0},2^{1_0},3^{+1},4^{-1})$ and mass dimension $6$. For simplicity, we did not consider a priori the ones with insertions of $p_4$ and those proportional to any insertion of $M_1$ and $M_2$. We are showing both the unbolded and the bolded versions of the graphs to make the one-to-one correspondence evident. The two framed graphs correspond to our basis after taking into account momentum conservation.}
	\label{fig:unboldinggraphs}
\end{figure}
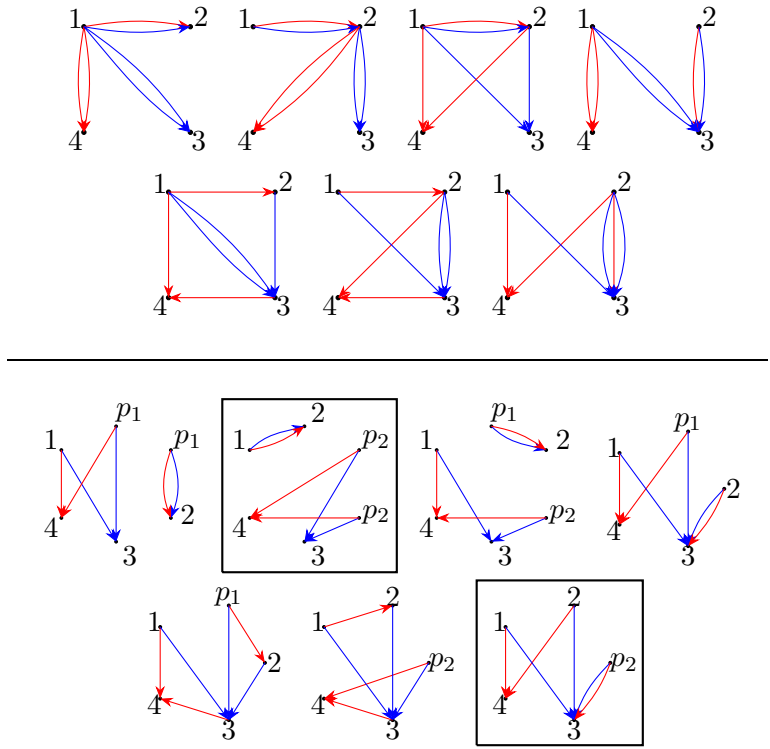

In the previous section, we have shown that choosing carefully the arrangement of the momentum vertices in the circular embedding is essential to guarantee that any polynomial structure can be written as a linear combination of structures in our basis which have the same number of explicit mass powers or higher. This suggests that there could be a way to classify independent spinor structures with fixed powers of mass, \textit{i.e.} a classification modulo equations of motion, excluding all the graphs where any momentum vertex is connected to the corresponding spin vertex. The key observation is the following: \textit{unbolded} graphs, \textit{i.e.} graphs for which we do not distinguish between momentum and spin vertices, are in \textit{one-to-one correspondence} with planar ``massive''-\textit{bolded} graphs for which no momentum vertex is linked to the corresponding spin vertex and between themselves, and the edges are not crossing. An example of this is illustrated in Figure~\ref{fig:unboldinggraphs}.

When we deal with massive structures, we have to introduce a notion of \textit{transversality}, because the spins characterise our structures only partially. Indeed, when we consider, for example, a spin-1 particle, its polarisation tensor (that we define to be dimensionless) could in principle be defined in several ways:
\begin{equation}
	\label{eq:polarisation}
	\frac{\lambda_\alpha^{(I} \lambda_{\beta}^{J)}}{m}\ , \qquad \frac{\lambda_\alpha^{(I} \tlambda_{\dot{\alpha}}^{J)}}{M}\ , \qquad \frac{\tlambda_{\dot{\alpha}}^{(I} \tlambda_{\dot{\beta}}^{J)}}{\widetilde{m}}\ ,
\end{equation}
which correspond to transversality $-1$, $0$, and $+1$ in our notation. In general, the transversality can take the values $C = - J, -J+1,\dots , J$ and we will specify it as $J_C$. The set of transversalities and helicities identify the helicity category of the structure.

Any edge, linking a momentum vertex with its respective spin vertex, gives a power of the mass and changes the transversality of the structure:
\begin{equation}
	p_{i \alpha \dot{\alpha}} \widetilde{\lambda}_i^{\dot{\alpha} I} = \widetilde{m}_i \lambda_{i \alpha}^{I}\ , \qquad p_{i}^{\dot{\alpha} \alpha} \lambda_{i \alpha}^{I} = m_i \tlambda_{i}^{\dot{\alpha} I}\ ,
\end{equation}
or graphically
\begin{figure}[h!]
	\centering
	\begin{tikzpicture}[>=Stealth]
		\node at (0,0.25) {$i$};
		\node at (1,0.25) {$p_i$};
		\node at (1.5,0) {$=$};

		\draw [fill] (0,0) circle [radius = 0.02];
		\draw [fill] (1,0) circle [radius = 0.02];

		\draw [->,thick,red] (0,0) -- (1,0);
		\draw [dashed,thick,red] (0,0) -- (-0.4,-0.4);
		\draw [dashed,thick,blue] (0,0) -- (0,-0.55);
		\draw [dashed,thick,blue] (0,0) -- (0.4,-0.4);
		\draw [thick,blue] (1,0) -- (1,-0.55);
	\end{tikzpicture}
	\begin{tikzpicture}[>=Stealth]
		\node at (-0.75,0) {$m_i$};
		\node at (0,0.25) {$i$};
		\node at (0.75,0) {$,$};

		\draw [fill] (0,0) circle [radius = 0.02];

		\draw [dashed,thick,red] (0,0) -- (-0.4,-0.4);
		\draw [dashed,thick,blue] (0,0) -- (0,-0.55);
		\draw [dashed,thick,blue] (0,0) -- (0.4,-0.4);
		\draw [thick,blue] (0,0) -- (0.55,0);
	\end{tikzpicture}\qquad
	\begin{tikzpicture}[>=Stealth]
		\node at (0,0.25) {$i$};
		\node at (1,0.25) {$p_i$};
		\node at (1.5,0) {$=$};

		\draw [fill] (0,0) circle [radius = 0.02];
		\draw [fill] (1,0) circle [radius = 0.02];

		\draw [->,thick,blue] (0,0) -- (1,0);
		\draw [dashed,thick,red] (0,0) -- (-0.4,-0.4);
		\draw [dashed,thick,blue] (0,0) -- (0,-0.55);
		\draw [dashed,thick,blue] (0,0) -- (0.4,-0.4);
		\draw [thick,red] (1,0) -- (1,-0.55);
	\end{tikzpicture}
	\begin{tikzpicture}[>=Stealth]
		\node at (-0.75,0) {$\widetilde{m}_i$};
		\node at (0,0.25) {$i$};
		\node at (0.75,0) {$.$};

		\draw [fill] (0,0) circle [radius = 0.02];

		\draw [dashed,thick,red] (0,0) -- (-0.4,-0.4);
		\draw [dashed,thick,blue] (0,0) -- (0,-0.55);
		\draw [dashed,thick,blue] (0,0) -- (0.4,-0.4);
		\draw [thick,red] (0,0) -- (0.55,0);
	\end{tikzpicture}
\end{figure}

Indeed, given the valency of the spin vertex of the $i^{\rm th}$-particle $(v_a^i,v_s^i)$, then $J= v_a^i+v_s^i$ and $C=v_s^i-v_a^i$. When we classify the combinations in the helicity category $(\{i^{S_i}\}_{i=1,\dots,n})$ (with $S_i = {J_i}_{C_i}$ or $S_i = h_i$ for massive and massless particles, respectively) and mass dimension $\Delta$, the number of momentum insertions is $\Delta-\sum_{i=1}^{n} |S_i|$. Then, if we are interested in the \textit{kinematic basis}, in addition to the structures with the chosen transversality and no mass insertion (\textit{helicity category basis}), we need to consider also the terms in which the equations of motion change the transversality. We consider the example shown in Figure~\ref{fig:unboldinggraphs}, \textit{i.e.} $(1^{1_0},2^{1_0},3^{+1},4^{-1})_{6}$: we also need to classify $(1^{1_{+1}},2^{1_0},3^{+1},4^{-1})_{5}$, $(1^{1_{-1}},2^{1_0},3^{+1},4^{-1})_{5}$, $(1^{1_{0}},2^{1_{+1}},3^{+1},4^{-1})_{5}$,  $(1^{1_{0}},2^{1_{-1}},3^{+1},4^{-1})_{5}$ (multiplied by $\widetilde{m}_1$, $m_1$, $\widetilde{m}_2$, $m_2$ respectively) and $(1^{1_{+1}},2^{1_{+1}},3^{+1},4^{-1})_{4}$, $(1^{1_{-1}},2^{1_{+1}},3^{+1},4^{-1})_{4}$, $(1^{1_{+1}},2^{1_{-1}},3^{+1},4^{-1})_{4}$,  $(1^{1_{-1}},2^{1_{-1}},3^{+1},4^{-1})_{4}$ (all multiplied by $\widetilde{m}_1\, \widetilde{m}_2$, $m_1 \,\widetilde{m}_2$, $\widetilde{m}_1\, m_2$, $m_1\,m_2$)\footnote{We have already taken into account momentum conservation identities, which will be described in the next section. The reader might notice that in the example shown, with the fourth particle being massless, momentum conservation identities are identical to the fully massless case, once we consider the unbolded graphs.}:
\begin{equation}
	\begin{split}
		\{& \agl{\mathbf{1}}{\mathbf{2}}\sqr{\mathbf{1}}{\mathbf{2}} \langle 4 | p_2 | 3 ]^2,\, - \agl{\mathbf{1}}{4}\agl{\mathbf{2}}{4}\sqr{\mathbf{1}}{3}\sqr{\mathbf{2}}{3} \langle 3 | p_2 | 3 ],\,\\
		&-\widetilde{m}_1 \agl{\mathbf{1}}{\mathbf{2}} \agl{\mathbf{1}}{4} \sqr{\mathbf{2}}{3}\langle 4 | p_2 | 3 ],\,-m_1 \sqr{\mathbf{1}}{\mathbf{2}} \agl{\mathbf{2}}{4} \sqr{\mathbf{1}}{3}\langle 4 | p_2 | 3 ]\\
		&-\widetilde{m}_2 \agl{\mathbf{1}}{\mathbf{2}} \agl{\mathbf{2}}{4} \sqr{\mathbf{1}}{3}\langle 4 | p_2 | 3 ],\,
		-m_2 \sqr{\mathbf{1}}{\mathbf{2}} \agl{\mathbf{1}}{4} \sqr{\mathbf{2}}{3}\langle 4 | p_2 | 3 ],\\
		& \widetilde{m}_1 m_2 \agl{\mathbf{1}}{4}^2 \sqr{\mathbf{2}}{3}^2,m_1 \widetilde{m}_2 \agl{\mathbf{2}}{4}^2 \sqr{\mathbf{1}}{3}^2\}\ .
	\end{split}
\end{equation}
and the structures $(1^{1_0},2^{1_0},3^{+1},4^{-1})_4$, multiplied by both $M_1^2$ and $M_2^2$:
\begin{equation}
	\{ M_1^2 \agl{\mathbf{1}}{4}\agl{\mathbf{2}}{4} \sqr{\mathbf{1}}{3} \agl{\mathbf{2}}{3},M_2^2 \agl{\mathbf{1}}{4}\agl{\mathbf{2}}{4} \sqr{\mathbf{1}}{3} \agl{\mathbf{2}}{3}\}\ .
\end{equation}
Such terms are generated by the contractions $p_{i \alpha \dot{\alpha}} p_{i}^{\dot{\alpha} \beta} = M_i^2 \delta_\alpha^\beta$ and $ p_{i}^{\dot{\alpha} \alpha} p_{i \alpha \dot{\beta}} = M_i^2 \delta_{\dot{\beta}}^{\dot{\alpha}}$, or graphically
\begin{figure}[h!]
	\centering
	\begin{tikzpicture}[>=Stealth]
		\node at (0,0.25) {$p_i$};
		\node at (1,0.25) {$p_i$};
		\node at (0,-1.25) {$k$};
		\node at (1,-1.25) {$j$};
		\node at (1.5,-0.5) {$=$};
		\node at (2,-0.5) {$M_i^2$};
		\node at (2.5,-0.25) {$k$};
		\node at (3.5,-0.25) {$j$};
		\node at (3.8,-0.5) {$,$};

		\draw [fill] (0,0) circle [radius = 0.02];
		\draw [fill] (1,0) circle [radius = 0.02];
		\draw [fill] (0,-1) circle [radius = 0.02];
		\draw [fill] (1,-1) circle [radius = 0.02];
		\draw [fill] (2.5,-0.5) circle [radius =0.02];
		\draw [fill] (3.5,-0.5) circle [radius =0.02];

		\draw [->,thick,red] (0,0) -- (1,0);
		\draw [->,thick,blue] (0,0) -- (0,-1);
		\draw [->,thick,blue] (1,0) -- (1,-1);
		\draw [->,thick,blue] (3.5,-0.5) -- (2.5,-0.5);
	\end{tikzpicture}
	\begin{tikzpicture}[>=Stealth]
		\node at (0,0.25) {$p_i$};
		\node at (1,0.25) {$p_i$};
		\node at (0,-1.25) {$k$};
		\node at (1,-1.25) {$j$};
		\node at (1.5,-0.5) {$=$};
		\node at (2,-0.5) {$M_i^2$};
		\node at (2.5,-0.25) {$k$};
		\node at (3.5,-0.25) {$j$};
		\node at (3.8,-0.5) {$.$};

		\draw [fill] (0,0) circle [radius = 0.02];
		\draw [fill] (1,0) circle [radius = 0.02];
		\draw [fill] (0,-1) circle [radius = 0.02];
		\draw [fill] (1,-1) circle [radius = 0.02];
		\draw [fill] (2.5,-0.5) circle [radius =0.02];
		\draw [fill] (3.5,-0.5) circle [radius =0.02];

		\draw [->,thick,blue] (0,0) -- (1,0);
		\draw [->,thick,red] (0,0) -- (0,-1);
		\draw [->,thick,red] (1,0) -- (1,-1);
		\draw [->,thick,red] (3.5,-0.5) -- (2.5,-0.5);
	\end{tikzpicture}
\end{figure}

The total power of the masses cannot exceed the number of momentum insertions in the original structure. In particular, the maximum power of the $i^{\rm th}$-particle mass in the kinematic basis is
\begin{equation}
	\min \left\{ \max\left\{ |C_i - J_i|,|C_i + J_i|\right\},\Delta-\sum_{i=1}^{n} |S_i|\right\}\ .
\end{equation}

\subsection{Momentum Conversation}
\label{sec:massivemomcons}

When dealing with massive structures, momentum conservation identities involve more subtleties. First, we can always choose a particle whose momentum does never appear in the structure: for example, the $n^{\rm th}$-state. Then, we can write the remaining momentum conservation identities as
\begin{equation}
	\label{eq:massiveMomCons}
	\begin{split}
		p_{n-1} | n^I ] &= - \sum_{i=1}^{n-2} p_{i} | n^I ] - p_{n} | n^I ]\ ,\\
		\langle n^I| p_{n-1} & = - \sum_{i=1}^{n-2} \langle n^I| p_{i}  - \langle n^I| p_{n}\ ,\\
		\langle n^I| p_{1} | (n-1)^J ] & = - \sum_{i=2}^{n-2} \langle n^I| p_{i} | (n-1)^J ]\\
		&\hspace{-1cm}- \langle n^I| p_{n} | (n-1)^J ] - \langle n^I| p_{n-1} | (n-1)^J ]\ ,\\
		\langle (n-1)^I| p_{1} | n^J ] & = - \sum_{i=2}^{n-2} \langle (n-1)^I| p_{i} | n^J ]\\
		&\hspace{-1cm} - \langle (n-1)^I| p_{n} | n^J ] - \langle (n-1)^I| p_{n-1} | n^J ]\ ,\\
		2\,  p_{1}\cdot p_{n-1} &= M_n^2 - \sum_{i=1}^{n-2} \sum_{j=i+1}^{n-1} 2\, p_{i}\cdot p_{j} - \sum_{i=1}^{n-1} M_i^2\ ,
	\end{split}
\end{equation}
where the LG indices can be either contracted or not, or not be present at all (as the corresponding particle could be massless). We write these identities such that the terms with higher powers of the masses are independent, \textit{i.e.} they can only be written as a linear combination of the structures with the same or higher mass powers. This allows to effectively set them to zero and work modulo equations of motion ($\Deq$), as explained in the previous section. For example, some of the equations in \eqref{eq:massiveMomCons} look like
\begin{equation}
\label{eq:uptomasses1}
	\begin{split}
		p_{n-1} | n^I ] &\Deq - \sum_{i=1}^{n-2} p_{i} | n^I ] \ ,\\
		\langle (n-1)^I| p_{1} | n^J ] & \Deq - \sum_{i=2}^{n-2} \langle (n-1)^I| p_{i} | n^J ]\ ,\\
		2\,  p_{1}\cdot p_{n-1} &\Deq - \sum_{i=1}^{n-2} \sum_{j=i+1}^{n-1} 2\, p_{i}\cdot p_{j} \ ,
	\end{split}
\end{equation}
which resemble fully massless identities \eqref{eq:momconsCond}. Classifying the structures up to equations of motion means that we can put forward the identification
\begin{equation}
\label{eq:uptomasses2}
	\langle j^J| p_{i} | k^K ]\ \lambda_{i\, \alpha}^I \Deq \agl{j^J}{i^I}\ p_{i} |k^K ]_{ \alpha}\ ,
\end{equation}
and its ``conjugate''. This is equivalent to stating that the momentum conservation conditions for unbolded graphs are identical to the fully massless case, \textit{i.e.} that we should not distinguish between spin and momentum vertices but only keep track of the number of momentum insertions for each particle. In particular, if the $n^{\rm th}$ particle is either massless or $J_n \leq \frac{1}{2}$, momentum conservation identities for unbolded graphs are identical to the fully massless case, once we check if there are insertions of $p_{1}$ or $p_{n-1}$.

On the other hand, there is a subtlety when we consider fully massive structures or, in general, we choose to fully eliminate the momentum of a spin-$J$ massive particle whose transversality $C\neq -J,\, J$ and $J\geq 1$. In particular, in these cases, the structures $\langle n^I| p_{1} | (n-1)^J ]$ and $\langle (n-1)^I| p_{1} | n^J ]$ are not guaranteed to be planar because the spin vertex of the $n^{\rm th}$-particle has non-vanishing valency for the edges corresponding to both squares and angles, even if there are no momentum insertions associated to it. Nevertheless, these non-planar structures give additional momentum conservation constraints which should be taken into account.

An \textit{ad hoc} solution to overcome this problem is the following:
\begin{enumerate}[label=\arabic*),leftmargin=*,align=left]
	\item We classify and count the number $m$ of independent structures, or planar graphs after redefining the valencies of the vertices as
	      \begin{equation}
		      \begin{split}
			      (v_a^1,v_s^1) &\rightarrow (v_a^1-l,v_s^1-l)\ ,\\[.2em]
			      (v_a^{n-1},v_s^{n-1}) &\rightarrow (v_a^{n-1}-l_1,v_s^{n-1}-l_2)\ ,\\[.2em]
			      (v_a^n,v_s^n) &\rightarrow (v_a^n-l_2,v_s^n-l_1)\ ,
		      \end{split}
	      \end{equation}
	      where $l=l_1+l_2$. In this way, we classify the independent structures for which we \textit{factorise} the product $\langle (n-1)^{I_1}| p_{1} | n^{J_1} ]^{l_1} \langle n^{J_2}| p_{1} | (n-1)^{I_2} ]^{l_2}$.
	\item If we restore the \textit{factorised} edges in the graphs obtained in this classification as
	      \begin{equation}
		      \begin{split}
			      A_{i,\, 1\, n-1} &\rightarrow A_{i,\, 1\, n-1} + l_1\ ,\\[.2em]
			      A_{i,\, 1\, n} &\rightarrow A_{i,\, 1\, n} + l_2\ ,\\[.2em]
			      S_{i,\, 1\, n-1} &\rightarrow S_{i,\, 1\, n-1} + l_2\ ,\\[.2em]
			      S_{i,\, 1\, n} &\rightarrow S_{i,\, 1\, n} + l_1\ ,
		      \end{split}
	      \end{equation}
	      for $i=1,\dots ,m$, we obtain a series of planar and non-planar graphs. The structures corresponding to planar graphs in this classification must be removed from our basis.
	\item Non-planar structures must be treated separately. There is a unique source of non-planarity in these graphs and, using iteratively Schouten identities, we can write this non-planar structures as linear combination of planar ones. In particular, we find
	      \begin{equation}
		      \label{eq:massivesubtlety}
		      \centering
		      \begin{tikzpicture}[scale=1.6,>=Stealth]

			      \node at (-0.08,-0.08) {$n$};
			      \node at (-0.08,1.08) {$1$};
			      \node at (1.1,1.1) {$i$};
			      \node at (1.08,-0.08) {$n-1$};
			      \node at (1.25,0.5) {$=$};

			      \draw [fill] (0,0) circle [radius = 0.02];
			      \draw [fill] (1,0) circle [radius = 0.02];
			      \draw [fill] (0,1) circle [radius = 0.02];
			      \draw [fill] (1,1) circle [radius = 0.02];

			      \draw (1,0) [<-,thick,blue] -- (0.55,0.45);
			      \draw [thick, blue] (0.45,0.55) arc (135:-45:0.07);
			      \draw [thick, blue] (0.45,0.55) -- (0,1);
			      \draw [->,thick,blue](1,1) -- (0,0);

			      \draw [->,thick,red] (0,1) -- (0,0);

		      \end{tikzpicture}
		      \begin{tikzpicture}[scale=1.6,>=Stealth]

			      \node at (-0.08,-0.08) {$n$};
			      \node at (-0.08,1.08) {$1$};
			      \node at (1.1,1.1) {$i$};
			      \node at (1.08,-0.08) {$n-1$};
			      \node at (1.25,0.5) {$+$};

			      \draw [fill] (0,0) circle [radius = 0.02];
			      \draw [fill] (1,0) circle [radius = 0.02];
			      \draw [fill] (0,1) circle [radius = 0.02];
			      \draw [fill] (1,1) circle [radius = 0.02];

			      \draw [<-,thick,blue](1,0) -- (1,1);
			      \draw [->,thick,blue](0,1) to [out=-100,in=100] (0,0);

			      \draw [->,thick,red] (0,1) to [out=-80,in=80] (0,0);

		      \end{tikzpicture}
		      \begin{tikzpicture}[scale=1.6,>=Stealth]

			      \node at (-0.08,-0.08) {$n$};
			      \node at (-0.08,1.08) {$1$};
			      \node at (1.1,1.1) {$i$};
			      \node at (1.08,-0.08) {$n-1$};
			      \node at (1.25,0.5) {$,$};

			      \draw [fill] (0,0) circle [radius = 0.02];
			      \draw [fill] (1,0) circle [radius = 0.02];
			      \draw [fill] (0,1) circle [radius = 0.02];
			      \draw [fill] (1,1) circle [radius = 0.02];

			      \draw [->,thick,blue](0,1) -- (1,1);
			      \draw [->,thick,blue](1,0) -- (0,0);

			      \draw [->,thick,red] (0,1) -- (0,0);

		      \end{tikzpicture}
	      \end{equation}
	      and its ``conjugate''. Using momentum conservation on the LHS we can trade $p_1$ insertions with a sum of structures which do not depend on neither $p_1$, $p_{n-1}$ nor $p_n$. This means that each non-planar structure gives a linear constraint for terms appearing on the RHS.
	\item Then, we can \textit{discard} a number $m$ of graphs whose adjacency matrices satisfy the conditions
	      \begin{equation}
		      A_{1\, n} > 0\ ,\hspace{.25cm} S_{1\, n} >0 \ , \hspace{.25cm}\exists i \ \ {\rm s.t.} \ \ S_{i\, n-1} > 0\ ,
	      \end{equation}
	      or
	      \begin{equation}
		      A_{1\, n} > 0\ ,\hspace{.25cm} S_{n-1\, n} >0 \ , \hspace{.25cm}\exists i \ \ {\rm s.t.} \ \ S_{1\, i} > 0\ ,
	      \end{equation}
	      or their ``conjugates''.
\end{enumerate}

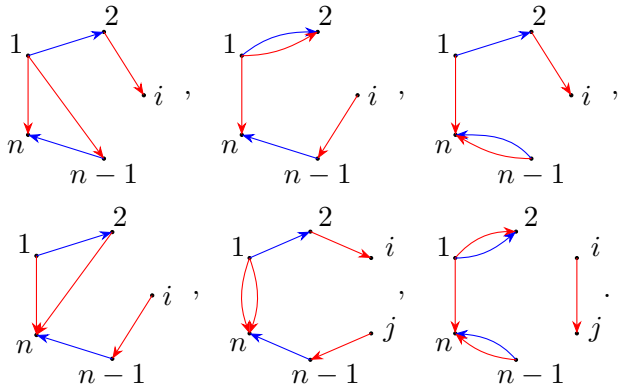
\begin{figure}[t]
	\centering
	\begin{tikzpicture}[scale=1.05,>=Stealth]

		\node at (-0.15,-0.15) {$n$};
		\node at (-0.15,1.15) {$1$};
		\node at (1.05,1.50) {$2$};
		\node at (1.64,0.5) {$i$};
		\node at (0.95,-0.5) {$n-1$};
		\node at (2,0.5) {$,$};

		\draw [fill] (0,0) circle [radius = 0.02];
		\draw [fill] (0,1) circle [radius = 0.02];
		\draw [fill] (0.95,1.3) circle [radius = 0.02];
		\draw [fill] (1.45,0.5) circle [radius = 0.02];
		\draw [fill] (0.95,-0.3) circle [radius = 0.02];

		\draw [->,blue] (0,1) -- (0.95,1.3);
		\draw [->,red] (0,1) -- (0,0);
		\draw [->,blue] (0.95,-0.3) -- (0,0);

		\draw [->,red] (0,1) -- (0.95,-0.3);
		\draw [->,red] (0.95,1.3) -- (1.45,0.5);

	\end{tikzpicture}
	\begin{tikzpicture}[scale=1.05,>=Stealth]

		\node at (-0.15,-0.15) {$n$};
		\node at (-0.15,1.15) {$1$};
		\node at (1.05,1.50) {$2$};
		\node at (1.64,0.5) {$i$};
		\node at (0.95,-0.5) {$n-1$};
		\node at (2,0.5) {$,$};

		\draw [fill] (0,0) circle [radius = 0.02];
		\draw [fill] (0,1) circle [radius = 0.02];
		\draw [fill] (0.95,1.3) circle [radius = 0.02];
		\draw [fill] (1.45,0.5) circle [radius = 0.02];
		\draw [fill] (0.95,-0.3) circle [radius = 0.02];

		\draw [->,blue] (0,1) to [out=40,in=180] (0.95,1.3);
		\draw [->,red] (0,1) -- (0,0);
		\draw [->,blue] (0.95,-0.3) -- (0,0);

		\draw [->,red] (0,1) to [out=0,in=210] (0.95,1.3);
		\draw [<-,red] (0.95,-0.3) -- (1.45,0.5);

	\end{tikzpicture}
	\begin{tikzpicture}[scale=1.05,>=Stealth]

		\node at (-0.15,-0.15) {$n$};
		\node at (-0.15,1.15) {$1$};
		\node at (1.05,1.50) {$2$};
		\node at (1.64,0.5) {$i$};
		\node at (0.95,-0.5) {$n-1$};
		\node at (2,0.5) {$,$};

		\draw [fill] (0,0) circle [radius = 0.02];
		\draw [fill] (0,1) circle [radius = 0.02];
		\draw [fill] (0.95,1.3) circle [radius = 0.02];
		\draw [fill] (1.45,0.5) circle [radius = 0.02];
		\draw [fill] (0.95,-0.3) circle [radius = 0.02];

		\draw [->,blue] (0,1) -- (0.95,1.3);
		\draw [->,red] (0,1) -- (0,0);
		\draw [->,blue] (0.95,-0.3) to [out=130,in=0] (0,0);

		\draw [->,red] (0.95,1.3) -- (1.45,0.5);
		\draw [->,red] (0.95,-0.3) to [out=180,in=-30] (0,0);

	\end{tikzpicture}\\
	\begin{tikzpicture}[scale=1.05,>=Stealth]

		\node at (-0.15,-0.15) {$n$};
		\node at (-0.15,1.15) {$1$};
		\node at (1.05,1.50) {$2$};
		\node at (1.64,0.5) {$i$};
		\node at (0.95,-0.5) {$n-1$};
		\node at (2,0.5) {$,$};

		\draw [fill] (0,0) circle [radius = 0.02];
		\draw [fill] (0,1) circle [radius = 0.02];
		\draw [fill] (0.95,1.3) circle [radius = 0.02];
		\draw [fill] (1.45,0.5) circle [radius = 0.02];
		\draw [fill] (0.95,-0.3) circle [radius = 0.02];

		\draw [->,blue] (0,1) -- (0.95,1.3);
		\draw [->,red] (0,1) -- (0,0);
		\draw [->,blue] (0.95,-0.3) -- (0,0);

		\draw [->,red] (1.45,0.5) -- (0.95,-0.3);
		\draw [->,red] (0.95,1.3) -- (0,0);

	\end{tikzpicture}
	\begin{tikzpicture}[scale=1,>=Stealth]

		\node at (-0.15,-0.15) {$n$};
		\node at (-0.15,1.15) {$1$};
		\node at (1,1.55) {$2$};
		\node at (1.85,1.15) {$i$};
		\node at (1.85,0.05) {$j$};
		\node at (1,-0.55) {$n-1$};
		\node at (2,0.5) {$,$};

		\draw [fill] (0,0) circle [radius = 0.02];
		\draw [fill] (0,1) circle [radius = 0.02];
		\draw [fill] (1.6,0) circle [radius = 0.02];
		\draw [fill] (1.6,1) circle [radius = 0.02];
		\draw [fill] (0.8,-0.35) circle [radius = 0.02];
		\draw [fill] (0.8,1.35) circle [radius = 0.02];

		\draw [->,red] (0,1) to [out=-110,in=110] (0,0);
		\draw [->,red] (0,1) to [out=-70,in=70] (0,0);
		\draw [->,blue] (0,1) -- (0.8,1.35);
		\draw [->,red] (0.8,1.35) -- (1.6,1);
		\draw [->,red] (1.6,0) -- (0.8,-0.35);
		\draw [->,blue] (0.8,-0.35) -- (0,0);

	\end{tikzpicture}
	\begin{tikzpicture}[scale=1,>=Stealth]

		\node at (-0.15,-0.15) {$n$};
		\node at (-0.15,1.15) {$1$};
		\node at (1,1.55) {$2$};
		\node at (1.85,1.15) {$i$};
		\node at (1.85,0.05) {$j$};
		\node at (1,-0.55) {$n-1$};
		\node at (2,0.5) {$.$};

		\draw [fill] (0,0) circle [radius = 0.02];
		\draw [fill] (0,1) circle [radius = 0.02];
		\draw [fill] (1.6,0) circle [radius = 0.02];
		\draw [fill] (1.6,1) circle [radius = 0.02];
		\draw [fill] (0.8,-0.35) circle [radius = 0.02];
		\draw [fill] (0.8,1.35) circle [radius = 0.02];

		\draw [->,red] (0,1) -- (0,0);
		\draw [->,red] (0,1) to [out=50,in=180] (0.8,1.35);
		\draw [->,blue] (0,1) to [out=0,in=220] (0.8,1.35);
		\draw [->,red] (1.6,1) -- (1.6,0);
		\draw [->,blue] (0.8,-0.35) to [out=130,in=-5] (0,0);
		\draw [->,red] (0.8,-0.35) to [out=175,in=-40] (0,0);

	\end{tikzpicture}
	\caption{Planar structures to eliminate corresponding to the factorised term $\agl{\mathbf{1}}{\mathbf{n}} \sqr{(\mathbf{n-1})}{\mathbf{n}} \langle (\mathbf{n-1}) | p_2 | \mathbf{1}]$.}
	\label{fig:extramassive}
\end{figure}

Surprisingly, if we consider cases in which $J_i \geq 1$ and $C_i \neq - J_i,\, J_i$ for $i=1,\, n-1,\, n$, there is an additional relation to take into account:
\begin{equation}
	\label{eq:extramassiveconstraint}
	\begin{split}
		\agl{\mathbf{1}}{\mathbf{n}} \sqr{(\mathbf{n-1})}{\mathbf{n}} &\langle (\mathbf{n-1}) | \sum_{i=2}^{n-2} p_i | \mathbf{1}] \Deq \sqr{\mathbf{1}}{\mathbf{n}} \agl{(\mathbf{n-1})}{\mathbf{n}} \langle \mathbf{1} | \sum_{i=2}^{n-2} p_i | \mathbf{n-1}]
	\end{split}
\end{equation}
In order for the terms in this additional relation to be independent from the momentum conservation conditions already considered, the momenta $p_1$, $p_{n-1}$, and $p_n$ must be massive and none of them can appear in the structures in \eqref{eq:extramassiveconstraint}. The algorithm to eliminate this additional constraint is identical to the one just presented, except for the factorised structure, which we choose to be $\agl{\mathbf{1}}{\mathbf{n}} \sqr{(\mathbf{n-1})}{\mathbf{n}} \langle (\mathbf{n-1}) | p_2 | \mathbf{1}]$, and the corresponding planar structures that we need to eliminate are shown in Figure~\ref{fig:extramassive}.

These algorithms give the basis of the independent kinematic structures we are looking for, modulo Schouten identities, momentum conservation, and equation of motion. To find the complete \textit{kinematic basis} we need to consider all the \textit{helicity categories bases} with lower mass dimensions multiplied by proper mass powers, as shown in Section~\ref{sec:equationofmotion}.

\subsection{A summary of the algorithm}

We now present a summary of the algorithm for minimal amplitudes, including massive particles. The general structure is the same presented for fully massless amplitudes, with few key differences.
\begin{enumerate}
	\item We start with an initial trivial assignment of vertex valencies determined by the field content of the operators and the distribution of momentum insertions (\textit{i.e.} mass dimensions).
	\item We are interested in the helicity category basis, then we will work up to terms with explicit powers of the masses, \textit{i.e.} we do not distinguish free spinors and momentum vertices in the graphs.
	\item We generate the corresponding structures which are kinematically independent\footnote{As in the massless case, this classification correspond to the \texttt{UniformMassStructures} function in the \texttt{MassiveGraphs} code. If we are also interested in the terms proportional to powers of the masses, \textit{i.e.} to the kinematic basis, the function to use is \texttt{IndependentSpinStructures}.}:
	      \begin{enumerate}
		      \item We classify all the planar graphs.
		      \item From this set of graphs, we exclude some of the graphs thanks to momentum conservation, as explained in Section~\ref{sec:massivemomcons}.
		      \item To each of the remaining graphs, we can associate a unique planar massive graph\footnote{For the interested reader, the bolding map at the level of graphs is not shown in this paper, but it can be found explicitly coded in the \texttt{MassiveGraphs} package, which is publicly available.}.
		      \item Using the massive generalisation of the map $\mathbb{M}$, we write down our basis of kinematically independent spinor structures in the helicity category basis.
	      \end{enumerate}
	\item When dealing with identical particles, we also need to decompose some structures which do not appear in our basis. Since our algorithm relies on the unbolding/bolding procedure at the level of graphs, we can find linear relations only up to terms with masses, like equations \eqref{eq:uptomasses1}, \eqref{eq:uptomasses2} or \eqref{eq:extramassiveconstraint}. The details can be extrapolated from point (4) in Section~\ref{sec:masslessAlgorithm} and the discussion of Section~\ref{sec:massivemomcons}. Even if such linear relations are not complete, the information they provide is enough, as the missing terms are proportional for minimal amplitude with the same field content, but a smaller mass dimension.
	\item After this decomposition we end up with linear combinations of terms in the chosen basis and transforming properly under permutations of the particles.
	\item Finally we check whether there are linear relations between such terms.
	\item The checks on the algorithm are exactly the same presented in Section~\ref{sec:checks}.
\end{enumerate}

\section{Applications}
\label{sec:applications}

In this section, we are going to present some applications of our method for the classification of irrelevant interactions in effective theories. The fully massless algorithm has already been used in \cite{AccettulliHuber:2021uoa} to list the SMEFT irrelevant operators up to dimension eight.

The method presented in this paper can be applied to any number of particles with arbitrary helicity and spin. In particular, we will show, as basic applications of our method, the classification of $D^{2n} F^4$ effective interactions in SU$(N)$ Yang-Mills theories, five-point effective interactions involving $W$, $Z$ and $\gamma$ vector bosons and spin-tidal interactions in gravity. In particular, in the last case, we will show how our method is related to the strategy presented in \cite{Durieux:2020gip}, explicitly showing the mass-complete relations relevant to the case considered.

We will briefly mention how to treat identical particles and colour structures, which have been extensively studied in the literature \cite{Durieux:2020gip,Henning:2015alf,Henning:2017fpj,Fonseca:2019yya,Li:2020gnx,Chowdhury:2020ddc}. We will stick to the strategy and the conventions presented in \cite{AccettulliHuber:2021uoa}.

All algorithms have been implement in the \texttt{Mathematica} package \href{https://github.com/StefanoDeAngelis/MassiveEFT-Operators}{\texttt{MassiveGraphs}}, which make use of the \href{https://github.com/StefanoDeAngelis/SpinorHelicity}{\texttt{SpinorHelicity}} package. More complicated examples can be solved using these codes.

\subsection{\texorpdfstring{$D^{2n}F^4$}{D2nF4} interactions in gauge theories}

We consider now a simple example in which all the particles are massless vector bosons in SU$(N)$ Yang-Mills theory with $N>3$. We will consider in order the three independent helicity configurations $(++++)$, $(+++-)$ and $(++--)$.

\subsubsection{All-plus configuration}

The algorithm provides us with a basis of kinematically independent structures which are compatible with the mass dimension $4+2n$ and the chosen helicity configuration. In particular, we find
\begin{equation}
	\begin{split}
		s_{12}^n \sqr{1}{2}^2\sqr{3}{4}^2&\, ,\  s_{12}^n \sqr{1}{4}^2\sqr{2}{3}^2\, ,\ s_{12}^n \sqr{1}{2}\sqr{2}{3}\sqr{3}{4}\sqr{4}{1}\, , \{s_{23}^{n-k} s_{12}^k \sqr{1}{4}^2\sqr{2}{3}^2 \}_{k=0,\dots n-1}\ ,
	\end{split}
\end{equation}
which correspond to the only $n+3$ graphs that meet all the requirements stated in Section~\ref{sec:massless}. A basis of independent colour structures for $N>3$ is
\begin{equation}
	\begin{split}
		\mathcal{C} &=\{\delta^{A_1A_4} \delta^{A_2A_3}\, ,\ \delta^{A_1A_3} \delta^{A_2A_4}\, ,\ \delta^{A_1A_2} \delta^{A_3A_4}\} \cup\{\tau^{A_1 A_i A_j A_k}\}_{(i,j,k)=\mathcal{P}_3(2,3,4)}\ ,
	\end{split}
\end{equation}
where $\mathcal{P}_3(2,3,4)$ corresponds to the permutations of $(2,3,4)$, and $\tau^{A_1 A_2 A_3 A_4}$ is the trace of four SU$(N)$ generators $\tau^{A}$.

These two bases must be combined to find the effective interactions we are looking for. At this point, we have $9\times (3+n)$ terms and, as we are dealing with identical particles, we need to sum over all the permutations of the external legs in these structures. For example, we can consider $\delta^{A_1A_4} \delta^{A_2A_3} s_{12}^n \sqr{1}{2}^2\sqr{3}{4}^2$:
\begin{equation}
	\begin{split}
		\mathrm{Sym}\, \delta^{A_1A_2} \delta^{A_3A_4} s_{12}^n \sqr{1}{2}^2\sqr{3}{4}^2 &\equiv \frac{1}{3} \delta^{A_1A_2} \delta^{A_3A_4} s_{12}^n \sqr{1}{2}^2\sqr{3}{4}^2 + \frac{1}{3}\delta^{A_1A_3} \delta^{A_2A_4} s_{13}^n \sqr{1}{3}^2\sqr{2}{4}^2\\
		&+ \frac{1}{3}\delta^{A_1A_4} \delta^{A_2A_3} s_{23}^n \sqr{1}{4}^2\sqr{2}{3}^2\ .
	\end{split}
\end{equation}
The structure $s_{13}^n \sqr{1}{3}^2\sqr{2}{4}^2$ does not belong to our basis and we have to rewrite it as a linear combination of elements of our basis. We can do this using the algorithm presented in Section~\ref{sec:massless} and we always verify such relations on rational kinematics, as presented in Appendix~\ref{sec:massivetwistors}. This means that, after symmetrisation, not all the $9\times (3+n)$ structures are kinematically independent. Indeed, it turns out that only $4+2\lfloor\frac{n}{2}\rfloor$ structures actually are. We are going to present a basis of effective interactions for $n\leq 8$, which we will denote by $\mathcal{B}_{4 + 2n}$:
\begin{align*}
	\mathcal{B}_{4}  & =(\sqr{1}{4}^2\sqr{2}{3}^2\times \mathcal{C}_1) \cup ( \sqr{1}{4}^2\sqr{2}{3}^2\times \mathcal{C}_2)\ ,    \\
	\mathcal{B}_{6}  & = s_{2 3}\times \mathcal{B}_{4}\ ,                                                                         \\
	\mathcal{B}_{8}  & = (s_{1 2}\times \mathcal{B}_{6} ) \cup ( s_{1 2}s_{2 3}\sqr{1}{4}^2\sqr{2}{3}^2\times \mathcal{C}_3  ),   \\
	\mathcal{B}_{10} & = s_{23}\times \mathcal{B}_{8}\ ,                                                                          \\
	\mathcal{B}_{12} & = (s_{12}\times \mathcal{B}_{10}  )\cup ( s_{12}s_{23}^3 \sqr{1}{4}^2\sqr{2}{3}^2 \times\mathcal{C}_1 )\ , \\
	\mathcal{B}_{14} & = s_{23}\times \mathcal{B}_{12}\ ,                                                                         \\
	\mathcal{B}_{16} & = (s_{12}^2 s_{23}^2\times \mathcal{B}_{8} ) \cup (s_{12}^2s_{23}^4\times \mathcal{B}_4 )\ ,               \\
	\mathcal{B}_{18} & = s_{23} \mathcal{B}_{16}\ ,                                                                               \\
	\mathcal{B}_{20} & = (s_{12}^4 s_{23}^4\times \mathcal{B}_{8} ) \cup ( s_{12}^3s_{23}^5 \times\mathcal{B}_8 )\ ,
\end{align*}
where
\begin{align*}
	\mathcal{C}_1 & = \{\delta^{A_1A_4} \delta^{A_2A_3},\tau^{A_1 A_2 A_3 A_4}\}\ ,  \\
	\mathcal{C}_2 & = \{ \delta^{A_1A_3} \delta^{A_2A_4},\tau^{A_1 A_2 A_4 A_3}\}\ , \\
	\mathcal{C}_3 & =\{ \delta^{A_1A_2} \delta^{A_3A_4},\tau^{A_1 A_3 A_2 A_4} \}\ ,
\end{align*}
and the symmetrisation is understood for each element in the lists. For example, the first element in $\mathcal{B}_4$ is
\begin{equation}
	\begin{split}
		\delta^{A_1A_4} \delta^{A_2A_3} \sqr{1}{4}^2\sqr{2}{3}^2 \to\ & \frac{1}{3}\delta^{A_1A_2} \delta^{A_3A_4} \sqr{1}{2}^2\sqr{3}{4}^2+\frac{1}{3}\delta^{A_1A_3} \delta^{A_2A_4} \sqr{1}{3}^2\sqr{2}{4}^2\\
		+&\frac{1}{3}\delta^{A_1A_4} \delta^{A_2A_3} \sqr{1}{4}^2\sqr{2}{3}^2\ .
	\end{split}
\end{equation}

\subsubsection{Single-minus configuration}
In this case, the basis of kinematically independent structures consists of $n$ elements:
\begin{equation}
	\{ s_{12}^{n-k-1} s_{23}^k \agl{2}{4}^2 \sqr{1}{2}^2 \sqr{2}{3}^2 \}_{k=0,\dots n-1}\ .
\end{equation}
After combining with the colour basis and symmetrising over the $(1,2,3)$ we end up with $\lfloor \frac{3n+1}{2}\rfloor$ independent contact terms:
\begin{align*}
	\mathcal{B}_{6}  & =(\agl{2}{4}^2 \sqr{1}{2}^2 \sqr{2}{3}^2\times \mathcal{C}_1)\ ,                                                                                                                           \\
	\mathcal{B}_{8}  & = (s_{2 3}\times \mathcal{B}_{6})\cup \{ \tau^{A_1 A_{2} A_{4} A_{3}}s_{2 3}\agl{2}{4}^2 \sqr{1}{2}^2 \sqr{2}{3}^2\}\ ,                                                                    \\
	\mathcal{B}_{10} & = s_{12} s_{23}\agl{2}{4}^2 \sqr{1}{2}^2 \sqr{2}{3}^2\times (\mathcal{C}_1 \cup \mathcal{C}_{2} ) \cup \{ \tau^{A_1 A_{3} A_4 A_2} s_{12} s_{23}\agl{2}{4}^2 \sqr{1}{2}^2 \sqr{2}{3}^2 \}, \\
	\mathcal{B}_{12} & = s_{23}\times \mathcal{B}_{10}\cup \{ \tau^{A_1 A_{3} A_{2} A_{4}}s_{1 2}s_{2 3}^2\agl{2}{4}^2 \sqr{1}{2}^2 \sqr{2}{3}^2\}\ ,                                                             \\
	\mathcal{B}_{14} & = s_{12}\times \mathcal{B}_{12}\cup s_{1 2}s_{2 3}^3\agl{2}{4}^2 \sqr{1}{2}^2 \sqr{2}{3}^2\times \mathcal{C}_1\ ,                                                                          \\
	\mathcal{B}_{16} & =s_{1 2}^2s_{2 3}^3\agl{2}{4}^2 \sqr{1}{2}^2 \sqr{2}{3}^2\times \mathcal{C}\ ,                                                                                                             \\
	\mathcal{B}_{18} & = s_{12}^2 s_{23}^2\times \mathcal{B}_{10}\cup s_{12} s_{23}^2\times \mathcal{B}_{12}\ ,                                                                                                   \\
	\mathcal{B}_{20} & = s_{1 2}s_{23}\times \mathcal{B}_{16} \cup s_{12}^2 s_{23}^4 \times \mathcal{B}_{8}\ ,
\end{align*}
where the symmetrisation is understood for each element in the lists.

\subsubsection{MHV configuration}
In this configuration the basis of kineamatically independent structures consists of $n+1$ elements:
\begin{equation}
	\begin{split}
		s_{12}^n \sqr{1}{2}^2 \agl{3}{4}^2 , \ s_{1 2}^{n-1} \sqr{1}{2}^2 \agl{3}{4} \langle 3 | p_2 p_1| 4\rangle,
		\{ s_{23}^{n-k-2} s_{12}^{k} \sqr{1}{2}^2 \langle 3 | p_2 p_1| 4\rangle^2\}_{k=0,\dots ,n-2}\ ,
	\end{split}
\end{equation}
where the negative powers of the Mandelstam invariants for $n=0,1$ mean that such a structure is not in the basis.

The number of effective interactions in the SU$(N)$ gauge theories is $4+\lfloor \frac{7n}{2}\rfloor$ and the choice of basis is given by
\begin{align*}
	\mathcal{B}_4    & = \sqr{1}{2}^2 \agl{3}{4}^2\times (\mathcal{C}_1\cup \mathcal{C}_3)\ ,                                                                                                                                               \\
	\mathcal{B}_6    & = s_{12}\times \mathcal{B}_4 \cup \sqr{1}{2}^2 \agl{3}{4} \langle 3 | p_2 p_1| 4\rangle\times \mathcal{C}_1^\prime\ ,                                                                                                \\
	\mathcal{B}_8    & = s_{12}^2\times \mathcal{B}_4 \cup \sqr{1}{2}^2 \langle 3 | p_2 p_1| 4\rangle^2 \times(\mathcal{C}_1 \cup \mathcal{C}_2\cup \mathcal{C}_3 \cup \{\tau^{A_1 A_4 A_2 A_3}\})\ ,                                       \\
	\mathcal{B}_{10} & = s_{12}\times \mathcal{B}_8 \cup s_{23} \sqr{1}{2}^2 \langle 3 | p_2 p_1| 4\rangle^2 \times \mathcal{C}_1^\prime\ ,                                                                                                 \\
	\mathcal{B}_{12} & = s_{12} \times \mathcal{B}_{10} \cup s_{23}^{2} \sqr{1}{2}^2 \langle 3 | p_2 p_1| 4\rangle^2\times (\mathcal{C}_1\cup \mathcal{C}_3)\ ,                                                                             \\
	\mathcal{B}_{14} & = s_{12} \times \mathcal{B}_{12} \cup s_{23}^{3} \sqr{1}{2}^2 \langle 3 | p_2 p_1| 4\rangle^2\times \mathcal{C}_1^\prime\ ,                                                                                          \\
	\mathcal{B}_{16} & = s_{12}^2 \times \mathcal{B}_{12} \cup s_{23}^{3} \sqr{1}{2}^2 \langle 3 | p_2 p_1| 4\rangle^2 \times(\mathcal{C}_1 \cup \mathcal{C}_2\cup \mathcal{C}_3 \cup \{\tau^{A_1 A_4 A_2 A_3}\})\ ,                        \\
	\mathcal{B}_{18} & = s_{12}^3 \times \mathcal{B}_{12} \cup s_{23}\times (\mathcal{B}_{16}\setminus s_{12}^2\times \mathcal{B}_{12}) \cup s_{12}^2 s_{23}^{3} \sqr{1}{2}^2 \langle 3 | p_2 p_1| 4\rangle^2 \times\mathcal{C}_1^\prime\ , \\
	\mathcal{B}_{20} & = s_{12}^4 \times \mathcal{B}_{12} \cup s_{23}^2\times (\mathcal{B}_{16}\setminus s_{12}^2\times \mathcal{B}_{12}) \cup s_{12}^3 s_{23}^{3} \sqr{1}{2}^2 \langle 3 | p_2 p_1| 4\rangle^2 \times\mathcal{C}_1^\prime  \\
	                 & \hspace{5.15cm}\cup s_{12}^3 s_{23}^{3} \sqr{1}{2}^2 \langle 3 | p_2 p_1| 4\rangle^2 \times(\mathcal{C}_1\cup \mathcal{C}_3) \ ,
\end{align*}
where
\begin{equation}
	\mathcal{C}_1^\prime = \mathcal{C}_1 \cup \{\tau^{A_1 A_3 A_2 A_4}\}\ .
\end{equation}
Clearly, the basis we found is not the most symmetric and recursive. In this section, we wanted to show how our method can systematically deal with this problem of classification. But it is easy to start from our basis and find more symmetric ones, as we will show explicitly in Section~\ref{sec:spintidal}.

\subsection{Five-point interactions between \texorpdfstring{$W$}{W}, \texorpdfstring{$Z$}{Z} and \texorpdfstring{$\gamma$}{gamma}}
\label{sec:WZgamma}
In this section we show how our algorithm can be applied beyond the results of \cite{Durieux:2020gip} (beyond four-point and purely massive amplitudes), classifying the effective interaction corresponding to dimension-6 operators at five-point, with massive (both charged and uncharged) and U$(1)$ massless vector bosons, which we call $W^\pm$, $Z$ and $\gamma$. Such operators can appear in various combinations: $D(W^+)^2(W^-)^2Z$, $DW^+W^-Z^3$, $D Z^5$, $F_\gamma (W^+)^2(W^-)^2$, $F_\gamma W^+W^-Z^2$ and $F_\gamma Z^4$. Now we will deal with the purely massive cases and the mixed case separately, as the purely kinematic structures are common in the two cases.

\subsubsection{Five-point massive effective interactions}
First, we need to classify all possible contact terms using the algorithm presented in Section~\ref{sec:massive}: we need terms with mass dimension 6 and in the helicity category $(1^{1_0},2^{1_0},3^{1_0},$ $4^{1_0},5^{1_0})$, and terms with mass dimension 5 in the various helicity categories $(1^{1_{\pm 1}},2^{1_0},$ $3^{1_0},4^{1_0},5^{1_0})$, plus the permutations needed to distinguish different particles. We are not going to write all the structures explicitly as they are of order $10^2$. As in the purely massless case, after symmetrising over identical particles we find structures that are not in our basis of kinematic independent structures. Nevertheless, we can use our algorithm to write down explicitly the decomposition of such structures. 
When doing so, it is important to distinguish between $m_i$, $\widetilde{m}_i$, and $M_i$, and only at the very end set $m_i = \widetilde{m}_i = M_i$, as explained in the Appendix~\ref{sec:massivetwistors}.

It is important to note that in the massive case, such relations contain terms across different helicity categories, which enter in the relations multiplied by powers of the masses. Then, the linear independence of the symmetrised terms can be checked by either setting all the masses to zero and working with a fixed helicity category, or by keeping the masses and considering the independence across the different categories in order of increasing mass dimension.

The number of independent contact terms in the three cases under consideration $D(W^+)^2(W^-)^2Z$, $DW^+W^-Z^3$, $D Z^5$ are, respectively, 20, 14 and 0:
\begin{align*}
	\mathcal{B}_6^{W^4Z} = & \{M_Z \agl{\mathbf{1}}{\mathbf{5}} \agl{\mathbf{2}}{\mathbf{3}} \agl{\mathbf{4}}{\mathbf{5}} \sqr{\mathbf{1}}{\mathbf{4}} \sqr{\mathbf{2}}{\mathbf{3}}, M_W \agl{\mathbf{1}}{\mathbf{5}} \agl{\mathbf{2}}{\mathbf{4}} \agl{\mathbf{3}}{\mathbf{4}} \sqr{\mathbf{1}}{\mathbf{5}} \sqr{\mathbf{2}}{\mathbf{3}},M_W \agl{\mathbf{1}}{\mathbf{4}} \agl{\mathbf{2}}{\mathbf{3}} \agl{\mathbf{4}}{\mathbf{5}} \sqr{\mathbf{1}}{\mathbf{5}} \sqr{\mathbf{2}}{\mathbf{3}},     \\
	                       & M_W \agl{\mathbf{1}}{\mathbf{2}} \agl{\mathbf{3}}{\mathbf{4}} \agl{\mathbf{4}}{\mathbf{5}} \sqr{\mathbf{1}}{\mathbf{5}} \sqr{\mathbf{2}}{\mathbf{3}},M_W \agl{\mathbf{2}}{\mathbf{5}} \agl{\mathbf{3}}{\mathbf{4}} \sqr{\mathbf{1}}{\mathbf{4}} \sqr{\mathbf{1}}{\mathbf{5}} \sqr{\mathbf{2}}{\mathbf{3}},M_W \agl{\mathbf{2}}{\mathbf{3}} \agl{\mathbf{4}}{\mathbf{5}} \sqr{\mathbf{1}}{\mathbf{4}} \sqr{\mathbf{1}}{\mathbf{5}} \sqr{\mathbf{2}}{\mathbf{3}},        \\
	                       & M_W \agl{\mathbf{1}}{\mathbf{3}} \agl{\mathbf{4}}{\mathbf{5}} \sqr{\mathbf{1}}{\mathbf{5}} \sqr{\mathbf{2}}{\mathbf{3}} \sqr{\mathbf{2}}{\mathbf{4}}, M_W \agl{\mathbf{1}}{\mathbf{2}} \agl{\mathbf{2}}{\mathbf{5}} \agl{\mathbf{3}}{\mathbf{4}} \sqr{\mathbf{1}}{\mathbf{5}} \sqr{\mathbf{3}}{\mathbf{4}},M_W \agl{\mathbf{1}}{\mathbf{5}} \agl{\mathbf{2}}{\mathbf{4}} \sqr{\mathbf{1}}{\mathbf{5}} \sqr{\mathbf{2}}{\mathbf{3}} \sqr{\mathbf{3}}{\mathbf{4}},       \\
	                       & M_W \agl{\mathbf{1}}{\mathbf{2}} \agl{\mathbf{4}}{\mathbf{5}} \sqr{\mathbf{1}}{\mathbf{5}} \sqr{\mathbf{2}}{\mathbf{3}} \sqr{\mathbf{3}}{\mathbf{4}}, M_W \agl{\mathbf{1}}{\mathbf{5}} \agl{\mathbf{2}}{\mathbf{3}} \agl{\mathbf{2}}{\mathbf{4}} \sqr{\mathbf{1}}{\mathbf{3}} \sqr{\mathbf{4}}{\mathbf{5}}, M_W \agl{\mathbf{1}}{\mathbf{2}} \agl{\mathbf{2}}{\mathbf{3}} \agl{\mathbf{4}}{\mathbf{5}} \sqr{\mathbf{1}}{\mathbf{3}} \sqr{\mathbf{4}}{\mathbf{5}},      \\
	                       & M_W \agl{\mathbf{1}}{\mathbf{5}} \agl{\mathbf{2}}{\mathbf{4}} \sqr{\mathbf{1}}{\mathbf{3}} \sqr{\mathbf{2}}{\mathbf{3}} \sqr{\mathbf{4}}{\mathbf{5}}, M_Z \agl{\mathbf{1}}{\mathbf{4}} \agl{\mathbf{2}}{\mathbf{3}} \sqr{\mathbf{1}}{\mathbf{5}} \sqr{\mathbf{2}}{\mathbf{3}}\sqr{\mathbf{4}}{\mathbf{5}},\agl{\mathbf{1}}{\mathbf{3}} \agl{\mathbf{4}}{\mathbf{5}} \sqr{\mathbf{1}}{\mathbf{5}} \sqr{\mathbf{3}}{\mathbf{4}} \langle \mathbf{2}|p_1|\mathbf{2}],      \\
	                       & \agl{\mathbf{1}}{\mathbf{3}} \agl{\mathbf{4}}{\mathbf{5}} \sqr{\mathbf{1}}{\mathbf{3}} \sqr{\mathbf{4}}{\mathbf{5}} \langle \mathbf{2}|p_1|\mathbf{2}],\agl{\mathbf{1}}{\mathbf{2}} \agl{\mathbf{4}}{\mathbf{5}} \sqr{\mathbf{1}}{\mathbf{5}} \sqr{\mathbf{2}}{\mathbf{4}} \langle \mathbf{3}|p_2|\mathbf{3}], \agl{\mathbf{1}}{\mathbf{5}} \agl{\mathbf{2}}{\mathbf{4}} \sqr{\mathbf{1}}{\mathbf{5}} \sqr{\mathbf{2}}{\mathbf{4}} \langle \mathbf{3}|p_2|\mathbf{3}], \\
	                       & \agl{\mathbf{1}}{\mathbf{5}} \agl{\mathbf{2}}{\mathbf{3}} \sqr{\mathbf{1}}{\mathbf{5}} \sqr{\mathbf{2}}{\mathbf{3}} \langle \mathbf{4}|p_3|\mathbf{4}], \agl{\mathbf{1}}{\mathbf{5}} \agl{\mathbf{2}}{\mathbf{3}} \sqr{\mathbf{2}}{\mathbf{3}} \sqr{\mathbf{4}}{\mathbf{5}} \langle \mathbf{4}|p_3|\mathbf{1}]\}\ ,
\end{align*}
and
\begin{align*}
	\mathcal{B}_6^{W^2Z^3} = & \{M_Z \agl{\mathbf{1}}{\mathbf{5}} \agl{\mathbf{2}}{\mathbf{5}} \agl{\mathbf{3}}{\mathbf{4}} \sqr{\mathbf{1}}{\mathbf{4}} \sqr{\mathbf{2}}{\mathbf{3}}, M_Z \agl{\mathbf{1}}{\mathbf{5}} \agl{\mathbf{2}}{\mathbf{3}} \agl{\mathbf{4}}{\mathbf{5}} \sqr{\mathbf{1}}{\mathbf{4}} \sqr{\mathbf{2}}{\mathbf{3}},M_Z \agl{\mathbf{1}}{\mathbf{2}}
	\agl{\mathbf{3}}{\mathbf{5}} \agl{\mathbf{4}}{\mathbf{5}} \sqr{\mathbf{1}}{\mathbf{4}} \sqr{\mathbf{2}}{\mathbf{3}},                                                                                                                                                                                                                                                                                                                                                                            \\
	                         & M_W \agl{\mathbf{2}}{\mathbf{5}} \agl{\mathbf{3}}{\mathbf{4}} \sqr{\mathbf{1}}{\mathbf{4}} \sqr{\mathbf{1}}{\mathbf{5}} \sqr{\mathbf{2}}{\mathbf{3}},M_W \agl{\mathbf{1}}{\mathbf{3}} \agl{\mathbf{4}}{\mathbf{5}} \sqr{\mathbf{1}}{\mathbf{5}} \sqr{\mathbf{2}}{\mathbf{3}} \sqr{\mathbf{2}}{\mathbf{4}}, M_W \agl{\mathbf{1}}{\mathbf{2}} \agl{\mathbf{2}}{\mathbf{5}} \agl{\mathbf{3}}{\mathbf{4}} \sqr{\mathbf{1}}{\mathbf{5}} \sqr{\mathbf{3}}{\mathbf{4}},     \\
	                         & M_Z \agl{\mathbf{1}}{\mathbf{5}} \agl{\mathbf{2}}{\mathbf{4}} \sqr{\mathbf{1}}{\mathbf{5}} \sqr{\mathbf{2}}{\mathbf{3}}
	\sqr{\mathbf{3}}{\mathbf{4}}, M_Z \agl{\mathbf{1}}{\mathbf{2}} \agl{\mathbf{4}}{\mathbf{5}} \sqr{\mathbf{1}}{\mathbf{5}} \sqr{\mathbf{2}}{\mathbf{3}} \sqr{\mathbf{3}}{\mathbf{4}},M_W \agl{\mathbf{1}}{\mathbf{4}} \agl{\mathbf{1}}{\mathbf{5}} \agl{\mathbf{2}}{\mathbf{3}} \sqr{\mathbf{2}}{\mathbf{5}} \sqr{\mathbf{3}}{\mathbf{4}},                                                                                                                                                        \\
	                         & M_Z \agl{\mathbf{1}}{\mathbf{5}} \agl{\mathbf{2}}{\mathbf{4}} \sqr{\mathbf{1}}{\mathbf{2}} \sqr{\mathbf{3}}{\mathbf{4}} \sqr{\mathbf{3}}{\mathbf{5}},\agl{\mathbf{1}}{\mathbf{3}} \agl{\mathbf{4}}{\mathbf{5}} \sqr{\mathbf{1}}{\mathbf{5}} \sqr{\mathbf{3}}{\mathbf{4}} \langle \mathbf{2}|p_1|\mathbf{2}], \agl{\mathbf{1}}{\mathbf{2}} \agl{\mathbf{4}}{\mathbf{5}} \sqr{\mathbf{1}}{\mathbf{2}} \sqr{\mathbf{4}}{\mathbf{5}} \langle \mathbf{3}|p_2|\mathbf{3}], \\
	                         & \agl{\mathbf{1}}{\mathbf{5}} \agl{\mathbf{2}}{\mathbf{4}} \sqr{\mathbf{1}}{\mathbf{5}} \sqr{\mathbf{2}}{\mathbf{4}} \langle \mathbf{3}|p_2|\mathbf{3}], \agl{\mathbf{1}}{\mathbf{5}} \agl{\mathbf{2}}{\mathbf{3}} \sqr{\mathbf{1}}{\mathbf{5}} \sqr{\mathbf{2}}{\mathbf{3}} \langle \mathbf{4}|p_3|\mathbf{4}]\}\ .
\end{align*}
All elements of these lists are understood to be properly symmetric under permutations: the elements of the former must be symmetric in $(1,2)$ and $(3,4)$ and the latter in $(3,4,5)$. For example, if we consider the first element of $\mathcal{B}_6^{W^4Z}$, we have:
\begin{align*}
	M_Z \agl{\mathbf{1}}{\mathbf{5}} \agl{\mathbf{2}}{\mathbf{3}} \agl{\mathbf{4}}{\mathbf{5}} \sqr{\mathbf{1}}{\mathbf{4}} \sqr{\mathbf{2}}{\mathbf{3}} \rightarrow\ \  & \frac{M_Z}{4} \agl{\mathbf{1}}{\mathbf{4}} \agl{\mathbf{2}}{\mathbf{5}} \agl{\mathbf{3}}{\mathbf{5}} \sqr{\mathbf{1}}{\mathbf{4}} \sqr{\mathbf{2}}{\mathbf{3}}+\frac{M_Z}{4} \agl{\mathbf{1}}{\mathbf{5}} \agl{\mathbf{2}}{\mathbf{3}} \agl{\mathbf{4}}{\mathbf{5}} \sqr{\mathbf{1}}{\mathbf{4}} \sqr{\mathbf{2}}{\mathbf{3}}    \\
	+                                                                                                                                                                    & \frac{M_Z}{4} \agl{\mathbf{1}}{\mathbf{5}} \agl{\mathbf{2}}{\mathbf{4}} \agl{\mathbf{3}}{\mathbf{5}} \sqr{\mathbf{1}}{\mathbf{3}} \sqr{\mathbf{2}}{\mathbf{4}}+\frac{M_Z}{4} \agl{\mathbf{1}}{\mathbf{3}} \agl{\mathbf{2}}{\mathbf{5}} \agl{\mathbf{4}}{\mathbf{5}} \sqr{\mathbf{1}}{\mathbf{3}} \sqr{\mathbf{2}}{\mathbf{4}}\ .
\end{align*}
Finally, keeping in mind the definition of the polarisation tensor for massive vectors \eqref{eq:polarisation}, we need to divide these polynomial structures by $M_W^4 M_Z$ and $M_W^2 M_Z^3$ to obtain contact terms with the correct mass dimension.

\subsubsection{\texorpdfstring{$F_\gamma W^4$}{FW4}, \texorpdfstring{$F_\gamma W^2Z^2$}{FW2Z2}, \texorpdfstring{$F_\gamma Z^4$}{FZ4} contact terms}

This example is easier to follow than the previous one because there is only one helicity category involved when we consider dimension-6 operators. The kinematic basis has only six elements:
\begin{equation}
	\begin{split}
		\{&\agl{\mathbf{1}}{\mathbf{2}} \agl{\mathbf{3}}{\mathbf{4}} \sqr{\mathbf{1}}{\mathbf{2}} \sqr{5}{\mathbf{3}} \sqr{5}{\mathbf{4}},\agl{\mathbf{1}}{\mathbf{2}} \agl{\mathbf{3}}{\mathbf{4}} \sqr{\mathbf{2}}{\mathbf{3}} \sqr{5}{\mathbf{1}} \sqr{5}{\mathbf{4}},\agl{\mathbf{1}}{\mathbf{2}} \agl{\mathbf{3}}{\mathbf{4}} \sqr{\mathbf{3}}{\mathbf{4}} \sqr{5}{\mathbf{1}} \sqr{5}{\mathbf{2}},\\
		&\agl{\mathbf{1}}{\mathbf{4}} \agl{\mathbf{2}}{\mathbf{3}} \sqr{\mathbf{1}}{\mathbf{2}} \sqr{5}{\mathbf{3}} \sqr{5}{\mathbf{4}}, \agl{\mathbf{1}}{\mathbf{4}} \agl{\mathbf{2}}{\mathbf{3}} \sqr{\mathbf{2}}{\mathbf{3}} \sqr{5}{\mathbf{1}} \sqr{5}{\mathbf{4}},\agl{\mathbf{1}}{\mathbf{4}} \agl{\mathbf{2}}{\mathbf{3}} \sqr{\mathbf{3}}{\mathbf{4}} \sqr{5}{\mathbf{1}} \sqr{5}{\mathbf{2}}\}\ .
	\end{split}
\end{equation}
After the proper symmetrisations, we find that the number of contact terms for the operators we are considering are 1, 3 and 0, respectively. In particular, we find
\begin{align*}
	\mathcal{B}_6^{\gamma W^4} = \{\frac{1}{M_W^4}\mathrm{Sym}_{(12)(34)}\agl{\mathbf{1}}{\mathbf{4}} \agl{\mathbf{2}}{\mathbf{3}} \sqr{\mathbf{2}}{\mathbf{3}} \sqr{5}{\mathbf{1}} \sqr{5}{\mathbf{4}}\}\ ,
\end{align*}
and
\begin{align*}
	\mathcal{B}_6^{\gamma W^2 Z^2} = \bigg\{ & \frac{1}{ M_Z^2 M_W^2}\mathrm{Sym}_{(34)}\agl{\mathbf{1}}{\mathbf{2}} \agl{\mathbf{3}}{\mathbf{4}} \sqr{\mathbf{1}}{\mathbf{2}} \sqr{5}{\mathbf{3}} \sqr{5}{\mathbf{4}}, \frac{1}{ M_Z^2 M_W^2}\mathrm{Sym}_{(34)}\agl{\mathbf{1}}{\mathbf{4}} \agl{\mathbf{2}}{\mathbf{3}} \sqr{\mathbf{1}}{\mathbf{2}} \sqr{5}{\mathbf{3}} \sqr{5}{\mathbf{4}}, \\
	                                         & \frac{1}{ M_Z^2 M_W^2}\mathrm{Sym}_{(34)}\agl{\mathbf{1}}{\mathbf{4}} \agl{\mathbf{2}}{\mathbf{3}} \sqr{\mathbf{2}}{\mathbf{3}} \sqr{5}{\mathbf{1}} \sqr{5}{\mathbf{4}}\bigg\}\ .
\end{align*}

\subsection{Spin-tidal interactions in gravitational EFTs}
\label{sec:spintidal}
In this section, we will apply our algorithm to classify the operators that encode spin-tidal interactions for spin $S_i=1$ in gravitational systems. This work has been carried out for $S_i=0,\frac{1}{2}$ in \cite{Haddad:2020que,Aoude:2020ygw}.
In this section, we will massage the original basis such that the terms appearing are easily recursive when going up in mass dimension. Furthermore, we will highlight the difference between our strategy and the existing procedure presented in \cite{Durieux:2020gip}. We will study the helicity category $(1^{1_0},2^{1_0},3^{+2},4^{+2})$ in detail and present the general result in Table~\ref{tab:spintidal}~and~\ref{tab:spintidalidentical}.

The minimal mass dimension for such helicity category is 6, and the basis is
\begin{equation}
	\mathcal{B}_6=\{\agl{\mathbf{1}}{\mathbf{2}}\sqr{\mathbf{1}}{\mathbf{2}}\sqr{3}{4}^4,\agl{\mathbf{1}}{\mathbf{2}}\sqr{\mathbf{1}}{4}\sqr{\mathbf{2}}{3}\sqr{3}{4}^3 \}\ .
\end{equation}
A \textit{helicity category basis} for higher mass dimensions is
\begin{equation}
	\begin{split}
		\mathcal{B}_{6+2n}\Deq\ &\{\agl{\mathbf{1}}{\mathbf{2}}\sqr{\mathbf{1}}{\mathbf{2}}\sqr{3}{4}^4 \tilde{s}_{1 2}^{n-k} \tilde{s}_{23}^k\}_{k=0,\dots,n} \cup \{ \agl{\mathbf{1}}{\mathbf{2}}\sqr{\mathbf{1}}{4}\sqr{\mathbf{2}}{3}\sqr{3}{4}^3 \tilde{s}_{23}^n\}\ ,
	\end{split}
\end{equation}
where $\tilde{s}_{i j} = s_{i j} - M_i^2 -M_j^2$ and $\Deq$ means that these basis have been found from the original basis from the algorithm Section~\ref{sec:massive}, working on the graphs modulo terms with powers of the masses (and different helicity categories). Terms of the form $\agl{\mathbf{1}}{\mathbf{2}}\sqr{\mathbf{1}}{4}\sqr{\mathbf{2}}{3}\sqr{3}{4}^3 \tilde{s}_{12}^k \tilde{s}_{23}^{n-k}$ with $k\neq 0$ are not kinematically independent from those in our basis. In particular, we have the (mass completed) relation:
\begin{equation}
	\begin{split}
		&\tilde{s}_{1 2} \agl{\mathbf{1}}{\mathbf{2}} \sqr{\mathbf{2}}{3} \sqr{\mathbf{1}}{4} \sqr{3}{4}^3 -\tilde{s}_{23} \agl{\mathbf{1}}{\mathbf{2}} \sqr{\mathbf{1}}{\mathbf{2}} \sqr{3}{4}^4= \\
		&\ \ - \widetilde{m}_2\agl{\mathbf{1}}{\mathbf{2}} \sqr{\mathbf{1}}{4} \sqr{3}{4}^3 \langle \mathbf{2}|p_1|3]-\widetilde{m}_1\agl{\mathbf{1}}{\mathbf{2}} \sqr{\mathbf{2}}{4} \sqr{3}{4}^3 \langle \mathbf{1}|p_2|3] +M_2^2 \agl{\mathbf{1}}{\mathbf{2}} \sqr{\mathbf{1}}{\mathbf{2}} \sqr{3}{4}^4\ .
	\end{split}
\end{equation}
In the strategy of \cite{Durieux:2020gip}, in the helicity categories $(1^{1_{-1}},2^{1_0},3^{+2},4^{+2})$ and $(1^{1_0},2^{1_{-1}},3^{+2},4^{+2})$ we would allow only (anti-symmetrised) spinor structures
\begin{equation}
	\widetilde{m}_2\agl{\mathbf{1}}{\mathbf{2}} \sqr{\mathbf{1}}{4} \sqr{3}{4}^3 \langle \mathbf{2}|p_1|3]-\widetilde{m}_1\agl{\mathbf{1}}{\mathbf{2}} \sqr{\mathbf{2}}{4} \sqr{3}{4}^3 \langle \mathbf{1}|p_2|3]\ ,
\end{equation}
such that we could consider $\tilde{s}_{1 2} \agl{\mathbf{1}}{2} \sqr{\mathbf{2}}{3} \sqr{\mathbf{1}}{4} \sqr{3}{4}^3$ and $\tilde{s}_{23} \agl{\mathbf{1}}{\mathbf{2}} \sqr{\mathbf{1}}{\mathbf{2}} \sqr{3}{4}^4$ as independent. On the other hand, our algorithm regards the two structures in the different helicity category as independent and excludes the terms mentioned above.

Until now, we have dealt with gravitons and spinning massive particles as different between themselves. The basis is further restricted when we consider identical particles:
\begin{equation}
	\mathcal{B}^{\rm id}_{6+4m} = \{\agl{\mathbf{1}}{\mathbf{2}}\sqr{\mathbf{1}}{\mathbf{2}}\sqr{3}{4}^4 \tilde{s}_{1 2}^{2m-2k} \tilde{s}_{23}^{2k}\}_{k=0,\dots,m}\ ,
\end{equation}
and
\begin{equation}
	\begin{split}
		\mathcal{B}^{\rm id}_{6+4m+2} =&  \{\agl{\mathbf{1}}{\mathbf{2}}\sqr{\mathbf{1}}{\mathbf{2}}\sqr{3}{4}^4 \tilde{s}_{1 2}^{2m+1-2k} \tilde{s}_{23}^{2k}\}_{k=0,\dots,m} \cup \{\agl{\mathbf{1}}{\mathbf{2}}\sqr{\mathbf{1}}{4}\sqr{\mathbf{2}}{3}\sqr{3}{4}^3 \tilde{s}_{23}^{2m+1}\}\ ,
	\end{split}
\end{equation}
where the elements of the bases are always understood as properly symmetrised.

\begin{table}[t]
	\centering
	\resizebox{\textwidth}{!}{
		\begin{tabular}{|c|c|c|}
			\hline
			Helicity category                       & $\mathcal{B}_{d_{\rm dim}}$                                                                                                                                                                                                                                                                                                & $\mathcal{B}_{d_{\rm dim}+2n}$                                                                                                                                                                                                                                                                                                                                                 \\ \hline
			$(1^{1_{+1}},2^{1_{+1}},3^{+2},4^{+2})$ & $\begin{array}{c}
					                                           \sqr{\mathbf{1}}{\mathbf{2}}^2 \sqr{3}{4}^4 \\ \sqr{\mathbf{1}}{4}^2 \sqr{\mathbf{2}}{3}^2 \sqr{3}{4}^4 \\ \sqr{\mathbf{1}}{\mathbf{2}} \sqr{\mathbf{1}}{4} \sqr{\mathbf{2}}{3} \sqr{3}{4}^4\end{array}$   & $\begin{array}{c} \{\sqr{\mathbf{1}}{\mathbf{2}}^2 \sqr{3}{4}^4 s_{12}^{n-k} s_{23}^k\}_{k=0,\dots,n} \\ \sqr{\mathbf{1}}{4}^2 \sqr{\mathbf{2}}{3}^2 \sqr{3}{4}^4 s_{12} s_{23}^{n-1} \\ \sqr{\mathbf{1}}{4}^2 \sqr{\mathbf{2}}{3}^2 \sqr{3}{4}^4 s_{23}^n \end{array}$                   \\ \hline
			$(1^{1_{+1}},2^{1_{0}},3^{+2},4^{+2})$  & $\begin{array}{c} \langle \mathbf{2} | p_1 | \mathbf{2} ] \sqr{\mathbf{1}}{3} \sqr{\mathbf{1}}{4} \sqr{3}{4}^3 \\ \langle \mathbf{2} | p_1 | 3 ] \sqr{\mathbf{1}}{4}^2 \sqr{\mathbf{2}}{3} \sqr{3}{4}^2 \end{array}$                                               & $\begin{array}{c} \{ \langle \mathbf{2} | p_1 | \mathbf{2} ] \sqr{\mathbf{1}}{3} \sqr{\mathbf{1}}{4} \sqr{3}{4}^3 s_{12}^{n-k} s_{23}^k\}_{k=0,\dots , n} \\ \langle \mathbf{2} | p_1 | 3 ] \sqr{\mathbf{1}}{4}^2 \sqr{\mathbf{2}}{3} \sqr{3}{4}^2 s_{23}^n \end{array}$                                               \\ \hline
			$(1^{1_{+1}},2^{1_{-1}},3^{+2},4^{+2})$ & $\langle \mathbf{2} | p_1 | 3 ]^2 \sqr{\mathbf{1}}{4}^2 \sqr{3}{4}^2$                                                                                                                                                                                                                                                      & $\{ \langle \mathbf{2} | p_1 | 3 ]^2 \sqr{\mathbf{1}}{4}^2 \sqr{3}{4}^2 s_{12}^{n-k} s_{23}^k\}_{k=0,\dots , n}$                                                                                                                                                                                                                                                               \\ \hline
			$(1^{1_{0}},2^{1_{+1}},3^{+2},4^{+2})$  & $\begin{array}{c} \langle \mathbf{1} | p_2 | 3 ] \sqr{\mathbf{1}}{\mathbf{2}} \sqr{\mathbf{2}}{4} \sqr{3}{4}^3 \\ \langle \mathbf{1} | p_2 | 3 ] \sqr{\mathbf{1}}{4} \sqr{\mathbf{2}}{3} \sqr{\mathbf{2}}{4} \sqr{3}{4}^2 \end{array}$ & $\begin{array}{c} \{ \langle \mathbf{1} | p_2 | 3 ] \sqr{\mathbf{1}}{\mathbf{2}} \sqr{\mathbf{2}}{4} \sqr{3}{4}^3 s_{12}^{n-k} s_{23}^k\}_{k=0,\dots , n} \\ \langle \mathbf{1} | p_2 | 3 ] \sqr{\mathbf{1}}{4} \sqr{\mathbf{2}}{3} \sqr{\mathbf{2}}{4} \sqr{3}{4}^2 s_{23}^n \end{array}$ \\ \hline
			$(1^{1_{0}},2^{1_{0}},3^{+2},4^{+2})$   & $\begin{array}{c} \agl{\mathbf{1}}{\mathbf{2}}\sqr{\mathbf{1}}{\mathbf{2}}\sqr{3}{4}^4 \\ \agl{\mathbf{1}}{\mathbf{2}}\sqr{\mathbf{1}}{4}\sqr{\mathbf{2}}{3}\sqr{3}{4}^3 \end{array}$                      & $\begin{array}{c}\{\agl{\mathbf{1}}{\mathbf{2}}\sqr{\mathbf{1}}{\mathbf{2}}\sqr{3}{4}^4 \tilde{s}_{1 2}^{n-k} \tilde{s}_{23}^k\}_{k=0,\dots,n} \\ \agl{\mathbf{1}}{\mathbf{2}}\sqr{\mathbf{1}}{4}\sqr{\mathbf{2}}{3}\sqr{3}{4}^3 \tilde{s}_{23}^n\end{array} $ \\ \hline
			$(1^{1_{0}},2^{1_{-1}},3^{+2},4^{+2})$  & $\langle \mathbf{2} | p_1 | 3 ] \agl{\mathbf{1}}{\mathbf{2}} \sqr{\mathbf{1}}{4} \sqr{3}{4}^3 $                                                                                                                                                                                                                            & $ \{\langle \mathbf{2} | p_1 | 3 ] \agl{\mathbf{1}}{\mathbf{2}} \sqr{\mathbf{1}}{4} \sqr{3}{4}^3 \tilde{s}_{1 2}^{n-k} \tilde{s}_{23}^k\}_{k=0,\dots,n}$                                                                                                                                                                                                                       \\ \hline
			$(1^{1_{-1}},2^{1_{+1}},3^{+2},4^{+2})$ & $\langle \mathbf{1} | p_2 | 3 ]^2 \sqr{\mathbf{2}}{4}^2 \sqr{3}{4}^2$                                                                                                                                                                                                                                                      & $\{ \langle \mathbf{1} | p_2 | 3 ]^2 \sqr{\mathbf{2}}{4}^2 \sqr{3}{4}^2 s_{12}^{n-k} s_{23}^k\}_{k=0,\dots , n}$                                                                                                                                                                                                                                                               \\ \hline
			$(1^{1_{-1}},2^{1_{0}},3^{+2},4^{+2})$  & $\langle \mathbf{1} | p_2 | 3 ] \agl{\mathbf{1}}{\mathbf{2}} \sqr{\mathbf{2}}{4} \sqr{3}{4}^3 $                                                                                                                                                                                                                            & $ \{\langle \mathbf{1} | p_2 | 3 ] \agl{\mathbf{1}}{\mathbf{2}} \sqr{\mathbf{2}}{4} \sqr{3}{4}^3 \tilde{s}_{1 2}^{n-k} \tilde{s}_{23}^k\}_{k=0,\dots,n}$                                                                                                                                                                                                                       \\ \hline
			$(1^{1_{-1}},2^{1_{-1}},3^{+2},4^{+2})$ & $\agl{\mathbf{1}}{\mathbf{2}}^2 \sqr{3}{4}^4$                                                                                                                                                                                                                                                                              & $ \{\agl{\mathbf{1}}{\mathbf{2}}^2 \sqr{3}{4}^4 \tilde{s}_{1 2}^{n-k} \tilde{s}_{23}^k\}_{k=0,\dots,n}$                                                                                                                                                                                                                                                                        \\ \hline
			$(1^{1_{+1}},2^{1_{+1}},3^{+2},4^{-2})$ & $\langle 4 | p_2 | 3 ]^4 \sqr{\mathbf{1}}{\mathbf{2}}^2$                                                                                                                                                                                                                                                                   & $\{ \langle 4 | p_2 | 3 ]^4 \sqr{\mathbf{1}}{\mathbf{2}}^2 \tilde{s}_{1 2}^{n-k} \tilde{s}_{23}^k\}_{k=0,\dots,n}$                                                                                                                                                                                                                                                             \\ \hline
			$(1^{1_{+1}},2^{1_{0}},3^{+2},4^{-2})$  & $\langle 4 | p_2 | 3 ]^3 \agl{\mathbf{2}}{4} \sqr{\mathbf{1}}{\mathbf{2}} \sqr{\mathbf{1}}{3}$                                                                                                                                                                                                                             & $\{ \langle 4 | p_2 | 3 ]^3 \agl{\mathbf{2}}{4} \sqr{\mathbf{1}}{\mathbf{2}} \sqr{\mathbf{1}}{3} \tilde{s}_{1 2}^{n-k} \tilde{s}_{23}^k\}_{k=0,\dots,n}$                                                                                                                                                                                                                       \\ \hline
			$(1^{1_{+1}},2^{1_{-1}},3^{+2},4^{-2})$ & $\langle 4 | p_2 | 3 ]^2 \agl{\mathbf{2}}{4}^2  \sqr{\mathbf{1}}{3}^2$                                                                                                                                                                                                                                                     & $\{ \langle 4 | p_2 | 3 ]^2 \agl{\mathbf{2}}{4}^2  \sqr{\mathbf{1}}{3}^2 \tilde{s}_{1 2}^{n-k} \tilde{s}_{23}^k\}_{k=0,\dots,n}$                                                                                                                                                                                                                                               \\ \hline
			$(1^{1_{0}},2^{1_{+1}},3^{+2},4^{-2})$  & $\langle 4 | p_2 | 3 ]^3 \agl{\mathbf{1}}{4} \sqr{\mathbf{1}}{\mathbf{2}} \sqr{\mathbf{2}}{3}$                                                                                                                                                                                                                             & $\{ \langle 4 | p_2 | 3 ]^3 \agl{\mathbf{1}}{4} \sqr{\mathbf{1}}{\mathbf{2}} \sqr{\mathbf{2}}{3} \tilde{s}_{1 2}^{n-k} \tilde{s}_{23}^k\}_{k=0,\dots,n}$                                                                                                                                                                                                                       \\ \hline
			$(1^{1_{0}},2^{1_{0}},3^{+2},4^{-2})$   & $\langle 4 | p_2 | 3 ]^2 \agl{\mathbf{1}}{4} \agl{\mathbf{2}}{4} \sqr{\mathbf{1}}{3} \sqr{\mathbf{2}}{3}$                                                                                                                                                                                                                  & $\{ \langle 4 | p_2 | 3 ]^2 \agl{\mathbf{1}}{4} \agl{\mathbf{2}}{4} \sqr{\mathbf{1}}{3} \sqr{\mathbf{2}}{3} \tilde{s}_{1 2}^{n-k} \tilde{s}_{23}^k\}_{k=0,\dots,n}$                                                                                                                                                                                                            \\ \hline
			$(1^{1_{0}},2^{1_{-1}},3^{+2},4^{-2})$  & $\langle 4 | p_2 | 3 ]^3 \agl{\mathbf{1}}{\mathbf{2}}  \agl{\mathbf{2}}{4} \sqr{\mathbf{1}}{3} $                                                                                                                                                                                                                           & $\{ \langle 4 | p_2 | 3 ]^3 \agl{\mathbf{1}}{\mathbf{2}} \agl{\mathbf{2}}{4} \sqr{\mathbf{1}}{3} \tilde{s}_{1 2}^{n-k} \tilde{s}_{23}^k\}_{k=0,\dots,n}$                                                                                                                                                                                                                       \\ \hline
			$(1^{1_{-1}},2^{1_{+1}},3^{+2},4^{-2})$ & $\langle 4 | p_2 | 3 ]^2 \agl{\mathbf{1}}{4}^2  \sqr{\mathbf{2}}{3}^2$                                                                                                                                                                                                                                                     & $\{ \langle 4 | p_2 | 3 ]^2 \agl{\mathbf{1}}{4}^2  \sqr{\mathbf{2}}{3}^2 \tilde{s}_{1 2}^{n-k} \tilde{s}_{23}^k\}_{k=0,\dots,n}$                                                                                                                                                                                                                                               \\ \hline
			$(1^{1_{-1}},2^{1_{0}},3^{+2},4^{-2})$  & $\langle 4 | p_2 | 3 ]^3 \agl{\mathbf{1}}{\mathbf{2}}  \agl{\mathbf{1}}{4} \sqr{\mathbf{2}}{3} $                                                                                                                                                                                                                           & $\{ \langle 4 | p_2 | 3 ]^3 \agl{\mathbf{1}}{\mathbf{2}} \agl{\mathbf{1}}{4} \sqr{\mathbf{2}}{3} \tilde{s}_{1 2}^{n-k} \tilde{s}_{23}^k\}_{k=0,\dots,n}$                                                                                                                                                                                                                       \\ \hline
			$(1^{1_{-1}},2^{1_{-1}},3^{+2},4^{-2})$ & $\langle 4 | p_2 | 3 ]^4 \agl{\mathbf{1}}{\mathbf{2}}^2$                                                                                                                                                                                                                                                                   & $\{ \langle 4 | p_2 | 3 ]^4 \agl{\mathbf{1}}{\mathbf{2}}^2 \tilde{s}_{1 2}^{n-k} \tilde{s}_{23}^k\}_{k=0,\dots,n}$                                                                                                                                                                                                                                                             \\ \hline
		\end{tabular}
	}
	\caption{The helicity category bases for the spin-tidal interactions at $S=1$.}
	\label{tab:spintidal}
\end{table}

\begin{table}[t]
	\centering
	\begin{tabular}{|c|l|}\hline
		Helicity category & $\mathcal{B}^{\rm id}_{d_{\min} +2n}$                                                                                                                                                                                                                                                                                                                                                                                                          \\ \hline
		$(++,++)$         & $\begin{array}{ll}
				                     \mathrm{Sym}_{(1,2)(3,4)} \sqr{\mathbf{1}}{\mathbf{2}}^2 \sqr{3}{4}^4 \tilde{s}_{1 2}^{n-2k} \tilde{s}_{23}^{2k} \qquad k=0,\dots ,\lfloor \frac{n}{2}\rfloor \\ \mathrm{Sym}_{(1,2)(3,4)} \sqr{\mathbf{1}}{4}^2 \sqr{\mathbf{2}}{3}^2 \sqr{3}{4}^4 s_{12} s_{23}^{n-1}
			                     \end{array}$                                                                                                                \\ \hline
		$(+0,++)$         & $\begin{array}{ll} \mathrm{Sym}_{(1,2)(3,4)} \langle \mathbf{2} | p_1 | \mathbf{2} ] \sqr{\mathbf{1}}{3} \sqr{\mathbf{1}}{4} \sqr{3}{4}^3 s_{12}^{n-k-1} s_{23}^{k+1} \qquad k=0,\dots ,\lfloor \frac{n}{2}\rfloor-1 \\ \mathrm{Sym}_{(1,2)(3,4)} \langle \mathbf{2} | p_1 | 3 ] \sqr{\mathbf{1}}{4}^2 \sqr{\mathbf{2}}{3} \sqr{3}{4}^2 s_{23}^n \qquad n \ \mathrm{even} \end{array}$ \\ \hline
		$(+-,++)$         & $\  \mathrm{Sym}_{(1,2)(3,4)} \langle \mathbf{2} | p_1 | 3 ]^2 \sqr{\mathbf{1}}{4}^2 \sqr{3}{4}^2 s_{12}^{n-k} s_{23}^k$ \qquad $k=0,\dots ,\lfloor \frac{n}{2}\rfloor$                                                                                                                                                                                                                                                                        \\ \hline
		$(0-,++)$         & $\ \mathrm{Sym}_{(1,2)(3,4)} \langle \mathbf{2} | p_1 | 3 ] \agl{\mathbf{1}}{\mathbf{2}} \sqr{\mathbf{1}}{4} \sqr{3}{4}^3 s_{12}^{n-k-1} s_{23}^{k+1} $\qquad $k=0,\dots ,\lfloor \frac{n}{2}\rfloor-1$                                                                                                                                                                                                                                        \\ \hline
		$(--,++)$         & $\ \mathrm{Sym}_{(1,2)(3,4)} \agl{\mathbf{1}}{\mathbf{2}}^2 \sqr{3}{4}^4 \tilde{s}_{1 2}^{n-2k} \tilde{s}_{23}^{2k} $\qquad $k=0,\dots ,\lfloor \frac{n}{2}\rfloor$                                                                                                                                                                                                                                                                            \\ \hline
		$(++,+-)$         & $\ \mathrm{Sym}_{(1,2)} \langle 4 | p_2 | 3 ]^4 \sqr{\mathbf{1}}{\mathbf{2}}^2 \tilde{s}_{1 2}^{n-2k} \tilde{s}_{23}^{2k} $\qquad $k=0,\dots ,\lfloor \frac{n}{2}\rfloor$                                                                                                                                                                                                                                                                      \\ \hline
		$(+0,+-)$         & $\ \mathrm{Sym}_{(1,2)} \langle 4 | p_2 | 3 ]^3 \agl{\mathbf{2}}{4} \sqr{\mathbf{1}}{\mathbf{2}} \sqr{\mathbf{1}}{3} \tilde{s}_{1 2}^{n-k} \tilde{s}_{23}^k$\qquad $k=0,\dots,n$                                                                                                                                                                                                                                                               \\ \hline
		$(+-,+-)$         & $\ \mathrm{Sym}_{(1,2)} \langle 4 | p_2 | 3 ]^2 \agl{\mathbf{2}}{4}^2  \sqr{\mathbf{1}}{3}^2 \tilde{s}_{1 2}^{n-k} \tilde{s}_{23}^k$\qquad $k=0,\dots,n$                                                                                                                                                                                                                                                                                       \\ \hline
		$(00,+-)$         & $\ \mathrm{Sym}_{(1,2)} \langle 4 | p_2 | 3 ]^2 \agl{\mathbf{1}}{4} \agl{\mathbf{2}}{4} \sqr{\mathbf{1}}{3} \sqr{\mathbf{2}}{3} \tilde{s}_{1 2}^{n-2k} \tilde{s}_{23}^{2k} $\qquad $k=0,\dots ,\lfloor \frac{n}{2}\rfloor$                                                                                                                                                                                                                     \\ \hline
		$(0-,+-)$         & $\ \mathrm{Sym}_{(1,2)} \langle 4 | p_2 | 3 ]^3 \agl{\mathbf{1}}{\mathbf{2}}  \agl{\mathbf{2}}{4} \sqr{\mathbf{1}}{3} \tilde{s}_{1 2}^{n-k} \tilde{s}_{23}^k$\qquad $k=0,\dots,n$                                                                                                                                                                                                                                                              \\ \hline
		$(--,+-)$         & $\ \mathrm{Sym}_{(1,2)} \langle 4 | p_2 | 3 ]^4 \agl{\mathbf{1}}{\mathbf{2}}^2 \tilde{s}_{1 2}^{n-2k} \tilde{s}_{23}^{2k} $\qquad $k=0,\dots ,\lfloor \frac{n}{2}\rfloor$                                                                                                                                                                                                                                                                      \\ \hline
	\end{tabular}
	\caption{The \textit{amplitude bases} for the different helicity categories, after taking into account that we are dealing with identical bosons. On the right-hand side, we wrote only the transversality and the helicity of each particle, separating massive and massless states with a comma.}
	\label{tab:spintidalidentical}
\end{table}


\section*{Acknowledgments}
I would like to thank Andreas Brandhuber, Gabriele Travaglini, Jung-Wook Kim and Kays Haddad for stimulating conversations, Manuel Accettulli Huber for useful tips for the implementation of the algorithms, Gauthier Durieux and Yael Shadmi for valuable comments on the algorithm, the code and comparisons with the strategy outlined in \cite{Durieux:2020gip}.
This work  was supported by the European Union's Horizon 2020 research and innovation programme under the Marie Sklodowska-Curie grant agreement No.~764850 ``SAGEX".

\section*{Added note}
Shortly after this preprint was posted on the arXiv, reference \cite{Dong:2022mcv} appeared. The authors presented a different method for the classification of amplitude bases in four dimensions, using Young Tableau techniques, extending and applying the algorithms presented in \cite{Dong:2021yak}. This method is built on rigorous group-theoretical statements and is completely general. Their bases are expected to be strictly related to those coming from the algorithms presented in this work, although the precise map needs further investigation.

\appendix

\section{The spinor helicity formalism: review and conventions}
\label{sec:spinorhelicity}

Most of the work in this paper has been done working in $(+---)$ signature, where the usual four-momenta can be converted to bispinors using Pauli matrices as a realisation of the isomorphism $so(1,3)\sim sl(2,\mathbb{C})$: $p_{\alpha \dot{\alpha}} = p_{\mu} \sigma^{\mu}_{\alpha \dot{\alpha}}$, $p^{\dot{\alpha} \alpha} = p_{\mu} \bar{\sigma}^{\mu \dot{\alpha}\alpha}$, where the Pauli matrices are $\sigma^{\mu}_{\alpha \dot{\alpha}}=(\mathbf{1},\vec{\sigma})$ and $\bar{\sigma}^{\mu \dot{\alpha}\alpha} = (\mathbf{1},-\vec{\sigma})$. The undotted and dotted indices transform in the fundamental and anti-fundamental representation of the SL$(2,\mathbb{C})$ group. These spinor indices are raised and lowered by the two-dimensional $\epsilon$-tensors, such that:
\begin{equation}
	\label{eq:epsilonconvention1}
	\epsilon_{\alpha \beta} \epsilon^{\beta \gamma} = \delta_{\alpha}^{\gamma}\ ,\qquad \epsilon_{\dot{\alpha} \dot{\beta}} \epsilon^{\dot{\beta} \dot{\gamma}} = \delta_{\dot{\alpha}}^{\dot{\gamma}}\ .
\end{equation}
For massless and massive momenta, we have
\begin{equation}
	\begin{split}
		\det p_{i\alpha \dot{\alpha}} = 0 \qquad &\Rightarrow \qquad p_{i \alpha \dot{\alpha}} \equiv \lambda_{i\alpha} \widetilde{\lambda}_{i\dot{\alpha}}\ ,\\
		\det p_{i\alpha \dot{\alpha}} = M_i^2 \qquad &\Rightarrow \qquad p_{i \alpha \dot{\alpha}} \equiv \lambda_{i\alpha}^{I} \widetilde{\lambda}_{i\dot{\alpha} I}\ ,
	\end{split}
\end{equation}
where $I$ is an index in the fundamental of SU$(2)$ (massive little group). Uniformly to \eqref{eq:epsilonconvention1}, SU$(2)$ indices are raised and lower by $\epsilon$-tensor defined such that
\begin{equation}
	\epsilon_{I J} \epsilon^{J K} = \delta_{I}^{K}\ .
\end{equation}
The two spinors are related by complex conjugation:
\begin{equation}
	\label{eq:complexconjugationSpinors}
	\left(\lambda_{\alpha}\right)^* = {\rm sign}(p^0)\, \widetilde{\lambda}_{\dot{\alpha}}\ ,\qquad \left(\lambda_{\alpha}^I\right)^* = {\rm sign}(p^0)\, \widetilde{\lambda}_{\dot{\alpha} I}
\end{equation}
The Lorentz invariants are defined as
\begin{equation}
	\agl{i}{j} = \langle i | | j \rangle \equiv \lambda_{i}^{\alpha} \lambda_{j \alpha}\ ,\qquad \sqr{i}{j} = [i | |j] \equiv \widetilde{\lambda}_{i \dot{\alpha}} \widetilde{\lambda}_{j}^{\dot{\alpha}}\ ,
\end{equation}
where the spinors in this definition can be either massless or massive, in which case we omitted the spinor indices. Spinors satisfy the Dirac equation:
\begin{equation}
	\begin{split}
		&p_i | i \rangle = 0\ ,\qquad p_i | i ] = 0\ ,\\
		&p_i | i^I \rangle = m_i | i^I ]\ ,\qquad p_i |i^I ] = \widetilde{m}_i | i^I \rangle\ ,
	\end{split}
\end{equation}
where $m_i = e^{i \alpha} M_i$ and $\widetilde{m}_i = e^{-i \alpha} M_i$, with $\alpha$ being a constant real number. This distinction is immaterial and we will set $\alpha = 0$ at the very end of the calculations, but it is relevant when we evaluate these structures numerically, as explained in detail in Appendix~\ref{sec:massivetwistors}. This is guaranteed is we define $m_i$ and $\widetilde{m}_i$ as
\begin{equation}
	\agl{i^I}{i^J} = - m_i \epsilon^{I J}\ ,\qquad \sqr{i^I}{i^J} = \widetilde{m}_i \epsilon^{I J}\ .
\end{equation}

\section{Momentum twistors and massive (rational) kinematics}
\label{sec:massivetwistors}

To verify that the structures in our basis are kinematically independent and to investigate the independence of the structures upon symmetrisation, we evaluated the polynomial structures over rational kinematics
\footnote{This procedure could be speeded up using finite field reconstruction, but we found this overkilling for the scope of this paper.}, which allow reconstructing rational functions without loss of precision \cite{vonManteuffel:2014ixa,Peraro:2016wsq}. Indeed, it is widely known that it is possible to generate rational kinematics satisfying both on-shell and momentum conservation conditions \cite{Badger:2013gxa,Badger:2016uuq}. This is possible, for example, by analytic continuation to $(++--)$ signature and by generating the kinematics in terms of the \textit{momentum twistors} variables introduced in \cite{Hodges:2009hk}. This construction was introduced for fully massless four-dimensional kinematics. The generalisation to the massive case is trivial.

The spinor helicity formalism in split signature is formally different from the one introduced in the previous section, but practically the same. The little group for massless and massive particles are $\mathbb{R}$ and SL$(2,\mathbb{R})$, respectively. Besides $m_i$ and $\widetilde{m}_i$ are two real numbers such that $m_i \widetilde{m}_i = M_i^2$. Condition \eqref{eq:complexconjugationSpinors} is lifted, and the dotted and undotted spinors are real and independent of each other.

Introducing spinor helicity variables automatically makes the momenta satisfy on-shell conditions, but the momentum conservation identity is a quadratic constraint on our kinematic variables. To make sure that the kinematics stays in the field of rational numbers, we need to rewrite this constraint in terms of linear equations. This is possible in four dimensions expressing the kinematic in terms of momentum-twistor variables. Momenta can be rewritten in terms of \textit{dual momentum variables} $x_i$:
\begin{equation}
	p_i = x_i - x_{i+1}\ ,
\end{equation}
which make momentum conservation between $n$ particles trivial:
\begin{equation}
	x_{n+1}=x_{1}\ .
\end{equation}
Then each massless momentum is associated with two null-separated points, and momentum conservation tells us that the set of dual variables forms a polygon. The on-shell condition defines a new variable $[\mu_i|$, through the \textit{incidence relation}
\begin{equation}
	[\mu_i| = \langle i | \,x_{i} = \langle i | x_{i+1}\ .
\end{equation}
Given a (randomly generated) set of pair of spinors $Z_i^A = \left( \lambda_{i}^{\alpha}, \mu_{i \,\dot{\alpha}} \right)$, named momentum twistor variables, such that $Z_{n+1} = Z_{1}$, and using the incidence relation, we can define the spinors $\widetilde{\lambda}_{i}^{\dot{\alpha}}$ through the \textit{dual twistor}
\begin{equation}
	W_{i A} = ( \widetilde{\mu}_{i\,\alpha}, \widetilde{\lambda}_{i}^{\dot{\alpha}} ) = \frac{\epsilon_{A B C D} Z_{i-1}^B Z_i^C Z_{i+1}^D}{\agl{(i-1)}{i} \agl{i}{(i+1)}}\ .
\end{equation}
Planar Mandelstam invariants can be written in terms of twistor variables:
\begin{equation}
	s_{i,i+1,\dots ,j-1} = (x_i - x_j)^2 = \frac{\epsilon_{A B C D} Z_{i-1}^A Z_i^B Z_{j-1}^C Z_{j}^D}{\agl{(i-1)}{i}\agl{(j-1)}{j}}\ .
\end{equation}

This procedure was introduced to generate a fully massless kinematics over the rational or finite fields. However, it can be generalised to the massive case once we decompose massive momenta into a couple of massless ones:
\begin{equation}
	p_{i \alpha \dot{\alpha}} = \lambda_{i \alpha}^{1} \tlambda_{i \dot{\alpha}\, 1} + \lambda_{i \alpha}^{2} \tlambda_{i \dot{\alpha}\, 2} \equiv k_{i \alpha \dot{\alpha}} + q_{i \alpha \dot{\alpha}}\ ,
\end{equation}
where $k^{\mu}$ and $q^\mu$ are two massless momenta such that $q_{i\, \alpha \dot{\alpha}}\, k_{i}^{\dot{\alpha} \alpha} = M_{i}^2$. Then if we are considering a scattering amplitude for $n$ massless and $m$ massive states, we need to randomly generate $n+2m$ twistor variables:
\begin{equation}
	\{Z_{i}^{I A}, Z_{j}^{A}\}\ ,
\end{equation}
where $i=1,\dots m$, $j=m+1,\dots n+m$ and $I=1,2$. By doing so, the masses are randomly generated as well:
\begin{equation}
	\label{eq:massTwistors}
	M_i^2 = \frac{\epsilon_{A B C D} Z_{i-1}^{2\, A} Z_i^{1\,B} Z_{i}^{2\,C} Z_{i+1}^{1\,D}}{\agl{(i-1)^2}{i^1}\agl{i^2}{(i+1)^1}}\ ,
\end{equation}
where $Z_{0}^{2\, A} = Z_{n+m}^A$ and $Z_{n+1}^{1\, D} = Z_{n+1}^D$. However, we might be interested in cases where some states have the same mass, like the examples considered in Section~\ref{sec:WZgamma}~and~\ref{sec:spintidal}. For example, we can consider $l$ particles with the same mass. In this case, we can generate a bi-twistor associated with one of these particles fully randomly, while for the others we can leave, for example, the component $\mu_{i\, \dot{2}}^{2}$ undetermined. These are fully fixed by $l-1$ linear equations requiring that the masses obtained from equation~\eqref{eq:massTwistors} must be equal to the one we generated randomly.



\section{The reduction the planar basis}
\label{sec:intoplanar}

In this section, we give details on the algorithm to decompose spinor structures corresponding to a non-planar graph into our basis, given by the set of structures related to planar graphs. Such decomposition amounts to repeatedly applying Schouten identities, which act separately on angle and square invariants. Then for simplicity, we are going to consider Lorentz invariant structures with only angle invariants ($S_{i j} = 0 \ \forall \, i,j$). The condition that identifies a crossing between the edges $(i,j)$ and $(k,l)$ is
\begin{equation}
	\label{eq:nonplanarityApp}
	A_{a b} \neq 0\ , \qquad A_{c d} \neq 0\ , \qquad a<c<b<d\ .
\end{equation}
Obviously, there are a finite number of edges (and crossings) associated with each vertex. We can consider the total number of crossings of the matrices $\mathbf{A}$ and $\mathbf{A} + \mathbf{E}^{(a,b)}_{(c,d)}$, which we call $n_\times$ and $n_\times^\prime$, respectively. Then, proving that
\begin{equation}
	n_\times^\prime - n_\times < 0\ ,
\end{equation}
is equivalent to the statement that every non-planar graph can be decomposed as a sum of planar ones in a finite number of steps. We have
\begin{equation}
	n_\times^\prime - n_\times = \sum_{i=1}^{n-1} \sum_{j=i+1}^{n} \sum_{l=j+1}^n \sum_{k=i+1}^{j-1} \left(A_{i j} E^{(a\,b)}_{(c\,d),\, k l}+E^{(a\,b)}_{(c\,d),\, i j} A_{k l}+E^{(a\,b)}_{(c\,d),\, i j} E^{(a\,b)}_{(c\,d),\, k l}\right)\ ,
\end{equation}
where
\begin{align}
	 & \sum_{i=1}^{n-1} \sum_{j=i+1}^{n} \sum_{l=j+1}^n \sum_{k=i+1}^{j-1} A_{i j} E^{(a\,b)}_{(c\,d),\, k l} = \left(- \sum_{i = c}^{b-1} \sum_{j = c+1}^{n} - \sum_{i = c+1}^{b} \sum_{j = d+1}^{n} + \sum_{i = a+1}^{b-1} \sum_{j = c+1}^{b}\right) A_{i j}\ ,             \\
	 & \sum_{i=1}^{n-1} \sum_{j=i+1}^{n} \sum_{l=j+1}^n \sum_{k=i+1}^{j-1} E^{(a\,b)}_{(c\,d),\, i j} A_{k l} = \left(-\sum_{i = 1}^{a - 1} \sum_{j = c}^{b - 1} -\sum_{i = 1}^{c - 1} \sum_{j = c + 1}^{b} + \sum_{i = c}^{b - 1} \sum_{j = b+1}^{d - 1}  \right) A_{i j}\ , \\
	 & \sum_{i=1}^{n-1} \sum_{j=i+1}^{n} \sum_{l=j+1}^n \sum_{k=i+1}^{j-1} E^{(a\,b)}_{(c\,d),\, i j} E^{(a\,b)}_{(c\,d),\, k l} = 1\ .
\end{align}
Summing these contributions, we find
\begin{equation}
	n_\times^\prime - n_\times \leq - A_{a b} - A_{c d} + 1 < 0\ .
\end{equation}
The same is true for the difference between the total number of crossings of $\mathbf{A}$ and $\mathbf{A} + \mathbf{F}^{(a,b)}_{(c,d)}$.

Then, we need to choose a recursive way of eliminating all the crossings. We select $a,b,c,d$ so that we have \eqref{eq:nonplanarityApp} and they are the smallest (in the selected order). We apply the decomposition in equation \eqref{eq:untyingcrossing} $\min\{A_{ab},A_{cd}\}$ times and we repeat this step until we end up with a sum of planar structures. Obviously, this choice is not always the fastest route, but the decomposition into planar graphs does not require an optimised strategy.

\bibliographystyle{JHEP}
\bibliography{papers}

\providecommand{\href}[2]{#2}\begingroup\raggedright\begin{thebibliography}{10}

\bibitem{Dixon:1996wi}
L.J.~Dixon, \emph{{Calculating scattering amplitudes efficiently}},  in
  \emph{{Theoretical Advanced Study Institute in Elementary Particle Physics
  (TASI 95): QCD and Beyond}}, pp.~539--584, 1, 1996
  [\href{https://arxiv.org/abs/hep-ph/9601359}{{\ttfamily hep-ph/9601359}}].

\bibitem{Dixon:2013uaa}
L.J.~Dixon, \emph{{A brief introduction to modern amplitude methods}},  in
  \emph{{Theoretical Advanced Study Institute in Elementary Particle Physics}:
  {Particle Physics: The Higgs Boson and Beyond}}, pp.~31--67, 2014,
  \href{https://doi.org/10.5170/CERN-2014-008.31}{DOI}
  [\href{https://arxiv.org/abs/1310.5353}{{\ttfamily 1310.5353}}].

\bibitem{Elvang:2013cua}
H.~Elvang and Y.-t.~Huang, \emph{{Scattering Amplitudes}},
  \href{https://arxiv.org/abs/1308.1697}{{\ttfamily 1308.1697}}.

\bibitem{Henn:2014yza}
J.M.~Henn and J.C.~Plefka, \emph{{Scattering Amplitudes in Gauge Theories}},
  vol.~883, Springer, Berlin (2014),
  \href{https://doi.org/10.1007/978-3-642-54022-6}{10.1007/978-3-642-54022-6}.

\bibitem{Cheung:2017pzi}
C.~Cheung, \emph{{TASI Lectures on Scattering Amplitudes}},  in
  \emph{{Proceedings, Theoretical Advanced Study Institute in Elementary
  Particle Physics : Anticipating the Next Discoveries in Particle Physics
  (TASI 2016)}: {Boulder, CO, USA, June 6-July 1, 2016}}, R.~Essig and I.~Low,
  eds., pp.~571--623 (2018),
  \href{https://doi.org/10.1142/9789813233348_0008}{DOI}
  [\href{https://arxiv.org/abs/1708.03872}{{\ttfamily 1708.03872}}].

\bibitem{Caron-Huot:2016cwu}
S.~Caron-Huot and M.~Wilhelm, \emph{{Renormalization group coefficients and the
  S-matrix}}, \href{https://doi.org/10.1007/JHEP12(2016)010}{\emph{JHEP}
  {\bfseries 12} (2016) 010}
  [\href{https://arxiv.org/abs/1607.06448}{{\ttfamily 1607.06448}}].

\bibitem{EliasMiro:2020tdv}
J.~Elias~Mir\'o, J.~Ingoldby and M.~Riembau, \emph{{EFT anomalous dimensions
  from the S-matrix}},
  \href{https://doi.org/10.1007/JHEP09(2020)163}{\emph{JHEP} {\bfseries 09}
  (2020) 163} [\href{https://arxiv.org/abs/2005.06983}{{\ttfamily
  2005.06983}}].

\bibitem{Baratella:2020lzz}
P.~Baratella, C.~Fernandez and A.~Pomarol, \emph{{Renormalization of
  Higher-Dimensional Operators from On-shell Amplitudes}},
  \href{https://doi.org/10.1016/j.nuclphysb.2020.115155}{\emph{Nucl. Phys. B}
  {\bfseries 959} (2020) 115155}
  [\href{https://arxiv.org/abs/2005.07129}{{\ttfamily 2005.07129}}].

\bibitem{Jiang:2020mhe}
M.~Jiang, T.~Ma and J.~Shu, \emph{{Renormalization Group Evolution from
  On-shell SMEFT}}, \href{https://doi.org/10.1007/JHEP01(2021)101}{\emph{JHEP}
  {\bfseries 01} (2021) 101}
  [\href{https://arxiv.org/abs/2005.10261}{{\ttfamily 2005.10261}}].

\bibitem{Bern:2020ikv}
Z.~Bern, J.~Parra-Martinez and E.~Sawyer, \emph{{Structure of two-loop SMEFT
  anomalous dimensions via on-shell methods}},
  \href{https://doi.org/10.1007/JHEP10(2020)211}{\emph{JHEP} {\bfseries 10}
  (2020) 211} [\href{https://arxiv.org/abs/2005.12917}{{\ttfamily
  2005.12917}}].

\bibitem{Baratella:2020dvw}
P.~Baratella, C.~Fernandez, B.~von Harling and A.~Pomarol, \emph{{Anomalous
  Dimensions of Effective Theories from Partial Waves}},
  \href{https://doi.org/10.1007/JHEP03(2021)287}{\emph{JHEP} {\bfseries 03}
  (2021) 287} [\href{https://arxiv.org/abs/2010.13809}{{\ttfamily
  2010.13809}}].

\bibitem{AccettulliHuber:2021uoa}
M.~Accettulli~Huber and S.~De~Angelis, \emph{{Standard Model EFTs via on-shell
  methods}}, \href{https://doi.org/10.1007/JHEP11(2021)221}{\emph{JHEP}
  {\bfseries 11} (2021) 221}
  [\href{https://arxiv.org/abs/2108.03669}{{\ttfamily 2108.03669}}].

\bibitem{Baratella:2021guc}
P.~Baratella, D.~Haslehner, M.~Ruhdorfer, J.~Serra and A.~Weiler, \emph{{RG of
  GR from On-shell Amplitudes}},
  \href{https://arxiv.org/abs/2109.06191}{{\ttfamily 2109.06191}}.

\bibitem{EliasMiro:2021jgu}
J.~Elias~Miro, C.~Fernandez, M.A.~Gumus and A.~Pomarol, \emph{{Gearing up for
  the next generation of LFV experiments, via on-shell methods}},
  \href{https://arxiv.org/abs/2112.12131}{{\ttfamily 2112.12131}}.

\bibitem{Cheung:2015aba}
C.~Cheung and C.-H.~Shen, \emph{{Nonrenormalization Theorems without
  Supersymmetry}},
  \href{https://doi.org/10.1103/PhysRevLett.115.071601}{\emph{Phys. Rev. Lett.}
  {\bfseries 115} (2015) 071601}
  [\href{https://arxiv.org/abs/1505.01844}{{\ttfamily 1505.01844}}].

\bibitem{Bern:2019wie}
Z.~Bern, J.~Parra-Martinez and E.~Sawyer, \emph{{Nonrenormalization and
  Operator Mixing via On-Shell Methods}},
  \href{https://doi.org/10.1103/PhysRevLett.124.051601}{\emph{Phys. Rev. Lett.}
  {\bfseries 124} (2020) 051601}
  [\href{https://arxiv.org/abs/1910.05831}{{\ttfamily 1910.05831}}].

\bibitem{Jiang:2020rwz}
M.~Jiang, J.~Shu, M.-L.~Xiao and Y.-H.~Zheng, \emph{{Partial Wave Amplitude
  Basis and Selection Rules in Effective Field Theories}},
  \href{https://doi.org/10.1103/PhysRevLett.126.011601}{\emph{Phys. Rev. Lett.}
  {\bfseries 126} (2021) 011601}
  [\href{https://arxiv.org/abs/2001.04481}{{\ttfamily 2001.04481}}].

\bibitem{Rose:2022njd}
L.D.~Rose, B.~von Harling and A.~Pomarol, \emph{{Wilson Coefficients and
  Natural Zeros from the On-Shell Viewpoint}},
  \href{https://arxiv.org/abs/2201.10572}{{\ttfamily 2201.10572}}.

\bibitem{Azatov:2016sqh}
A.~Azatov, R.~Contino, C.S.~Machado and F.~Riva, \emph{{Helicity selection
  rules and noninterference for BSM amplitudes}},
  \href{https://doi.org/10.1103/PhysRevD.95.065014}{\emph{Phys. Rev. D}
  {\bfseries 95} (2017) 065014}
  [\href{https://arxiv.org/abs/1607.05236}{{\ttfamily 1607.05236}}].

\bibitem{Neill:2013wsa}
D.~Neill and I.Z.~Rothstein, \emph{{Classical Space-Times from the S Matrix}},
  \href{https://doi.org/10.1016/j.nuclphysb.2013.09.007}{\emph{Nucl. Phys.}
  {\bfseries B877} (2013) 177}
  [\href{https://arxiv.org/abs/1304.7263}{{\ttfamily 1304.7263}}].

\bibitem{Bjerrum-Bohr:2013bxa}
N.E.J.~Bjerrum-Bohr, J.F.~Donoghue and P.~Vanhove, \emph{{On-shell Techniques
  and Universal Results in Quantum Gravity}},
  \href{https://doi.org/10.1007/JHEP02(2014)111}{\emph{JHEP} {\bfseries 02}
  (2014) 111} [\href{https://arxiv.org/abs/1309.0804}{{\ttfamily 1309.0804}}].

\bibitem{Cachazo:2017jef}
F.~Cachazo and A.~Guevara, \emph{{Leading Singularities and Classical
  Gravitational Scattering}},
  \href{https://doi.org/10.1007/JHEP02(2020)181}{\emph{JHEP} {\bfseries 02}
  (2020) 181} [\href{https://arxiv.org/abs/1705.10262}{{\ttfamily
  1705.10262}}].

\bibitem{Cheung:2018wkq}
C.~Cheung, I.Z.~Rothstein and M.P.~Solon, \emph{{From Scattering Amplitudes to
  Classical Potentials in the Post-Minkowskian Expansion}},
  \href{https://doi.org/10.1103/PhysRevLett.121.251101}{\emph{Phys. Rev. Lett.}
  {\bfseries 121} (2018) 251101}
  [\href{https://arxiv.org/abs/1808.02489}{{\ttfamily 1808.02489}}].

\bibitem{Kosower:2018adc}
D.A.~Kosower, B.~Maybee and D.~O'Connell, \emph{{Amplitudes, Observables, and
  Classical Scattering}},
  \href{https://doi.org/10.1007/JHEP02(2019)137}{\emph{JHEP} {\bfseries 02}
  (2019) 137} [\href{https://arxiv.org/abs/1811.10950}{{\ttfamily
  1811.10950}}].

\bibitem{Bern:2019nnu}
Z.~Bern, C.~Cheung, R.~Roiban, C.-H.~Shen, M.P.~Solon and M.~Zeng,
  \emph{{Scattering Amplitudes and the Conservative Hamiltonian for Binary
  Systems at Third Post-Minkowskian Order}},
  \href{https://doi.org/10.1103/PhysRevLett.122.201603}{\emph{Phys. Rev. Lett.}
  {\bfseries 122} (2019) 201603}
  [\href{https://arxiv.org/abs/1901.04424}{{\ttfamily 1901.04424}}].

\bibitem{Bern:2019crd}
Z.~Bern, C.~Cheung, R.~Roiban, C.-H.~Shen, M.P.~Solon and M.~Zeng, \emph{{Black
  Hole Binary Dynamics from the Double Copy and Effective Theory}},
  \href{https://doi.org/10.1007/JHEP10(2019)206}{\emph{JHEP} {\bfseries 10}
  (2019) 206} [\href{https://arxiv.org/abs/1908.01493}{{\ttfamily
  1908.01493}}].

\bibitem{Parra-Martinez:2020dzs}
J.~Parra-Martinez, M.S.~Ruf and M.~Zeng, \emph{{Extremal black hole scattering
  at $\mathcal{O}(G^3)$: graviton dominance, eikonal exponentiation, and
  differential equations}},
  \href{https://doi.org/10.1007/JHEP11(2020)023}{\emph{JHEP} {\bfseries 11}
  (2020) 023} [\href{https://arxiv.org/abs/2005.04236}{{\ttfamily
  2005.04236}}].

\bibitem{Bjerrum-Bohr:2021din}
N.E.J.~Bjerrum-Bohr, P.H.~Damgaard, L.~Plant\'e and P.~Vanhove, \emph{{The
  amplitude for classical gravitational scattering at third Post-Minkowskian
  order}}, \href{https://doi.org/10.1007/JHEP08(2021)172}{\emph{JHEP}
  {\bfseries 08} (2021) 172}
  [\href{https://arxiv.org/abs/2105.05218}{{\ttfamily 2105.05218}}].

\bibitem{Brandhuber:2021eyq}
A.~Brandhuber, G.~Chen, G.~Travaglini and C.~Wen, \emph{{Classical
  gravitational scattering from a gauge-invariant double copy}},
  \href{https://doi.org/10.1007/JHEP10(2021)118}{\emph{JHEP} {\bfseries 10}
  (2021) 118} [\href{https://arxiv.org/abs/2108.04216}{{\ttfamily
  2108.04216}}].

\bibitem{Bern:2021dqo}
Z.~Bern, J.~Parra-Martinez, R.~Roiban, M.S.~Ruf, C.-H.~Shen, M.P.~Solon et~al.,
  \emph{{Scattering Amplitudes and Conservative Binary Dynamics at ${\cal
  O}(G^4)$}}, \href{https://doi.org/10.1103/PhysRevLett.126.171601}{\emph{Phys.
  Rev. Lett.} {\bfseries 126} (2021) 171601}
  [\href{https://arxiv.org/abs/2101.07254}{{\ttfamily 2101.07254}}].

\bibitem{Guevara:2017csg}
A.~Guevara, \emph{{Holomorphic Classical Limit for Spin Effects in
  Gravitational and Electromagnetic Scattering}},
  \href{https://doi.org/10.1007/JHEP04(2019)033}{\emph{JHEP} {\bfseries 04}
  (2019) 033} [\href{https://arxiv.org/abs/1706.02314}{{\ttfamily
  1706.02314}}].

\bibitem{Chung:2019duq}
M.-Z.~Chung, Y.-T.~Huang and J.-W.~Kim, \emph{{Classical potential for general
  spinning bodies}}, \href{https://doi.org/10.1007/JHEP09(2020)074}{\emph{JHEP}
  {\bfseries 09} (2020) 074}
  [\href{https://arxiv.org/abs/1908.08463}{{\ttfamily 1908.08463}}].

\bibitem{Maybee:2019jus}
B.~Maybee, D.~O'Connell and J.~Vines, \emph{{Observables and amplitudes for
  spinning particles and black holes}},
  \href{https://doi.org/10.1007/JHEP12(2019)156}{\emph{JHEP} {\bfseries 12}
  (2019) 156} [\href{https://arxiv.org/abs/1906.09260}{{\ttfamily
  1906.09260}}].

\bibitem{Bern:2020buy}
Z.~Bern, A.~Luna, R.~Roiban, C.-H.~Shen and M.~Zeng, \emph{{Spinning black hole
  binary dynamics, scattering amplitudes, and effective field theory}},
  \href{https://doi.org/10.1103/PhysRevD.104.065014}{\emph{Phys. Rev. D}
  {\bfseries 104} (2021) 065014}
  [\href{https://arxiv.org/abs/2005.03071}{{\ttfamily 2005.03071}}].

\bibitem{Kosmopoulos:2021zoq}
D.~Kosmopoulos and A.~Luna, \emph{{Quadratic-in-spin Hamiltonian at $
  \mathcal{O} $(G$^{2}$) from scattering amplitudes}},
  \href{https://doi.org/10.1007/JHEP07(2021)037}{\emph{JHEP} {\bfseries 07}
  (2021) 037} [\href{https://arxiv.org/abs/2102.10137}{{\ttfamily
  2102.10137}}].

\bibitem{Chiodaroli:2021eug}
M.~Chiodaroli, H.~Johansson and P.~Pichini, \emph{{Compton Black-Hole
  Scattering for $s \leq 5/2$}},
  \href{https://arxiv.org/abs/2107.14779}{{\ttfamily 2107.14779}}.

\bibitem{Aoude:2021oqj}
R.~Aoude and A.~Ochirov, \emph{{Classical observables from coherent-spin
  amplitudes}}, \href{https://doi.org/10.1007/JHEP10(2021)008}{\emph{JHEP}
  {\bfseries 10} (2021) 008}
  [\href{https://arxiv.org/abs/2108.01649}{{\ttfamily 2108.01649}}].

\bibitem{Haddad:2021znf}
K.~Haddad, \emph{{Exponentiation of the leading eikonal phase with spin}},
  \href{https://doi.org/10.1103/PhysRevD.105.026004}{\emph{Phys. Rev. D}
  {\bfseries 105} (2022) 026004}
  [\href{https://arxiv.org/abs/2109.04427}{{\ttfamily 2109.04427}}].

\bibitem{Cheung:2020sdj}
C.~Cheung and M.P.~Solon, \emph{{Tidal Effects in the Post-Minkowskian
  Expansion}},
  \href{https://doi.org/10.1103/PhysRevLett.125.191601}{\emph{Phys. Rev. Lett.}
  {\bfseries 125} (2020) 191601}
  [\href{https://arxiv.org/abs/2006.06665}{{\ttfamily 2006.06665}}].

\bibitem{Haddad:2020que}
K.~Haddad and A.~Helset, \emph{{Tidal effects in quantum field theory}},
  \href{https://doi.org/10.1007/JHEP12(2020)024}{\emph{JHEP} {\bfseries 12}
  (2020) 024} [\href{https://arxiv.org/abs/2008.04920}{{\ttfamily
  2008.04920}}].

\bibitem{Bern:2020uwk}
Z.~Bern, J.~Parra-Martinez, R.~Roiban, E.~Sawyer and C.-H.~Shen, \emph{{Leading
  Nonlinear Tidal Effects and Scattering Amplitudes}},
  \href{https://doi.org/10.1007/JHEP05(2021)188}{\emph{JHEP} {\bfseries 05}
  (2021) 188} [\href{https://arxiv.org/abs/2010.08559}{{\ttfamily
  2010.08559}}].

\bibitem{Kim:2020dif}
J.-W.~Kim and M.~Shim, \emph{{Quantum corrections to tidal Love number for
  Schwarzschild black holes}},
  \href{https://doi.org/10.1103/PhysRevD.104.046022}{\emph{Phys. Rev. D}
  {\bfseries 104} (2021) 046022}
  [\href{https://arxiv.org/abs/2011.03337}{{\ttfamily 2011.03337}}].

\bibitem{Aoude:2020ygw}
R.~Aoude, K.~Haddad and A.~Helset, \emph{{Tidal effects for spinning
  particles}}, \href{https://doi.org/10.1007/JHEP03(2021)097}{\emph{JHEP}
  {\bfseries 03} (2021) 097}
  [\href{https://arxiv.org/abs/2012.05256}{{\ttfamily 2012.05256}}].

\bibitem{Brandhuber:2019qpg}
A.~Brandhuber and G.~Travaglini, \emph{{On higher-derivative effects on the
  gravitational potential and particle bending}},
  \href{https://doi.org/10.1007/JHEP01(2020)010}{\emph{JHEP} {\bfseries 01}
  (2020) 010} [\href{https://arxiv.org/abs/1905.05657}{{\ttfamily
  1905.05657}}].

\bibitem{AccettulliHuber:2020oou}
M.~Accettulli~Huber, A.~Brandhuber, S.~De~Angelis and G.~Travaglini,
  \emph{{Eikonal phase matrix, deflection angle and time delay in effective
  field theories of gravity}},
  \href{https://doi.org/10.1103/PhysRevD.102.046014}{\emph{Phys. Rev. D}
  {\bfseries 102} (2020) 046014}
  [\href{https://arxiv.org/abs/2006.02375}{{\ttfamily 2006.02375}}].

\bibitem{Grzadkowski:2010es}
B.~Grzadkowski, M.~Iskrzynski, M.~Misiak and J.~Rosiek, \emph{{Dimension-Six
  Terms in the Standard Model Lagrangian}},
  \href{https://doi.org/10.1007/JHEP10(2010)085}{\emph{JHEP} {\bfseries 10}
  (2010) 085} [\href{https://arxiv.org/abs/1008.4884}{{\ttfamily 1008.4884}}].

\bibitem{Lehman:2015via}
L.~Lehman and A.~Martin, \emph{{Hilbert Series for Constructing Lagrangians:
  expanding the phenomenologist's toolbox}},
  \href{https://doi.org/10.1103/PhysRevD.91.105014}{\emph{Phys. Rev. D}
  {\bfseries 91} (2015) 105014}
  [\href{https://arxiv.org/abs/1503.07537}{{\ttfamily 1503.07537}}].

\bibitem{Henning:2015alf}
B.~Henning, X.~Lu, T.~Melia and H.~Murayama, \emph{{2, 84, 30, 993, 560, 15456,
  11962, 261485, ...: Higher dimension operators in the SM EFT}},
  \href{https://doi.org/10.1007/JHEP08(2017)016}{\emph{JHEP} {\bfseries 08}
  (2017) 016} [\href{https://arxiv.org/abs/1512.03433}{{\ttfamily
  1512.03433}}].

\bibitem{Henning:2015daa}
B.~Henning, X.~Lu, T.~Melia and H.~Murayama, \emph{{Hilbert series and operator
  bases with derivatives in effective field theories}},
  \href{https://doi.org/10.1007/s00220-015-2518-2}{\emph{Commun. Math. Phys.}
  {\bfseries 347} (2016) 363}
  [\href{https://arxiv.org/abs/1507.07240}{{\ttfamily 1507.07240}}].

\bibitem{Henning:2017fpj}
B.~Henning, X.~Lu, T.~Melia and H.~Murayama, \emph{{Operator bases,
  $S$-matrices, and their partition functions}},
  \href{https://doi.org/10.1007/JHEP10(2017)199}{\emph{JHEP} {\bfseries 10}
  (2017) 199} [\href{https://arxiv.org/abs/1706.08520}{{\ttfamily
  1706.08520}}].

\bibitem{Cheung:2016drk}
C.~Cheung, K.~Kampf, J.~Novotny, C.-H.~Shen and J.~Trnka, \emph{{A Periodic
  Table of Effective Field Theories}},
  \href{https://doi.org/10.1007/JHEP02(2017)020}{\emph{JHEP} {\bfseries 02}
  (2017) 020} [\href{https://arxiv.org/abs/1611.03137}{{\ttfamily
  1611.03137}}].

\bibitem{Shadmi:2018xan}
Y.~Shadmi and Y.~Weiss, \emph{{Effective Field Theory Amplitudes the On-Shell
  Way: Scalar and Vector Couplings to Gluons}},
  \href{https://doi.org/10.1007/JHEP02(2019)165}{\emph{JHEP} {\bfseries 02}
  (2019) 165} [\href{https://arxiv.org/abs/1809.09644}{{\ttfamily
  1809.09644}}].

\bibitem{Britto:2005fq}
R.~Britto, F.~Cachazo, B.~Feng and E.~Witten, \emph{{Direct proof of tree-level
  recursion relation in Yang-Mills theory}},
  \href{https://doi.org/10.1103/PhysRevLett.94.181602}{\emph{Phys. Rev. Lett.}
  {\bfseries 94} (2005) 181602}
  [\href{https://arxiv.org/abs/hep-th/0501052}{{\ttfamily hep-th/0501052}}].

\bibitem{Risager:2005vk}
K.~Risager, \emph{{A Direct proof of the CSW rules}},
  \href{https://doi.org/10.1088/1126-6708/2005/12/003}{\emph{JHEP} {\bfseries
  12} (2005) 003} [\href{https://arxiv.org/abs/hep-th/0508206}{{\ttfamily
  hep-th/0508206}}].

\bibitem{Arkani-Hamed:2008bsc}
N.~Arkani-Hamed and J.~Kaplan, \emph{{On Tree Amplitudes in Gauge Theory and
  Gravity}}, \href{https://doi.org/10.1088/1126-6708/2008/04/076}{\emph{JHEP}
  {\bfseries 04} (2008) 076} [\href{https://arxiv.org/abs/0801.2385}{{\ttfamily
  0801.2385}}].

\bibitem{Cohen:2010mi}
T.~Cohen, H.~Elvang and M.~Kiermaier, \emph{{On-shell constructibility of tree
  amplitudes in general field theories}},
  \href{https://doi.org/10.1007/JHEP04(2011)053}{\emph{JHEP} {\bfseries 04}
  (2011) 053} [\href{https://arxiv.org/abs/1010.0257}{{\ttfamily 1010.0257}}].

\bibitem{Cheung:2015ota}
C.~Cheung, K.~Kampf, J.~Novotny, C.-H.~Shen and J.~Trnka, \emph{{On-Shell
  Recursion Relations for Effective Field Theories}},
  \href{https://doi.org/10.1103/PhysRevLett.116.041601}{\emph{Phys. Rev. Lett.}
  {\bfseries 116} (2016) 041601}
  [\href{https://arxiv.org/abs/1509.03309}{{\ttfamily 1509.03309}}].

\bibitem{Falkowski:2020aso}
A.~Falkowski and C.S.~Machado, \emph{{Soft Matters, or the Recursions with
  Massive Spinors}}, \href{https://doi.org/10.1007/JHEP05(2021)238}{\emph{JHEP}
  {\bfseries 05} (2021) 238}
  [\href{https://arxiv.org/abs/2005.08981}{{\ttfamily 2005.08981}}].

\bibitem{Bern:1994cg}
Z.~Bern, L.J.~Dixon, D.C.~Dunbar and D.A.~Kosower, \emph{{Fusing gauge theory
  tree amplitudes into loop amplitudes}},
  \href{https://doi.org/10.1016/0550-3213(94)00488-Z}{\emph{Nucl. Phys.}
  {\bfseries B435} (1995) 59}
  [\href{https://arxiv.org/abs/hep-ph/9409265}{{\ttfamily hep-ph/9409265}}].

\bibitem{Bern:1994zx}
Z.~Bern, L.J.~Dixon, D.C.~Dunbar and D.A.~Kosower, \emph{{One loop $n$-point
  gauge theory amplitudes, unitarity and collinear limits}},
  \href{https://doi.org/10.1016/0550-3213(94)90179-1}{\emph{Nucl. Phys.}
  {\bfseries B425} (1994) 217}
  [\href{https://arxiv.org/abs/hep-ph/9403226}{{\ttfamily hep-ph/9403226}}].

\bibitem{Britto:2004nc}
R.~Britto, F.~Cachazo and B.~Feng, \emph{{Generalized unitarity and one-loop
  amplitudes in N=4 super-Yang-Mills}},
  \href{https://doi.org/10.1016/j.nuclphysb.2005.07.014}{\emph{Nucl. Phys.}
  {\bfseries B725} (2005) 275}
  [\href{https://arxiv.org/abs/hep-th/0412103}{{\ttfamily hep-th/0412103}}].

\bibitem{Bern:2004cz}
Z.~Bern, L.J.~Dixon and D.A.~Kosower, \emph{{Two-loop g ---\ensuremath{>} gg
  splitting amplitudes in QCD}},
  \href{https://doi.org/10.1088/1126-6708/2004/08/012}{\emph{JHEP} {\bfseries
  08} (2004) 012} [\href{https://arxiv.org/abs/hep-ph/0404293}{{\ttfamily
  hep-ph/0404293}}].

\bibitem{Mastrolia:2008jb}
P.~Mastrolia, G.~Ossola, C.G.~Papadopoulos and R.~Pittau, \emph{{Optimizing the
  Reduction of One-Loop Amplitudes}},
  \href{https://doi.org/10.1088/1126-6708/2008/06/030}{\emph{JHEP} {\bfseries
  06} (2008) 030} [\href{https://arxiv.org/abs/0803.3964}{{\ttfamily
  0803.3964}}].

\bibitem{Forde:2007mi}
D.~Forde, \emph{{Direct extraction of one-loop integral coefficients}},
  \href{https://doi.org/10.1103/PhysRevD.75.125019}{\emph{Phys. Rev.}
  {\bfseries D75} (2007) 125019}
  [\href{https://arxiv.org/abs/0704.1835}{{\ttfamily 0704.1835}}].

\bibitem{Badger:2008cm}
S.D.~Badger, \emph{{Direct Extraction Of One Loop Rational Terms}},
  \href{https://doi.org/10.1088/1126-6708/2009/01/049}{\emph{JHEP} {\bfseries
  01} (2009) 049} [\href{https://arxiv.org/abs/0806.4600}{{\ttfamily
  0806.4600}}].

\bibitem{DeCausmaecker:1981jtq}
P.~De~Causmaecker, R.~Gastmans, W.~Troost and T.T.~Wu, \emph{{Multiple
  Bremsstrahlung in Gauge Theories at High-Energies. 1. General Formalism for
  Quantum Electrodynamics}},
  \href{https://doi.org/10.1016/0550-3213(82)90488-6}{\emph{Nucl. Phys. B}
  {\bfseries 206} (1982) 53}.

\bibitem{Berends:1981uq}
F.A.~Berends, R.~Kleiss, P.~De~Causmaecker, R.~Gastmans, W.~Troost and T.T.~Wu,
  \emph{{Multiple Bremsstrahlung in Gauge Theories at High-Energies. 2. Single
  Bremsstrahlung}},
  \href{https://doi.org/10.1016/0550-3213(82)90489-8}{\emph{Nucl. Phys. B}
  {\bfseries 206} (1982) 61}.

\bibitem{Kleiss:1985yh}
R.~Kleiss and W.J.~Stirling, \emph{{Spinor Techniques for Calculating p anti-p
  ---\ensuremath{>} W+- / Z0 + Jets}},
  \href{https://doi.org/10.1016/0550-3213(85)90285-8}{\emph{Nucl. Phys. B}
  {\bfseries 262} (1985) 235}.

\bibitem{Xu:1986xb}
Z.~Xu, D.-H.~Zhang and L.~Chang, \emph{{Helicity Amplitudes for Multiple
  Bremsstrahlung in Massless Nonabelian Gauge Theories}},
  \href{https://doi.org/10.1016/0550-3213(87)90479-2}{\emph{Nucl. Phys. B}
  {\bfseries 291} (1987) 392}.

\bibitem{Arkani-Hamed:2017jhn}
N.~Arkani-Hamed, T.-C.~Huang and Y.-t.~Huang, \emph{{Scattering amplitudes for
  all masses and spins}},
  \href{https://doi.org/10.1007/JHEP11(2021)070}{\emph{JHEP} {\bfseries 11}
  (2021) 070} [\href{https://arxiv.org/abs/1709.04891}{{\ttfamily
  1709.04891}}].

\bibitem{Ma:2019gtx}
T.~Ma, J.~Shu and M.-L.~Xiao, \emph{{Standard Model Effective Field Theory from
  On-shell Amplitudes}},  \href{https://arxiv.org/abs/1902.06752}{{\ttfamily
  1902.06752}}.

\bibitem{Aoude:2019tzn}
R.~Aoude and C.S.~Machado, \emph{{The Rise of SMEFT On-shell Amplitudes}},
  \href{https://doi.org/10.1007/JHEP12(2019)058}{\emph{JHEP} {\bfseries 12}
  (2019) 058} [\href{https://arxiv.org/abs/1905.11433}{{\ttfamily
  1905.11433}}].

\bibitem{Durieux:2019eor}
G.~Durieux, T.~Kitahara, Y.~Shadmi and Y.~Weiss, \emph{{The electroweak
  effective field theory from on-shell amplitudes}},
  \href{https://doi.org/10.1007/JHEP01(2020)119}{\emph{JHEP} {\bfseries 01}
  (2020) 119} [\href{https://arxiv.org/abs/1909.10551}{{\ttfamily
  1909.10551}}].

\bibitem{Chowdhury:2019kaq}
S.D.~Chowdhury, A.~Gadde, T.~Gopalka, I.~Halder, L.~Janagal and S.~Minwalla,
  \emph{{Classifying and constraining local four photon and four graviton
  S-matrices}}, \href{https://doi.org/10.1007/JHEP02(2020)114}{\emph{JHEP}
  {\bfseries 02} (2020) 114}
  [\href{https://arxiv.org/abs/1910.14392}{{\ttfamily 1910.14392}}].

\bibitem{Falkowski:2019zdo}
A.~Falkowski, \emph{{Bases of massless EFTs via momentum twistors}},
  \href{https://arxiv.org/abs/1912.07865}{{\ttfamily 1912.07865}}.

\bibitem{Durieux:2019siw}
G.~Durieux and C.S.~Machado, \emph{{Enumerating higher-dimensional operators
  with on-shell amplitudes}},
  \href{https://doi.org/10.1103/PhysRevD.101.095021}{\emph{Phys. Rev. D}
  {\bfseries 101} (2020) 095021}
  [\href{https://arxiv.org/abs/1912.08827}{{\ttfamily 1912.08827}}].

\bibitem{Chakraborty:2020rxf}
S.~Chakraborty, S.D.~Chowdhury, T.~Gopalka, S.~Kundu, S.~Minwalla and
  A.~Mishra, \emph{{Classification of all 3 particle S-matrices quadratic in
  photons or gravitons}},
  \href{https://doi.org/10.1007/JHEP04(2020)110}{\emph{JHEP} {\bfseries 04}
  (2020) 110} [\href{https://arxiv.org/abs/2001.07117}{{\ttfamily
  2001.07117}}].

\bibitem{Li:2020gnx}
H.-L.~Li, Z.~Ren, J.~Shu, M.-L.~Xiao, J.-H.~Yu and Y.-H.~Zheng, \emph{{Complete
  set of dimension-eight operators in the standard model effective field
  theory}}, \href{https://doi.org/10.1103/PhysRevD.104.015026}{\emph{Phys. Rev.
  D} {\bfseries 104} (2021) 015026}
  [\href{https://arxiv.org/abs/2005.00008}{{\ttfamily 2005.00008}}].

\bibitem{Li:2020xlh}
H.-L.~Li, Z.~Ren, M.-L.~Xiao, J.-H.~Yu and Y.-H.~Zheng, \emph{{Complete Set of
  Dimension-9 Operators in the Standard Model Effective Field Theory}},
  \href{https://arxiv.org/abs/2007.07899}{{\ttfamily 2007.07899}}.

\bibitem{Durieux:2020gip}
G.~Durieux, T.~Kitahara, C.S.~Machado, Y.~Shadmi and Y.~Weiss,
  \emph{{Constructing massive on-shell contact terms}},
  \href{https://doi.org/10.1007/JHEP12(2020)175}{\emph{JHEP} {\bfseries 12}
  (2020) 175} [\href{https://arxiv.org/abs/2008.09652}{{\ttfamily
  2008.09652}}].

\bibitem{Falkowski:2020fsu}
A.~Falkowski, G.~Isabella and C.S.~Machado, \emph{{On-shell effective theory
  for higher-spin dark matter}},
  \href{https://doi.org/10.21468/SciPostPhys.10.5.101}{\emph{SciPost Phys.}
  {\bfseries 10} (2021) 101}
  [\href{https://arxiv.org/abs/2011.05339}{{\ttfamily 2011.05339}}].

\bibitem{Li:2022tec}
H.-L.~Li, Z.~Ren, M.-L.~Xiao, J.-H.~Yu and Y.-H.~Zheng, \emph{{Operators For
  Generic Effective Field Theory at any Dimension: On-shell Amplitude Basis
  Construction}},  \href{https://arxiv.org/abs/2201.04639}{{\ttfamily
  2201.04639}}.

\bibitem{Balkin:2021dko}
R.~Balkin, G.~Durieux, T.~Kitahara, Y.~Shadmi and Y.~Weiss, \emph{{On-shell
  Higgsing for EFTs}},  \href{https://arxiv.org/abs/2112.09688}{{\ttfamily
  2112.09688}}.

\bibitem{Dong:2021yak}
Z.-Y.~Dong, T.~Ma and J.~Shu, \emph{{Constructing on-shell operator basis for
  all masses and spins}},  \href{https://arxiv.org/abs/2103.15837}{{\ttfamily
  2103.15837}}.

\bibitem{Benincasa:2007xk}
P.~Benincasa and F.~Cachazo, \emph{{Consistency Conditions on the S-Matrix of
  Massless Particles}},  \href{https://arxiv.org/abs/0705.4305}{{\ttfamily
  0705.4305}}.

\bibitem{Costa:2011mg}
M.S.~Costa, J.~Penedones, D.~Poland and S.~Rychkov, \emph{{Spinning Conformal
  Correlators}}, \href{https://doi.org/10.1007/JHEP11(2011)071}{\emph{JHEP}
  {\bfseries 11} (2011) 071} [\href{https://arxiv.org/abs/1107.3554}{{\ttfamily
  1107.3554}}].

\bibitem{Fonseca:2019yya}
R.M.~Fonseca, \emph{{Enumerating the operators of an effective field theory}},
  \href{https://doi.org/10.1103/PhysRevD.101.035040}{\emph{Phys. Rev. D}
  {\bfseries 101} (2020) 035040}
  [\href{https://arxiv.org/abs/1907.12584}{{\ttfamily 1907.12584}}].

\bibitem{Chowdhury:2020ddc}
S.D.~Chowdhury and A.~Gadde, \emph{{Classification of four-point local gluon
  S-matrices}}, \href{https://doi.org/10.1007/JHEP01(2021)104}{\emph{JHEP}
  {\bfseries 01} (2021) 104}
  [\href{https://arxiv.org/abs/2006.12458}{{\ttfamily 2006.12458}}].

\bibitem{Dong:2022mcv}
Z.-Y.~Dong, T.~Ma, J.~Shu and Y.-H.~Zheng, \emph{{Constructing Generic
  Effective Field Theory for All Masses and Spins}},
  \href{https://arxiv.org/abs/2202.08350}{{\ttfamily 2202.08350}}.

\bibitem{vonManteuffel:2014ixa}
A.~von Manteuffel and R.M.~Schabinger, \emph{{A novel approach to integration
  by parts reduction}},
  \href{https://doi.org/10.1016/j.physletb.2015.03.029}{\emph{Phys. Lett. B}
  {\bfseries 744} (2015) 101}
  [\href{https://arxiv.org/abs/1406.4513}{{\ttfamily 1406.4513}}].

\bibitem{Peraro:2016wsq}
T.~Peraro, \emph{{Scattering amplitudes over finite fields and multivariate
  functional reconstruction}},
  \href{https://doi.org/10.1007/JHEP12(2016)030}{\emph{JHEP} {\bfseries 12}
  (2016) 030} [\href{https://arxiv.org/abs/1608.01902}{{\ttfamily
  1608.01902}}].

\bibitem{Badger:2013gxa}
S.~Badger, H.~Frellesvig and Y.~Zhang, \emph{{A Two-Loop Five-Gluon Helicity
  Amplitude in QCD}},
  \href{https://doi.org/10.1007/JHEP12(2013)045}{\emph{JHEP} {\bfseries 12}
  (2013) 045} [\href{https://arxiv.org/abs/1310.1051}{{\ttfamily 1310.1051}}].

\bibitem{Badger:2016uuq}
S.~Badger, \emph{{Automating QCD amplitudes with on-shell methods}},
  \href{https://doi.org/10.1088/1742-6596/762/1/012057}{\emph{J. Phys. Conf.
  Ser.} {\bfseries 762} (2016) 012057}
  [\href{https://arxiv.org/abs/1605.02172}{{\ttfamily 1605.02172}}].

\bibitem{Hodges:2009hk}
A.~Hodges, \emph{{Eliminating spurious poles from gauge-theoretic amplitudes}},
  \href{https://doi.org/10.1007/JHEP05(2013)135}{\emph{JHEP} {\bfseries 05}
  (2013) 135} [\href{https://arxiv.org/abs/0905.1473}{{\ttfamily 0905.1473}}].

\end{thebibliography}\endgroup

\end{document}